%% file: survey.tex
\documentclass[journal]{IEEEtran}
\input{packages.tex}

\begin{document}
%
\title{Survey and Systematization of Secure Device Pairing}
%
%
%

\author{Mikhail~Fomichev,
        Flor~\'{A}lvarez,
        Daniel~Steinmetzer,
        Paul~Gardner-Stephen,
        and~Matthias~Hollick 
\thanks{M. Fomichev, F. \'{A}lvarez, D. Steinmetzer and M. Hollick are with the Department
of Computer Science, Technische Universit{\"a}t Darmstadt, Darmstadt,
Germany e-mail: (\{mfomichev, falvarez, dsteinmetzer, mhollick\}@seemoo.tu-darmstadt.de).}
\thanks{P. Gardner-Stephen is with the School of Computer Science, Engineering and Mathematics, 
Flinders University, Adelaide, Australia e-mail: (paul.gardner-stephen@flinders.edu.au).}
\thanks{\copyright{} 2017 IEEE. Personal use of this material is permitted. Permission from IEEE must be obtained for all other uses, in any current or future media, including reprinting/republishing this material for advertising or promotional purposes, creating new collective works, for resale or redistribution to servers or lists, or reuse of any copyrighted component of this work in other works. The official version can be found at \url{http://dx.doi.org/10.1109/COMST.2017.2748278}}
}

\maketitle

\begin{abstract}
\input{section/abstract.tex}
\end{abstract}

\glsresetall


%
\IEEEpeerreviewmaketitle

\input{section/introduction.tex}
\input{section/related_work.tex}

\input{section/background.tex}
\input{section/physical_channels.tex}
\input{section/hci_channels.tex}
\input{section/application_classes.tex}
\input{section/challenges.tex}
\input{section/conclusion.tex}

\input{section/acknowledgment.tex}




%
\bibliographystyle{IEEEtran}
\bibliography{References}

\end{document}

%% file: packages.tex
\usepackage{cite}
\usepackage{csquotes}

\usepackage{outlines}

\ifCLASSINFOpdf
   \usepackage[pdftex]{graphicx}
   \graphicspath{{../pdf/}{../jpeg/}}
   \DeclareGraphicsExtensions{.pdf,.jpeg,.png}
\else
   \usepackage[dvips]{graphicx}
   \graphicspath{{../eps/}}
   \DeclareGraphicsExtensions{.eps}
\fi
\ifCLASSOPTIONcompsoc
  \usepackage[caption=false,font=normalsize,labelfont=sf,textfont=sf]{subfig}
  \usepackage{subcaption}
\else
  \usepackage[caption=false,font=footnotesize]{subfig}
\usepackage[acronym, nowarn]{glossaries}
\makeglossaries
\input{abbreviations.tex}

\usepackage{todonotes}
\usepackage{tabularx}
\usepackage[flushleft]{threeparttable}
\usepackage{array,booktabs,ragged2e}

\usepackage{tikz}
\usetikzlibrary{calc, positioning, arrows, shapes, shadows, trees, arrows.meta, 
decorations.pathreplacing, hobby, backgrounds} 
\usepackage{adjustbox}
\usepackage{epstopdf}
\usepackage{multirow}
\usepackage{wasysym}
\usepackage{gensymb}
\usepackage{amssymb}
\usepackage{pifont}
\usepackage{enumitem}

\usepackage{amsthm}
\usepackage{hyperref}
\usepackage[hyphenbreaks]{breakurl}

\theoremstyle{definition}

\newcommand*\circled[1]{\tikz[baseline=(char.base)]{%
            \node[shape=circle,draw,inner sep=2pt] (char) {#1};}}
                      
\newcommand*\encircle[1]{\tikz[baseline=(char.base)]{%
            \node[shape=circle,draw,inner sep=0.5pt] (char) {#1};}}

\pgfdeclaredecoration{dashsoliddouble}{initial}{
  \state{initial}[width=\pgfdecoratedinputsegmentlength]{
    \pgfmathsetlengthmacro\lw{.3pt+.5\pgflinewidth}
    \begin{pgfscope}
      \pgfpathmoveto{\pgfpoint{0pt}{\lw}}%
      \pgfpathlineto{\pgfpoint{\pgfdecoratedinputsegmentlength}{\lw}}%
      \pgfmathtruncatemacro\dashnum{%
        round((\pgfdecoratedinputsegmentlength-3pt)/6pt)
      }
      \pgfmathsetmacro\dashscale{%
        \pgfdecoratedinputsegmentlength/(\dashnum*6pt + 3pt)
      }
      \pgfmathsetlengthmacro\dashunit{3pt*\dashscale}
      \pgfsetdash{{\dashunit}{\dashunit}}{0pt}
      \pgfusepath{stroke}
      \pgfsetdash{}{0pt}
      \pgfpathmoveto{\pgfpoint{0pt}{-\lw}}%
      \pgfpathlineto{\pgfpoint{\pgfdecoratedinputsegmentlength}{-\lw}}%
      \pgfusepath{stroke}
    \end{pgfscope}
  }
}  

\tikzset{
  -|-/.style={
    to path={
      (\tikztostart) -| ($(\tikztostart)!#1!(\tikztotarget)$) |- (\tikztotarget)
      \tikztonodes
    }
  },
  -|-/.default=0.5,
  |-|/.style={
    to path={
      (\tikztostart) |- ($(\tikztostart)!#1!(\tikztotarget)$) -| (\tikztotarget)
      \tikztonodes
    }
  },
  |-|/.default=0.5,
}

%% file: abbreviations.tex
\newacronym{sdp}{\textsc{SDP}}{Secure Device Pairing}
\newacronym{phy}{\textsc{PHY}}{Physical}
\newacronym{nfc}{\textsc{NFC}}{Near Field Communication}
\newacronym{oob}{\textsc{OoB}}{out-of-band}
\newacronym{iot}{\textsc{IoT}}{Internet of Things}
\newacronym{mitm}{\textsc{MITM}}{man-in-the-middle}
\newacronym{hci}{\textsc{HCI}}{Human-computer interaction}
\newacronym{los}{\textsc{LoS}}{line-of-sight}
\newacronym{lan}{\textsc{LAN}}{local area network}
\newacronym{ap}{\textsc{AP}}{access point}
\newacronym{dos}{\textsc{DoS}}{denial-of-service}
\newacronym{wps}{\textsc{WPS}}{Wi-Fi Protected Setup}
\newacronym{pbc}{\textsc{PBC}}{Push Button Configuration}
\newacronym{dh}{\textsc{DH}}{Diffie-Hellman}
\newacronym{tep}{\textsc{TEP}}{tamper-evident pairing}
\newacronym{tea}{\textsc{TEA}}{Tamper Evident Announcement}
\newacronym{rfid}{\textsc{RFID}}{Radio-frequency Identification}
\newacronym{vlc}{\textsc{VLC}}{Visible Light Communication}
\newacronym{wsn}{\textsc{WSN}}{wireless sensor networks}
\newacronym{irda}{\textsc{IrDA}}{Infrared Data Association}
\newacronym{imd}{\textsc{IMD}}{implantable medical devices}
\newacronym{rss}{\textsc{RSS}}{received signal strength}
\newacronym{zia}{\textsc{ZIA}}{zero-interaction authentication}
\newacronym{csi}{\textsc{CSI}}{Channel state information}
\newacronym{mana}{\textsc{MANA}}{Manual authentication}
\newacronym{mac}{\textsc{MAC}}{message authentication code}
\newacronym{sas}{\textsc{SAS}}{short authentication strings}
\newacronym{ecc}{\textsc{ECC}}{elliptic curve cryptography}
\newacronym{pfs}{\textsc{PFS}}{perfect forward secrecy}

%% file: section/abstract.tex
\gls{sdp} schemes have been developed to facilitate secure communications among 
smart devices, both personal mobile devices and \gls{iot} devices.  
Comparison and assessment of \gls{sdp} schemes is troublesome, because each scheme
makes different assumptions about out-of-band channels and adversary models,
and are driven by their particular use-cases.
A conceptual model that
facilitates meaningful comparison among \gls{sdp} schemes is missing.
We provide such a model.
In this article, we survey and analyze a wide range of \gls{sdp} schemes that
are described in the literature, including a number that have been adopted as
standards.
A system model and consistent terminology for \gls{sdp} schemes are built on the
foundation of this
survey, which are then used to classify existing \gls{sdp} schemes into a
taxonomy that, for the first time, enables their meaningful comparison and
analysis.
The existing \gls{sdp} schemes are analyzed using this model, revealing
common systemic security weaknesses among the surveyed \gls{sdp} schemes that
should become priority areas for future \gls{sdp} research, such as improving
the integration of privacy requirements into the design of \gls{sdp} schemes.
Our results allow \gls{sdp} scheme designers to create
schemes that are more easily comparable with one another, and to assist the
prevention of persisting the weaknesses common to the current generation of
\gls{sdp} schemes.

%% file: section/introduction.tex

\section{Introduction}
\label{sec:intro}
In recent years, the advances in automation \cite{Automation:2016} and a rapid
growth of the consumer electronics market \cite{Electronics:2015} have
resulted in a tremendous increase in the number of smart devices and personal
gadgets. For example, it is estimated that the number of interconnected
\gls{iot} devices used in a great variety of applications will reach 24
billion by 2020 \cite{Gubbi:2013}. Authenticating a plethora of devices in
such a dynamic setting to provide secure communications is a challenge that
has not yet been fully addressed \cite{Jing:2014}. This stems from the highly
distributed and diverse nature of the \gls{iot} environment which makes it
impractical to apply traditional approaches for establishing secure
communications such as Public Key Infrastructure \cite{Sicari:2015}.

\gls{sdp} was proposed as an approach to introduce security into the
ubiquitous computing environment where devices pair in an ad-hoc manner
\cite{Stajano:1999}. Specifically, two parties that have never met each other
and would like to bootstrap a secure communication channel need to perform key
exchange and authentication procedures \cite{Balfanz:2002}. The latter is
particularly difficult in the ad-hoc scenario, because two pairing devices do
not have any prior security context or a common point of trust
\cite{Kumar:2009}. This aspect, in addition to wireless nature of ubiquitous
computing makes device pairing vulnerable to \gls{mitm} attacks
\cite{Pathan:2010}. Traditionally, the \gls{mitm} threat was considered as one
of the core challenges in \gls{sdp} \cite{Mirzadeh:2014}. In order to overcome
\gls{mitm} attacks the use of auxiliary, so-called ``\gls{oob}" channels was
proposed \cite{Kumar:2009, Mirzadeh:2014}. Such channels aim to provide
authenticity and even confidentiality to ensure that pairing is performed only
between the intended devices, that is, no \gls{mitm} has intermediated. 

Since the initial idea was put forward \cite{Stajano:1999}, numerous pairing
schemes utilizing various \gls{oob} channels have been proposed both in
academia and ``in the wild''. A great variety of suggested pairing schemes
were studied and evaluated with respect to security \cite{Kainda:2009,
Mirzadeh:2014}, usability \cite{Kumar:2009, Kobsa:2009} and user interaction
\cite{Chong:2014}. The prior work on \gls{sdp} mainly considered two distinct
use-cases: \textit{a)} pairing among personal gadgets of a single user
\cite{Chong:2012}, and \textit{b)} pairing  devices of different users, for
example, smartphones \cite{Uzun:2011}.

With the advent of the \gls{iot}, \gls{sdp} has become one of the viable mechanisms 
to introduce security to this diverse and distributed environment \cite{Sadeghi:2015}. 
The applicability of \gls{sdp} to the \gls{iot} has already been demonstrated with different communication technologies, 
such as Wi-Fi and Bluetooth standardizing a number of various pairing schemes \cite{WPS:NFC, Bluetooth:les}. 
Since Wi-Fi and Bluetooth serve as the backbone of centralized and ad-hoc device communications in the \gls{iot}, 
this clearly indicates the utmost importance of \gls{sdp} in the \gls{iot}.
Additionally, recent research has clearly demonstrated that \gls{sdp} can be successfully applied to secure
an important class of the IoT devices such as wearables\cite{Miettinen:2014, Ahmed:2015, Wang:2016, Yang:2016, Li:2016, Liang:2017, Schurmann:2017}, for example, smartwatches, fitness trackers, smartglasses, etc.  
Thus, we are convinced that the role of \gls{sdp} as an ad-hoc security mechanism in the \gls{iot} 
is as important as its centralized counterpart, 
where a dedicated trusted server manages the key exchange and authentication procedures between the connected devices \cite{Sicari:2015}. 

The complex \gls{iot} domain compounds the complexity of \gls{sdp}, as well as
increasing ambiguity, because of two reasons. First, the available hardware
capabilities such as wireless radio interfaces, sensing functionality and
computational power vary significantly among different platforms
\cite{Grubert:2016}. Second, the interaction patterns among the devices as
well as between human operators and devices have become more intricate. For
instance, two devices can perform pairing without any human involvement at all
\cite{Varshavsky:2007, Schurmann:2013, Miettinen:2014, Zhao:2016} or a user device can
communicate with a third party device or infrastructure without explicit
consent \cite{Rostami:2013, Truong:2014}. These growing trends consider more
heterogeneous settings and user-device interactions. Thus, adversary models
and assumptions made for pairing here are different as compared to the
``classic pairing'' cases mentioned above. 

The sound comparison of various pairing schemes is not straightforward.
Several extensive surveys were conducted in order to analyze numerous pairing
schemes from different viewpoints \cite{Kumar:2009, Kobsa:2009, Kainda:2009,
Mirzadeh:2014, Chong:2014}. Interestingly, all those studies come to a common
conclusion: no universal pairing approach exists. Moreover, the comparative
analysis cannot be accurately aligned and justified even over a single metric,
for example, security or usability, since there is no common ground on what
information should be provided about a pairing scheme to make an assessment.
There are two reasons why such incompatibility occurs. First, available
hardware interfaces have been traditionally considered as one of the main
arguments for introducing yet another pairing scheme \cite{Soriente:2007,
Soriente:2008, Prasad:2008, Ahmed:2015}. From this starting point, the
selection of \gls{oob} channels as well as justification for user interaction
modes and use-case scenarios were made. Correspondingly, many proposed pairing
solutions focused on specific issues in a restricted setting and were rather
disconnected from the results of previous endeavors. Second, the study on user
perception of \gls{sdp} revealed that a choice of a particular pairing scheme
is context and environment dependent \cite{Ion:2010}. Hence, employing
different human-centric models widens the gap between properties deemed
relevant for pairing which causes controversy \cite{Ion:2010}. 

Two other issues that introduce disparity to the field relate to a concept of
\gls{oob} channel which is a cornerstone in \gls{sdp}. The first problem is a
lack of common understanding as to what constitutes an \gls{oob} channel.
There were dozens of different alternatives proposed
\cite{Kumar:2009,Chong:2014} which is a direct consequence of the
aforementioned design incentives behind many pairing schemes. Several attempts
\cite{Mirzadeh:2014, Nguyen:2014} to categorize \gls{oob} channels applied
mixed terminology and overlapping adversary capabilities which did not yield
the desired clarification. The second issue is erroneous assumptions about the
security of \gls{oob} channels which resulted in numerous attacks on various
pairing schemes \cite{Haataja:2010, Studer:2011, Halevi:2013, Anand:2015}. 

As it can be seen, the field of \gls{sdp} is rather fragmented. The lack of
coherent understanding of underlying key concepts has lead to poor design
decisions. That, in turn, resulted in a myriad of pairing solutions which only
focus on specific goals, use different vocabulary and rely on unrealistic
security assumptions. Hence, we are motivated to systematize knowledge in the
field of \gls{sdp} in order to identify the most difficult problems and
facilitate further research on this crucial topic. We make the following
specific contributions:

\begin{outline}
\1 A system model and consistent terminology that facilitates precise description and reasoning about \gls{sdp} schemes, by considering the three components:
\2 \gls{phy} channels;
\2 \gls{hci} channels; and
\2 Application classes.
\1 Classification of the existing \gls{sdp} schemes using this model.
\1 Identification and analysis of systemic security weaknesses commonly found in such schemes, revealing areas where future \gls{sdp} research is required.
\1 Revelation of the rarity with which privacy is considered among current \gls{sdp} schemes.
\1 Principles for designing robust \gls{sdp} schemes.

\end{outline}

The reminder of this article is organized as follows. 
In Section \ref{sec:literature} we review existing surveys on \gls{sdp} and highlight the contributions of our work. 
In Section \ref{sec:background} we present our consistent terminology and system model
for \gls{sdp}, and derive an \gls{sdp} taxonomy based on this model. 
We survey \gls{phy} communication channels along with the representative
pairing schemes which utilize those channels in Section \ref{sec:phy}. 
Section \ref{sec:hci} follows, where we review the \gls{hci} channels applied
to device pairing, as well as, the corresponding pairing schemes. 
In Section \ref{sec:app_classes} we discuss the application classes and
classify the surveyed pairing schemes with respect to their application
classes. 
We outline the open research challenges and provide the future perspective on
the field of \gls{sdp} in Section \ref{sec: challenges}. 
We conclude by summarizing our findings in Section \ref{sec:conclusion}.   

%% file: section/related_work.tex

\section{Related Work}
\label{sec:literature}
In the literature several surveys have investigated different aspects of \gls{sdp}.
Kumar et al. \cite{Kumar:2009} presented the first comparative study to quantify usability and security of various pairing schemes. 
Our work reveals that quantitative comparison of different \gls{sdp} schemes is questionable due to the previously taken design decisions, and 
we qualitatively address the design aspects of \gls{sdp} to enable meaningful comparison of different \gls{sdp} schemes. 
Two other studies from Kobsa et al. \cite{Kobsa:2009} and Kainda et al. \cite{Kainda:2009} focused more closely on usability and the role of user actions to achieve security in \gls{sdp}.
Our work has wider scope, because we consider the role of the user as one of the fundamental design aspects of \gls{sdp}, in addition to physical communication media and particular use cases.

The work of Mirzadeh et al. \cite{Mirzadeh:2014} provided an extensive survey on security and performance of different cryptographic protocols used in various \gls{sdp} schemes, in addition to presenting classification of \gls{oob} channels. In our work we devise a more fine-grained classification of communication channels in \gls{sdp} by differentiating between \gls{phy} and \gls{hci}  channels, and focus on security issues of those channels instead of cryptographic protocols. Since security weaknesses of various communication channels have resulted in numerous successful attacks on different \gls{sdp} schemes, we consider our qualitative analysis of those channels as a novel contribution. In their survey Chong et al. \cite{Chong:2014} presented different modes of user interaction for \gls{sdp} and analyzed a vast number of \gls{sdp} schemes using this taxonomy. We refine their findings to classify \gls{hci} channels and, additionally, present a set of common security and usability properties to coherently analyze those channels and the \gls{sdp} schemes relying on them, which has not been done before.

In our survey, we focus on \gls{sdp} schemes proposed for two (or several) devices and consider multi-device \gls{sdp} outside the scope of this article. In comparison to the prior work, our survey is innovative in three aspects. 
First, we devise a novel system model for \gls{sdp}, which addresses security weaknesses of the existing generation of \gls{sdp} schemes. Second, we propose a new approach to design \gls{sdp} schemes, which enables their meaningful comparison. Third, we provide a deep insight into a current state of \gls{sdp} from the point of \gls{phy} channels, \gls{hci} channels and application classes, as well as present an overview of \gls{sdp} challenges and perspectives in light of the upcoming \gls{iot}. 

In this section, we have reviewed the related work on \gls{sdp} and highlighted the contributions of our survey. 
In the next section, we present our system model and taxonomy.

%% file: section/background.tex

\section{System Model and Taxonomy}
\label{sec:background}
In this section we first give a high-level overview of a generalized pairing
procedure together with widely-used notations. 
Second, we address ambiguity in current terminology by providing clear definitions 
to describe \gls{sdp}. 
Third, we present a system model that illustrates the notion and properties of
communication channels as well as facilitates a more unified approach towards the
design of pairing schemes. 
Fourth, we discuss threats in \gls{sdp} with respect to our system model.
We conclude by explaining our taxonomy, which is used to systematize and
evaluate proposed pairing schemes.

\subsection{The Generalized Pairing Procedure}

\begin{figure}[htb!]
	\centering
	\resizebox{0.5\textwidth}{!}{%
	\begin{tikzpicture}
    
    \node[inner sep=0pt] (device1) at (0,0){\hspace*{1.15cm}\includegraphics[scale=0.5]{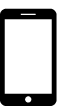}};
    \draw (-0.5, 0) node[inner sep=0pt, scale = 0.5] {\tiny \textit{D1}};
    
   	\node[inner sep=0pt] (device2) at (0,0){\hspace*{-1cm}\includegraphics[scale=0.5]{gfx/device}}; 
   	\draw (0.57, 0) node[inner sep=0pt, scale = 0.5] {\tiny \textit{D2}};

	\node[inner sep=0pt, rotate=-60] (wifi1) at (-0.3,0.23){\includegraphics[scale=0.15]{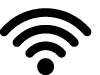}};
	\node[inner sep=0pt, rotate=-90] (hand1) at (-0.5,0.33){\includegraphics[scale=0.15]{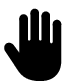}};
	
    \node[inner sep=0pt, rotate=60] (wifi2) at (0.36,0.23){\includegraphics[scale=0.15]{gfx/wifi}};
    \node[inner sep=0pt, rotate=90] (hand2) at (0.56,0.33){\reflectbox{\includegraphics[scale=0.15]{gfx/hand}}};
	
	\node[inner sep=0pt, scale=0.2] (stage1) at (0.029,0.31) {\circled{1}};
	\node[inner sep=0pt, scale=0.2] (stage2) at (0.029,0.05) {\circled{2}};
	\node[inner sep=0pt, scale=0.2] (stage3) at (0.029,-0.34) {\circled{3}};
	
	\draw [black, thin, <->,>={Latex[length=1mm]}] (-0.36,-0.03) -- (0.42,-0.02); 
	
	\node[inner sep=0pt] (dummy1) at (-0.52,-0.25) {}; 
    \node[inner sep=0pt] (dummy2) at (0.59,-0.25) {}; 
	
	\draw [black, thin, densely dotted, <->,>={Latex[length=0.55mm]}] (dummy1) to [out=-30, in=210] (dummy2);
    
  	\end{tikzpicture}
    }%
    \caption{Generalized pairing procedure}
    \label{fig:pair_overview}
\end{figure}

Traditionally, the pairing procedure has been considered as depicted in Figure
\ref{fig:pair_overview}. The scenario consists of two devices, \textit{D1} and
\textit{D2}, which do not share any prior knowledge and would like to pair. 
That is, two devices need to exchange some secret
information, ensuring it came from the correct party, and is not obtained by
any third party. In order to achieve pairing, three steps need to be followed:
\encircle{1} \textit{discovery}, \encircle{2} \textit{secret exchange} and \encircle{3}
\textit{verification}. In the first step, \textit{D1} and
\textit{D2} become aware of each other, which can happen either automatically,
for example, Bluetooth discovery, or with user assistance, for example,
physical contact. During the second step both devices exchange some
cryptographic material, for example, public keys, or a password, which can
later be used to establish secure communication. In the final step, both
parties verify the obtained secrets, to ensure that the process has not
been compromised by an attacker.

To provide a better understanding of the interactions presented in Figure
\ref{fig:pair_overview}, we examine the commonly used notation for \gls{sdp}.
Three main terms are commonly used in the literature: \textit{(a) in-band
channel}, \textit{(b) out-of-band channel} and \textit{(c) user interaction}.
By employing the generalized pairing procedure shown above, we demonstrate how
those concepts apply using a well-known example \cite{Balfanz:2002}. Two
devices discover each other, after having been brought together physically by
a user \textit{(c)}. Then they exchange hashes of their public keys over an
auxiliary channel \textit{(b)}, followed by a mutual transfer of the
corresponding public keys over a wireless radio link \textit{(a)}. Of course,
the given example illustrates just one possible case of how the pairing flow
can be implemented. There are other variants, for example, where the discovery
can happen without user interaction as in \cite{Miettinen:2014}, or the
secret key is first transmitted via the in-band channel, and subsequently
verified via the \gls{oob} channel as in \cite{Goodrich:2006}.

To gain a deeper understanding of the major pairing concepts, it is important
to specify the characteristics of in-band and \gls{oob} channels that have been
traditionally discussed by the research community. The pioneering work of
Balfanz et al. \cite{Balfanz:2002} stated two related properties that an \gls{oob}
channel should possess: \textit{demonstrative identification} and
\textit{authenticity}, and also that \textit{confidentiality} should not be assumed.

It is authenticity which is the defining
characteristic of an \gls{oob} channel: it is the infeasibility of forging
communications over an \gls{oob} channel, without being detected,
that makes \gls{oob} communications so valuable in \gls{sdp}.
In practice, this implies that
\gls{oob} channels must possess demonstrative identification, that is,
it must be easy to demonstrate that the \gls{oob} communication is occuring between
the intended parties, for example, by showing the display of a device to another
user. Demonstrative identification, thus, implies that the devices must be brought
sufficiently close to one another to allow their mutual positive identification
by their users.
While \gls{oob} channels should not be assumed to offer confidentiality, a number 
of the surveyed \gls{sdp} schemes depended on the \gls{oob} channels being confidential.

The in-band channel, in contrast, has been generally regarded as a communication channel with relaxed security
characteristics. That is, it refers to a wireless radio link which is
easily accessible by a powerful attacker \cite{Dolev:1983} and, thus, deemed as
inherently \textit{insecure}.

\begin{figure*}[htb!]
	\centering
	\begin{adjustbox}{width=\textwidth}
 	\begin{tikzpicture}
 
  	\node[inner sep=0pt] (device1) at (0,0){\hspace*{1.5cm}\includegraphics[scale=1]{gfx/device}};
	\draw (0.74, 0) node {\tiny \textit{D1}};
	
  	\node[inner sep=0pt] (device2) at (5,0){\hspace*{-3.5cm}\includegraphics[scale=1]{gfx/device}};  
	\draw (3.24, 0) node {\tiny \textit{D2}};
    
   	\draw (1.95, 0.42) node [inner sep=0pt] {\tiny \textsc{Physical}}; 
  
   	\draw [black, very thin, <->,>=latex] (1.05,0.33) -- (2.92,0.33);  
   	\draw [black, very thin, <->,>=latex] (1.05,0.1) -- (2.92,0.1);
   	\node[text width=1cm, black] at (2.35,-0.12) {\footnotesize {...}};
   	\draw [black, very thin, <->,>=latex] (1.05,-0.34) -- (2.92,-0.34);
   	
   	\node[inner sep=0pt, scale=0.38] (int1) at (1.95,0.219) {\circled{1}};
   
   	\node[inner sep=0pt] (users) at (1.95,-1){\includegraphics[scale=0.5]{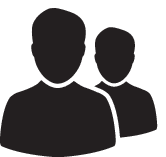}};
   	
	\draw (0.95, -0.97) node [inner sep=0pt] {\tiny \textsc{HCI}};  
	\draw (2.97, -0.97) node [inner sep=0pt] {\tiny \textsc{HCI}}; 	
   	
   	\draw [black, double, very thin, <->,>=stealth] (1.53,-1.15) -- (0.74,-0.53); 
   	\draw [black, double, very thin, <->,>=stealth] (2.38,-1.15) -- (3.24,-0.53); 
   	
   	\node[inner sep=0pt, scale=0.38] (int21) at (1.28,-0.72) {\circled{3}}; 
   	\node[inner sep=0pt, scale=0.38] (int22) at (2.68,-0.72) {\circled{3}}; 
   
   	\node[
    	xshift = -0.6cm,
   		inner sep=0pt,
        cloud, 
        cloud puffs = 10,
    	minimum width=2cm,
   		minimum height=1.5cm,
        draw,
        scale=0.5,
    ] (env1) at (0,-0.06) {};
    
    \node[inner sep=0pt] at (env1.center){\includegraphics[scale=0.15]{gfx/wifi}};
    \node[inner sep=0pt] at ([xshift=0.6em, yshift=0.6em]env1.south){\includegraphics[scale=0.23]{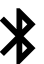}};
    \node[inner sep=0pt] at ([xshift=-0.6em, yshift=-0.6em]env1.north){\includegraphics[scale=0.015]{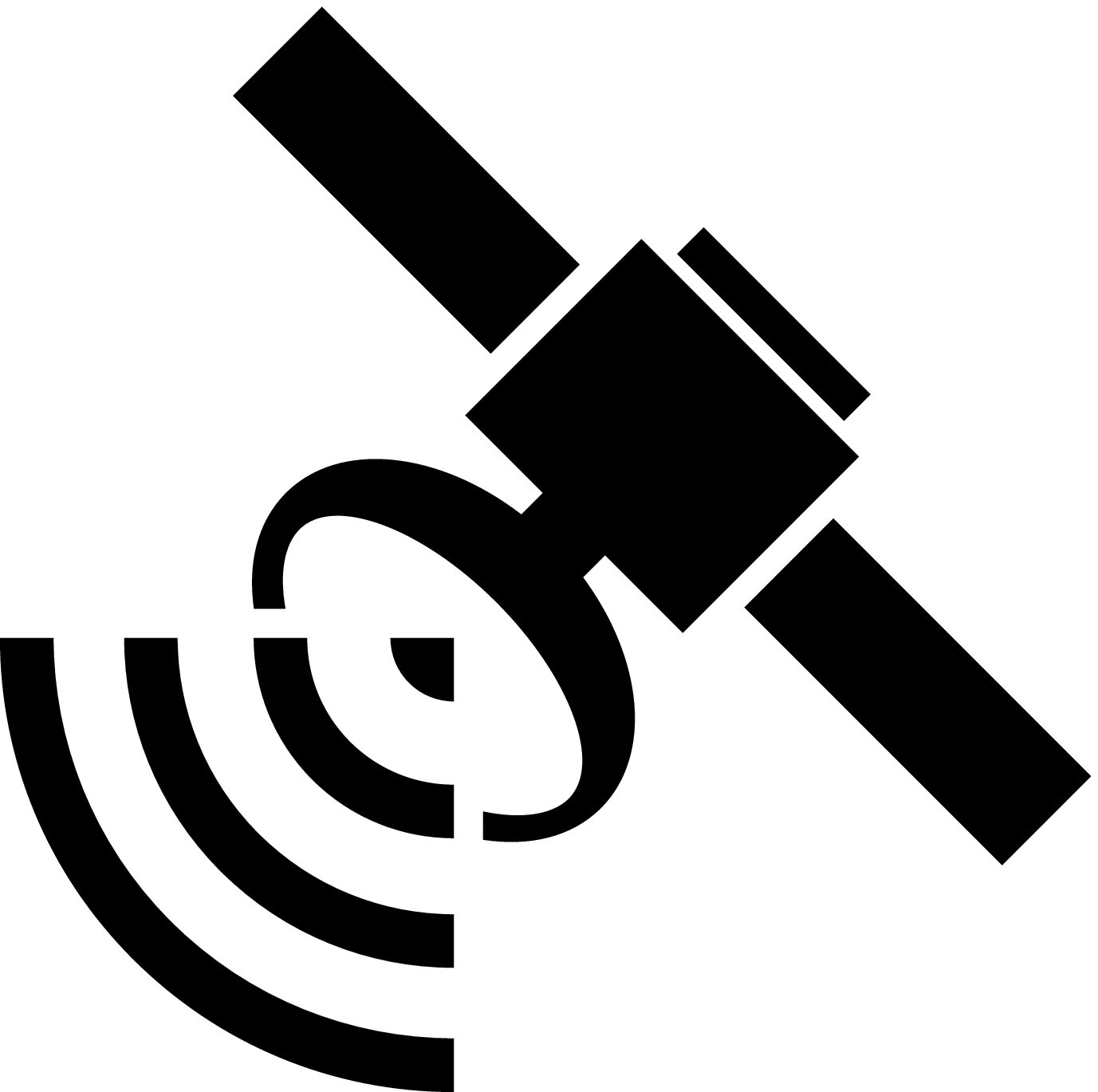}};
    \node[inner sep=0pt] at ([xshift=-0.62em, yshift=0.5em]env1.east){\includegraphics[scale=0.15]{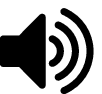}};
    \node[inner sep=0pt] at ([xshift=0.63em, yshift=-0.43em]env1.west){\includegraphics[scale=0.014]{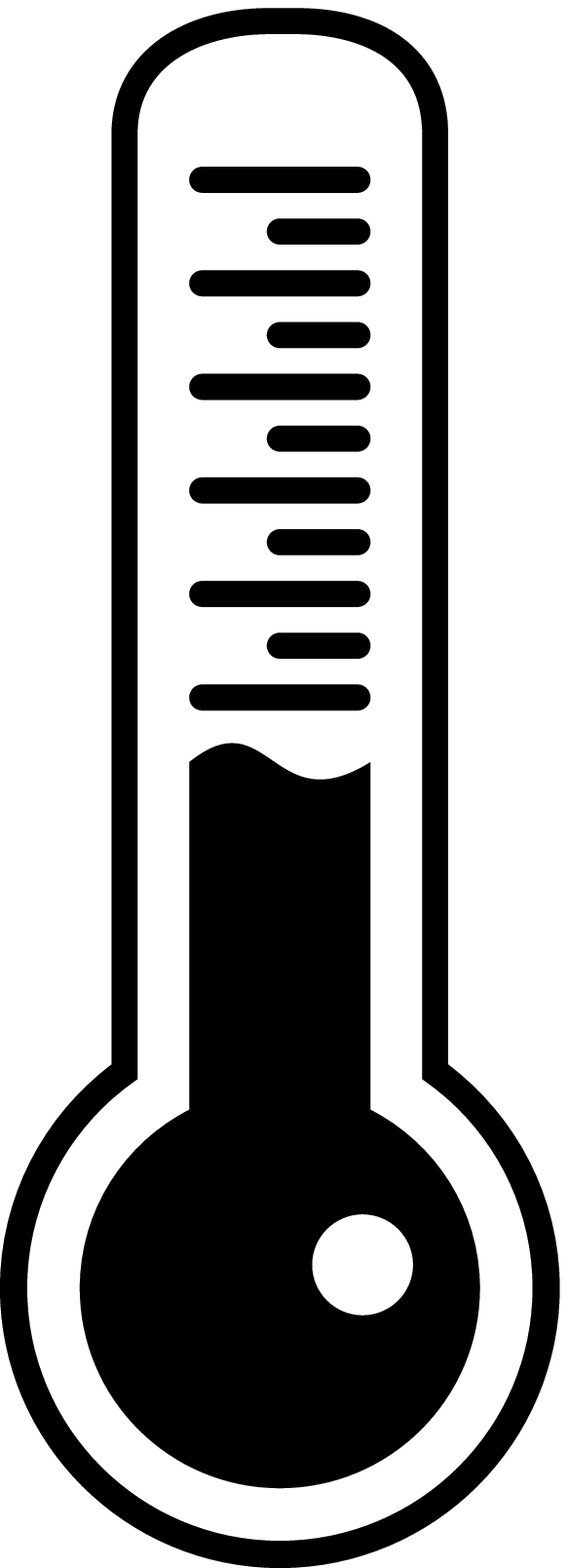}};
   
    \node[
   		inner sep=0pt,
        cloud, 
        cloud puffs = 10,
    	minimum width=2cm,
   		minimum height=1.5cm,
        draw,
        scale=0.5,
    ] (env2) at (4.6,-0.06) {};

    \node[inner sep=0pt] at (env2.center){\includegraphics[scale=0.15]{gfx/wifi}};
    \node[inner sep=0pt] at ([xshift=0.6em, yshift=0.6em]env2.south){\includegraphics[scale=0.23]{gfx/bluetooth}};
    \node[inner sep=0pt] at ([xshift=-0.6em, yshift=-0.6em]env2.north){\includegraphics[scale=0.015]{gfx/gps}};
    \node[inner sep=0pt] at ([xshift=-0.62em, yshift=0.5em]env2.east){\includegraphics[scale=0.15]{gfx/audio}};
    \node[inner sep=0pt] at ([xshift=0.63em, yshift=-0.43em]env2.west){\includegraphics[scale=0.014]{gfx/temperature}};
   
	\draw (3.97, 0.42) node [inner sep=0pt] {\tiny \textsc{Physical}};
   	\draw (0, 0.42) node [inner sep=0pt] {\tiny \textsc{Physical}};  	
   	 
   	\draw [black, densely dotted, thin, ->, >=triangle 60] (-0.1,-0.06) -- (0.42,-0.06);
   	\draw [black, densely dotted, thin, ->, >=triangle 60] (4.12,-0.06) -- (3.55,-0.06);
   	
   	\node[inner sep=0pt, scale=0.38] (int21) at (0.075,0.1) {\circled{2}}; 
   	\node[inner sep=0pt, scale=0.38] (int22) at (3.95,0.1) {\circled{2}}; 
   
 \end{tikzpicture}
\end{adjustbox}
\caption{System model for secure device pairing}
\label{fig:sys_model}
\end{figure*}

So far, we have discussed the core components of \gls{sdp} along with their
prime purposes and vital properties. Yet, there are no precise definitions of
the \gls{oob} channel and user interaction which are common in the field. We
consider this point to be the principle weakness of the existing terminology.
Many researchers have described the \gls{oob} channel to be a side \gls{phy}
channel, which is either human-perceptible and/or directly controlled by a user
\cite{Kumar:2009}. Nevertheless, there is a number of pairing schemes that
intrinsically rely on user actions to accomplish pairing
\cite{Kainda:2009}. One example of the latter is Secure Simple
Pairing \cite{Bluetooth:2007} which is a de facto standard for
connecting Bluetooth devices securely. Consequently, such disparity results in
a situation where one part of the community only considers physical media as
the \gls{oob} instance while neglecting the user-mediated channels, and vice
versa. In addition, communication channels differ in fundamental ways, hence
assumptions about media and attacker models vary significantly, 
and are not straightforward to align. 

Another issue is that the boundary
between the human-mediated \gls{oob} channels and user interaction is often blurred. 
That is, the latter is a
more general term that can include the former. However, the essential purpose
of the \gls{oob} channel is to provide some form of data authenticity.
Specifically, a human operator can assist in initiating device pairing during
the discovery step, for example, by co-locating devices, aligning them or
enabling physical contact. Yet, we argue that only explicit actions which
directly affect the security of the pairing scheme should be considered as
the \gls{oob} channel.

\subsection{Defining Secure Device Pairing Terminology}
\label{subsec:sdp_def}
A specific challenge to the comparison and analysis of \gls{sdp} schemes is
the lack of accepted terminology covering such schemes.
We therefore present the terminology that we use in the remainder of this
article, both for clarity of explanation here, 
and as a suggestion for a common vocabulary to facilitate communications among
practitioners in the future. 

\begin{itemize}
\item{\textit{Pairing} refers to the establishment of a secure communication channel
between two or more devices.}

\item{An \textit{application class} represents a particular pairing scenario that is
determined by the degree of involvement and level of control that a user has
over the pairing devices.  An application class covers use-cases that
share broadly similar security threats and objectives.}

\item{An \textit{\gls{sdp} scheme} consists of the procedures, cryptographic protocols
and the motivating application class required to securely pair devices.}

\item{An \textit{\gls{sdp} method} or \textit{\gls{sdp} procedure} is
the sequence of actions required to execute an \gls{sdp} scheme.
While considering method and procedure interchangeable, we avoid the synonym
protocol, because of the  strong association of this word with cryptographic
protocols.}

\item{A \textit{party} is someone or something who controls of one or
more devices that participate in an \gls{sdp} procedure.}

\item{A \textit{security domain} is the set of devices, data, policies and intentions that
a single party controls.  That is, every device belongs to a
security domain, but there may be more than one security domain involved in a
given application class.}

\item{A \textit{channel} is a means by which communications occur in an \gls{sdp}
scheme, whether over a physical medium, or through an \gls{hci}.}

\item{An \textit{HCI channel} is a means of communication where a user acts as the channel
by which the communications occurs by undertaking some form of interaction
with the devices involved. This could take the form, for example, of a user reading information from
the display of two devices, and entering confirmation that they match into one of those
devices. }

\item{A \textit{PHY channel} is a communication channel that allows data
transmission or acquisition over a physical medium. \gls{phy} channels can be described
by their objective physical characteristics and where the information is not transferred by a user,
that is, it is not an \gls{hci} channel.}

\end{itemize}

\begin{figure*}[htb!]
\centering
\begin{adjustbox}{width=\textwidth}
 \begin{tikzpicture} [every node/.style = {shape = rectangle,
                                         rounded corners,
                                         draw,
                                         thick,
                                         minimum width = 2.7cm, 
                                         minimum height = 1cm,
                                         align = center,
                                         text = black,
                                         font=\small},
                     edge/.style  = { -,
                                         ultra thick,
                                         thick,
                                         black,
                                         shorten >= 1pt, 
                                         shorten <= 1pt, 
                                         }]

	\node(sdp) at (0,0) {\textsc{\textbf{Secure Device}} \\ \textsc{\textbf{Pairing}}};
	
		\node(chn)  at (-5,-2) {\textsc{\textbf{Channels}}}; 
		
			\node(phy) at (-6.5,-4) {\textsc{\textbf{PHY}}};
			
			\node(hci) at (-3.5,-4) {\textsc{\textbf{HCI}}};
			
			\draw[edge] (chn.south) to[|-|] (phy.north); 
  			\draw[edge] (chn.south) to[|-|] (hci.north);
		
		\node(apd) at (5,-2) {\textsc{\textbf{Application}} \\ \textsc{\textbf{Classes}}}; 
			
				\node(prv) at (0.5,-4) {\textsc{\textbf{Private}}};  
       			\node(pub) at (3.5,-4) {\textsc{\textbf{Public}}};
       		    \node(soc) at (6.5,-4) {\textsc{\textbf{Social}}};
	       		\node(una) at (9.5,-4) {\textsc{\textbf{Unattended}}};
	       		
				\draw[edge] (apd.south) to[|-|] (prv.north); 
  				\draw[edge] (apd.south) to[|-|] (pub.north);
  				\draw[edge] (apd.south) to[|-|] (soc.north); 
  				\draw[edge] (apd.south) to[|-|] (una.north);
	       		
		\draw[edge] (sdp.south) to[|-|] (chn.north); 
  		\draw[edge] (sdp.south) to[|-|] (apd.north);
	
 \end{tikzpicture}
\end{adjustbox}
\caption{Taxonomy of secure device pairing}
\label{fig:taxonomy}
\end{figure*}
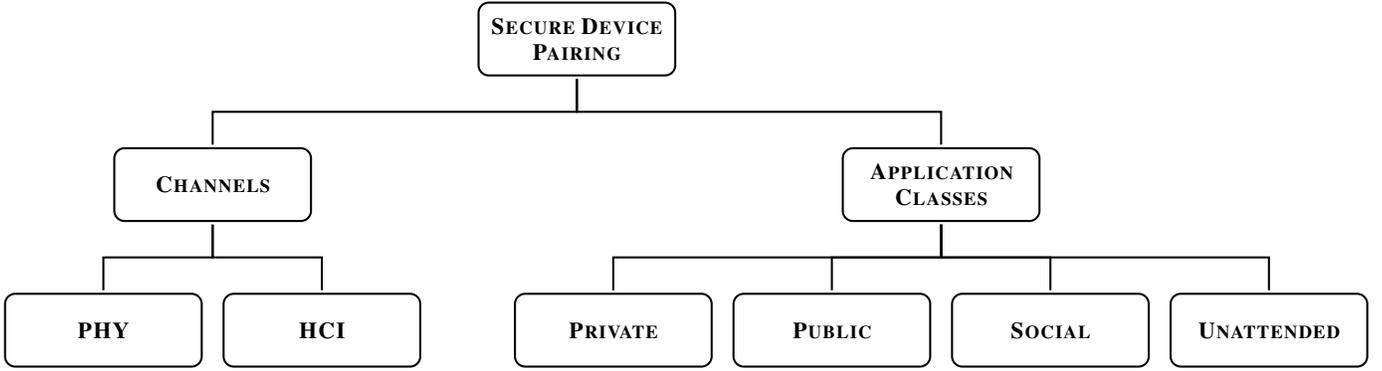

\subsection{System Model}
To address the issues in \gls{sdp} mentioned previously, we introduce our
system model depicted in Figure \ref{fig:sys_model}. The objectives of our
approach are threefold. First, it takes into account a set of diverse
interactions that appear in the context of \gls{iot}. Second, it aims to
resolve the ambiguities in the pairing concepts which are currently present in
the field. Third, our model facilitates a more unified procedure for the
pairing design.   

Our system model contains three main components. Particularly, there are
\textit{(a) two devices} to be paired \textit{D1} and \textit{D2}, \textit{(b)
a varying number of users} and \textit{(c) the ambient environment} in which
device pairing takes place. In addition, several types of distinctive
interactions can happen between those elements. First, \textit{D1} and
\textit{D2} can communicate with each other by means of various wireless
technologies such as Wi-Fi, Bluetooth, etc. \encircle{1}. Second, a device
can obtain the information about the ambient environment such as temperature,
location, etc. via its sensing capabilities \encircle{2}. Third, the
connection between a human operator and the respective device is established in
a form of \gls{hci} \encircle{3}. We further consider the relationship
between a user and a pairing device.  Specifically, a human operator can either
control both devices involved in \gls{sdp}, a single one or none
at all. Implied in the system model is the purpose for which the devices are being
paired, that is, the use-case.

From the above, we can consider an \gls{sdp} scheme as consisting of
the automated communications between two devices over conventional \gls{phy}
channels, plus the human-mediated communications between the devices over \gls{hci}
channels. Pairing of devices always occurs for a purpose, that is, it happens
within the context of an application class.
We, therefore, use three key concepts as the foundation for our system
model:
\begin{itemize}
\item{\gls{phy} channels.}
\item{\gls{hci} channels.}
\item{Application classes.}
\end{itemize}
Figure \ref{fig:taxonomy} illustrates the relationship between those concepts, that is,
we consider the channels, both \gls{phy} and \gls{hci} to be orthogonal to the application 
classes. 

\begin{figure}[htb!]
	\centering
    \resizebox{0.5\textwidth}{!}{%
    \begin{tikzpicture}[baseline,decoration=brace]
        
	\draw[black, semithick] (0,0) rectangle (2,5) node[pos=.5] {\large \textit{Device}};
	
	\node[inner sep=0pt] (phy1) at (3.55,5) {}; 
    \node[inner sep=0pt] (phy2) at (3.55,0.1) {}; 
    
    \draw[decorate, transform canvas={xshift=0.2em}, thick] (phy1.south west) -- node[right=2pt] {\textsc{Phy}} (phy2.north west);

	\node[inner sep=0pt, scale=0.85] at (2.8,4.9) {\footnotesize \textsc{Wi-Fi}};
    \draw[black, semithick, <->,>=stealth] (2.05,4.75) -- (3.5,4.75);
    
    \node[inner sep=0pt, scale=0.85] at (2.8,4.45) {\footnotesize \textsc{Bluetooth}};
    \draw[black, semithick, <->,>=stealth] (2.05,4.3) -- (3.5,4.3);
    
    \node[inner sep=0pt, scale=0.85] at (2.78,4) {\footnotesize \textsc{mm-Waves}};
    \draw[black, semithick, <->,>=stealth] (2.05,3.85) -- (3.5,3.85);
    
    \node[inner sep=0pt, scale=0.85] at (2.8,3.55) {\footnotesize \textsc{RFID}};
    \draw[black, semithick, <->,>=stealth] (2.05,3.4) -- (3.5,3.4);
    
    \node[inner sep=0pt, scale=0.85] at (2.8,3.1) {\footnotesize \textsc{NFC}};
    \draw[black, semithick, <->,>=stealth] (2.05,2.95) -- (3.5,2.95);
    
    \node[inner sep=0pt, scale=0.85] at (2.8,2.65) {\footnotesize \textsc{VLC}};
    \draw[black, semithick, <->,>=stealth] (2.05,2.5) -- (3.5,2.5);
    
    \node[inner sep=0pt, scale=0.85] at (2.8,2.2) {\footnotesize \textsc{Visual}}; 
    \draw[black, semithick, <->,>=stealth] (2.05,2.05) -- (3.5,2.05);
    
    \node[inner sep=0pt, scale=0.85] at (2.8,1.75) {\footnotesize \textsc{IrDA}}; 
    \draw[black, semithick, <->,>=stealth] (2.05,1.6) -- (3.5,1.6);
    
	\node[inner sep=0pt, scale=0.85] at (2.8,1.3) {\footnotesize \textsc{Sound}}; 
    \draw[black, semithick, <->,>=stealth] (2.05,1.15) -- (3.5,1.15);
	
	\node[inner sep=0pt, scale=0.85] at (2.8,0.85) {\footnotesize \textsc{Haptic}};
    \draw[black, semithick, ->,>=stealth] (2.05,0.7) -- (3.5,0.7);
    
    \node[inner sep=0pt, scale=0.85] at (2.8,0.4) {\footnotesize \textsc{Sensing}};
    \draw[black, semithick, <-,>=stealth] (2.05,0.25) -- (3.5,0.25);
    
	\node[inner sep=0pt] (hci1) at (-1.55,4.9) {}; 
    \node[inner sep=0pt] (hci2) at (-1.55,0.12) {}; 
    
    \draw[decorate, transform canvas={xshift=0em}, thick] (hci2.south west) -- node[left=2pt] {\textsc{HCI}} (hci1.north west);
    
    \node[inner sep=0pt, scale=0.85] at (-0.75,4.45) {\footnotesize \textsc{Relay}};
    \draw[black, semithick, <->,>=stealth] (-1.5,4.3) -- (-0.05,4.3);
    
    \node[inner sep=0pt, scale=0.85] at (-0.75,3.55) {\footnotesize \textsc{Comparison}};
    \draw[black, semithick, <-,>=stealth] (-1.5,3.4) -- (-0.05,3.4);
    
    \node[inner sep=0pt, scale=0.85] at (-0.8,2.65) {\footnotesize \textsc{Generation}};
    \draw[black, semithick, ->,>=stealth] (-1.5,2.5) -- (-0.05,2.5);
    
    \node[inner sep=0pt, scale=0.85] (hci4) at (-0.8,0.4) {\footnotesize \textsc{Handling}};
    \draw[black, semithick, dashed, ->,>=stealth] (-1.5,0.25) -- (-0.05,0.25);
    
  	\end{tikzpicture}
    }%
    \caption{Pairing device with \gls{phy} and \gls{hci} channels. This model is independent of application classes.}
    \label{fig:entity_channels}
\end{figure}
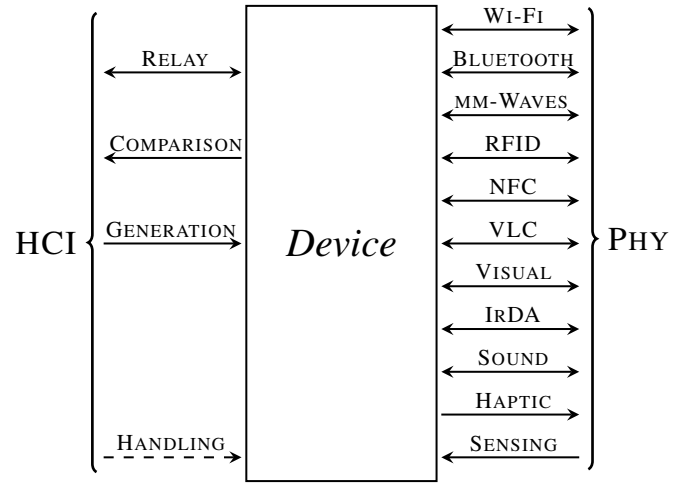

The first two concepts specify two fundamentally different types of
interactions that can be utilized by a pairing scheme.
With this in mind, we further analyze \gls{phy} and \gls{hci} channels
independently, to identify the most important features of each class, and
expose the trade-offs involved.
To account for both types of interactions, we understand ``device'' to mean any
physical device that possesses one or more communication channels that can be
used to connect to the outside world.
\gls{sdp} is achieved using some set of such channels.
Figure \ref{fig:entity_channels} depicts an abstract visualization of such a device concept, 
including a comprehensive list of \gls{phy} and \gls{hci}
channels.
A rigorous discussion covering each channel category in detail is provided in
Section \ref{sec:phy}, for \gls{phy} channels, and Section \ref{sec:hci} for \gls{hci}
channels.     

As for the application classes, we identify four different cases which cater to
classify all the pairing schemes proposed up-to-date. The categories are as
follows: \textit{(a) private}, \textit{(b) public}, \textit{(c) social} and
\textit{(d) unattended}. 
The private class corresponds to a ``classic pairing'' case, where a single user
either owns or directly controls two devices that ought to be
paired. 
The public class is related to a single user possessing
one device, where the user performs the pairing with some
third party infrastructure, for example, a payment terminal,
over which she has no control. 
The social class incorporates
two users who would like to securely pair their corresponding
devices. 
The unattended class deals with the case where
two devices belonging to the same ownership domain, for
example, owned by the same person or organization, pair with
no user involvement.

For each application class, we present distinct interaction patterns,
demonstrated by instantiating our system model (Section \ref{sec:app_classes}).
Furthermore, we identify commonalities in the form of adversary capabilities,
as well as security and usability implications that have to be taken into
account for a particular application class. 
Consequently, it is possible to determine a set of common
security and usability properties shared by a 
group of pairing schemes that have been designed with a specific application class in mind.
Section \ref{sec:app_classes} explores the potential for such application classes to facilitate the
design of better, more coherent \gls{sdp} schemes.

\subsection{Worked Example}
To demonstrate the relevance of our system model we consider the security of an important class of the \gls{iot} devices such as \gls{imd}. 
For example, the case of Dick Cheney has raised further awareness of life-threatening repercussions stemming from the compromise of a pacemaker \cite{Cheney:2013}. 
Unfortunately, the current state of security and privacy in existing \gls{imd} is still very immature \cite{Marin:2016}. 

To enable secure communications between the implant and a medical programmer or a base station, used for configuring or reading records from the implant, \gls{sdp} can be applied. 
When designing a pairing scheme for such a use case it is important to understand that user interaction is restricted, that is, a user cannot be actively involved in the pairing process. 
Thus, an \gls{imd} and the maintenance equipment must pair autonomously, which implies the unattended application class. 
The actual implementation of such pairing schemes is, of course, an open question but it is already clear that pairing paradigms typical for the unattended application class, that is, a notion of physical proximity or the use of contextual information can be applied. 
Furthermore, communication channels used for \gls{sdp} can be more reasonably selected. For example, since the user interaction is limited, using \gls{hci} channels as \gls{oob} does not seem feasible, thus, the use of \gls{phy} channels is justified. 
Given the critical nature of \gls{imd}, it is highly favorable to select either some short-range and restricted \gls{phy} channel such as 60 GHz, or employ some means of sensing as an \gls{oob} channel for \gls{sdp}. 

As it can be seen, starting a design process with the application class has laid the foundation for further design steps and established a common ground for \gls{sdp} schemes targeting \gls{imd}. The selection of communication channels in \gls{sdp} is also better justified, and corresponds to security countermeasures proposed to mitigate attacks on \gls{imd}, that is, the application of selective jamming \cite{Marin:2016} vs. the use of short-range and restricted \gls{phy} channel such as 60 GHz.

\subsection{Overview of Threats}
The formulation of a detailed adversary model is beyond the scope of this article.  
However, a general overview of relevant threats is
still required to meaningfully compare different \gls{sdp} schemes.
We consider two pairing devices D1 and D2, as depicted in Figure
\ref{fig:sys_model}, and assume that they are not compromised with malware and
controlled directly by their respective owners.  
The goal of an adversary is always to undermine \gls{sdp}. 

We focus primarily on attacks against authenticity and confidentiality, as they
are the most relevant to the \gls{sdp} process. 
We consider two broadly representative classes of adversaries seeking to
undermine \gls{sdp}: first, those who want to attack via \gls{phy}
communications channels, and second, those who wish to attack via \gls{hci}
channels. 

\subsubsection{Attacks on \gls{phy} Channels}
An adversary who exploits \gls{phy} channels can mount many different attacks. The majority of these attacks are considered particularly severe in \gls{sdp}, as they undermine the basic assumptions of authenticity and confidentiality that \gls{sdp} seeks
to establish.
\gls{phy} channels used for data transmission are especially vulnerable to attacks on confidentiality such as 
\gls{mitm} and eavesdropping attacks, while \gls{phy} channels used for data
acquisition, for example, environmental sensing, are susceptible to attacks that undermine authenticity such as relay attacks. For example, the adversary can reproduce the relevant sensor readings, by manipulating the temperature or humidity to match that of a remote location.

Regarding availability and integrity properties, the adversary can jam or
otherwise disturb the communication media which can result in \gls{dos}
attacks. 
While \gls{dos} is a legitimate security issue, because it can prevent
\gls{sdp}, it cannot lead to a false sense of security, that is, where two
users mistakenly believe that their devices have securely paired. Hence, 
\gls{dos} can deny availability, but not subvert authenticity or confidentiality.
This stands in contrast to \gls{mitm}, eavesdropping and relay attacks where the devices 
will behave as though they have paired securely, when, in fact, they have not. 

More sophisticated attacks such as selective jamming, disturbing signal parts or bit-flipping can force retransmissions and increase adversary's chances of compromising \gls{sdp}. 
Time-based attacks, for instance, replaying previously captured packages or delaying messages in transmission can also impair the \gls{sdp} process. 
In our survey, we are primarily concerned with attacks on \gls{phy} channels 
that can subvert authenticity and confidentiality, that is, \gls{mitm}, eavesdropping and relay attacks. More exotic attacks on various \gls{phy} channels are discussed in the respective sections where relevant. 

\subsubsection{Attacks on \gls{hci} Channels}
There are two main classes of \gls{hci} adversaries: \textit{(a)} an
external attacker, that is, someone who is not a legitimate pairing party and
\textit{(b)} an internal attacker, that is, one of the pairing participants.

External adversaries aim to violate the authenticity and confidentiality of
the \gls{hci} channel by observing user interaction, in order to be able to
covertly participate in the communications between the pairing devices. 
For example, an external \gls{hci} adversary may monitor the
\gls{hci} during a pairing event, in order to also derive the cryptographic
material being exchanged during the pairing process.

Internal adversaries, on the other hand, already have such access, but seek to
use the process of pairing in order to extract sensitive data from the other
participant's device.  This may take the form of social engineering. When the
attacker's primary objective is to participate in the pairing, in order to
undermine the privacy of the other party, we label them as honest-but-curious
adversaries. For example, an honest-but-curious attacker may seek to obtain
the telephone number of the other party, contrary to their wishes.

\subsection{Taxonomy}
We propose a taxonomy built upon the three key concepts that have been introduced
and described above: \gls{phy} channels, \gls{hci} channels and application
classes, which makes the following two contributions to the field of \gls{sdp}.
First, it provides systematization based on these three key concepts drawn from
the design space of \gls{sdp}. 
Second, it enables qualitative assessment and meaningful comparison of
different pairing schemes. 
The structure of our taxonomy is given in Figure \ref{fig:taxonomy}. 

In order
to investigate a device's channels we adopt the following framework. First, we
identify the most important characteristics that are relevant to a
communication channel. For \gls{phy} channels, such parameters are
measurable and objective, whereas the \gls{hci} channels are represented by
more subjective metrics. Second, we focus on three sets of properties that are
vital in the context of \gls{sdp}, namely: security, usability and
ease-of-adoption. Third, based on this structure we review the existing pairing
schemes to reveal how these properties are addressed by a particular scheme
and what are the trade-offs. With regard to application classes, we first
provide a thorough description of each class followed by a discussion on its
specific security and usability implications. Second, we map the proposed
pairing schemes to the corresponding application classes to provide a more systematic
overview of the current state in the field.

In this section, we have presented and motivated our system model, as well as provided the taxonomy 
which we use to survey existing \gls{sdp} schemes. 
In the next section, we review \gls{phy} channels and the corresponding pairing schemes. 

%% file: section/physical_channels.tex

\section{Physical Channels}
\label{sec:phy}
The \gls{phy} channels listed in Figure \ref{fig:entity_channels} allow a device to
communicate with other devices as well as interact with the ambient
environment. We base our analysis on several aspects to conduct the meticulous
investigation of \gls{phy} channels. Specifically, we consider \textit{channel
characteristics}, \textit{known attacks} and \textit{ease-of-adoption} to
compare different types of communication channels. With regard to
security, we identify a set of channel properties that have a direct security
impact, namely they indicate the amount of effort necessary to intercept the
pairing process. Furthermore, we discuss pairing schemes that utilize each
surveyed channel to show how it is employed to achieve pairing. 
Finally, we summarize the most important findings, presented in Table \ref{tab:phy_channels}, and discuss key
issues from our study of \gls{phy} channels. 

\subsection{Channel Characteristics}
The physical nature of wireless communication channels is described by
different properties. 
The most important ones are the \textit{frequency range} of transmitted
electromagnetic or mechanical waves and the achievable \textit{data rates}. In
fact, the data rate is defined by technical specifications of a communication 
protocol, such as the available bandwidth, coding and modulation schemes, rather than the underlying \gls{phy} channel.   
Hence, we consider this as the amount of transmitted
information in a unit of time using common state of the art protocols. The
frequency and bandwidth of a particular channel comes along with certain propagation
characteristics, such as \textit{coverage}, human \textit{perceptibility},
\textit{penetration}, and \textit{line-of-sight} propagation. We put these
properties into security perceptive by refining the observations made by
Balfanz et al. \cite{Balfanz:2002} on location-limited channels.

\subsubsection{Coverage}
Defines the maximum nominal distance at which a signal on a \gls{phy} channel
can be successfully received, that is, differentiated from noise. Naturally,
increasing the sensitivity of receivers leads to wider coverage, which makes
it a questionable security property. However, the amount of effort and cost
required to receive a signal far outside the nominal range are high. For an
adversary, a \gls{phy} channel with smaller coverage is harder to access and thus, attack. 

\subsubsection{Perceptibility}
Specifies whether a user can perceive the fact of data transmission through
major human senses such as sight, hearing and touch \cite{Dix:2009}. This
property can have both advantageous and harmful repercussions. On the one
hand, a benign user can be alerted if some unexpected interaction occurs. On
the other hand, an attacker can easily observe such type of communication
without any specialized equipment. Nevertheless, this trade-off can be
leveraged by secure protocol design so that the benefits of human
perceptibility greatly outweigh the risks. 

\subsubsection{Penetration}
The propagation properties of a particular channel depend on the underlying
physics. Both electromagnetic and mechanical waves are subject to diffraction,
reflection, refraction, scattering and absorption \cite{Straw:2003,
Terzic:2013}. With these effects a signal can be partially or entirely blocked
which hinders the communication. Penetration characterizes the ability of the
signal to propagate trough solid obstacles such as walls, doors, and
furniture. Thus, a communication channel with high blockage, that is, low
penetration, effectively limits the operation range of the channel, thus 
hampering the attacker's ability to access it.

\subsubsection{\gls{los}}
Signals propagate throughout the environment and find multiple paths from the
transmitter to the receiver over several reflections, diffractions, and refractions.
As was stated above, channels are differently affected by these
factors. 
Generally speaking, with higher frequency diffraction and refraction become less significant.
With low multi-path components a channel becomes dependent on \gls{los} which is a direct path 
between a transmitter and a receiver without obstruction. 
Correspondingly, non-line-of-sight communication does not require an obstruction-free path between a transmitter and a receiver. 
\gls{los} enables predominantly directional communication which hinders the ability of the adversary 
to stealthily intercept the channel from outside the main transmission beam.

\begin{table*}[!htb]
  \caption{Summary of \gls{phy} channels}
  \label{tab:phy_channels}
  \centering
  \begin{threeparttable}
  \rotatebox{0}{\scriptsize{
		\begin{tabularx}{\textwidth}{p{1.80cm}p{0.75cm}p{1.2cm}p{1.1cm}p{0.8cm}p{0.2cm}p{0.75cm}p{0.2cm}p{2.85cm}p{0.2cm}p{3cm}}
		\hline
		\hline

		\multicolumn{1}{c}{\shortstack{\\ \textbf{\gls{phy} Channel} }}& 
		\multicolumn{7}{c}{\shortstack{\\ \textbf{Channel Characteristics} }}&
		\multicolumn{1}{c}{\shortstack{\\ \textbf{Attacks}}}&
		\multicolumn{1}{c}{\shortstack{\\ \textbf{Common} \\ \textbf{Interface}}}&
		\multicolumn{1}{c}{\shortstack{\\ \textbf{Pairing Scheme} }}\\
		\hline

		&
		\rotatebox{55}{\textbf{Description		}}& 		
		\rotatebox{55}{\shortstack{\\ \textbf{Frequency} \\ \textbf{Range}}}& 		
		\rotatebox{55}{\shortstack{\\ \textbf{Typical} \\ \textbf{Data Rates}}}&
		\rotatebox{55}{\shortstack{\\ \textbf{Typical} \\ \textbf{Coverage}}}&
		\rotatebox{55}{\textbf{Perceptibility	}}&		
		\rotatebox{55}{\shortstack{\\ \phantom{\textbf{Penetration}} \\ \textbf{Penetration}}}&		
		\rotatebox{55}{\textbf{Line-of-sight}} 
		&
		& 
		& \\
		\hline
		
		Wi-Fi \newline
		(\textsection \ref{subsubsec:wifi}, pp. \pageref{subsubsec:wifi}) & 
		Wireless Radio Comm. &
		2.40 GHz\textendash \newline\phantom{I}2.48 GHz, \newline 
		5.03 GHz\textendash \newline\phantom{I}5.83 GHz &
		1 Mbit/s\textendash \newline
		\phantom{I}1 Gbit/s  &
		30 m{\textendash} \newline\phantom{I}250 m &
		\Circle &
		\textcolor{white}{a}High & 
		\Circle &
		\textcolor{white}{aaa}Eavesdropping \cite{Aime:2007} \newline
		\textcolor{white}{aaa}\gls{mitm} \cite{Aime:2007, Fund:2016} \newline
		\textcolor{white}{aaa}Jamming \cite{Bayraktaroglu:2013, Schulz:2017} \newline
		\textcolor{white}{aaa}\gls{dos} \cite{Lu:2010, Schulz:2017} &
    	\shortstack{\textcolor{white}{\textbf{aa}} \CIRCLE} & 
    	Push Button Configuration \cite{PBC:2006} \newline 
    	Integrity codes \cite{Capkun:2008} \newline
    	Tamper-evident pairing \cite{Gollakota:2011} \\ 
    	
		\hline
    	
    	Bluetooth \newline
	    (\textsection \ref{subsubsec:bluetooth}, pp. \pageref{subsubsec:bluetooth}) &  
		Wireless Radio Comm. &
		2.40 GHz\textendash \newline\phantom{I}2.48 GHz &
		1 Mbit/s\textendash \newline\phantom{I}24 Mbit/s &
		1 m\textendash \newline\phantom{I}100 m &
		\Circle &
		\textcolor{white}{a}High & 
		\Circle &
		\textcolor{white}{aaa}Eavesdropping \cite{Dunning:2010} \newline 
		\textcolor{white}{aaa}\gls{mitm} \cite{Dunning:2010, Haataja:2010} \newline
		\textcolor{white}{aaa}\gls{dos} \cite{Dunning:2010} &
    	\shortstack{\textcolor{white}{\textbf{aa}} \CIRCLE} &
        Just Works \cite{Bluetooth:2007} \\
    	
    	\hline
    	
    	mm-Waves \newline
		(\textsection \ref{subsubsec:mm-waves}, pp. \pageref{subsubsec:mm-waves}) & 
		Wireless Radio Comm. &
		57 GHz\textendash \newline\phantom{I}64 GHz &
		6.75 Gbit/s &
		10 m &
		\Circle &
		\textcolor{white}{a}Low & 
		\CIRCLE &
		\textcolor{white}{aaa}Eavesdropping\cite{Steinmetzer:2015} &
    	\shortstack{\textcolor{white}{\textbf{aa}} \Circle} &
    	Push Button Configuration \cite{PBC:2006} \\
    	
    	\hline
    	
    	RFID \newline
		(\textsection \ref{subsubsec:rfid}, pp. \pageref{subsubsec:rfid}) &  
		Wireless Radio Comm. &
		120 kHz\textendash \newline\phantom{I}150 kHz,\newline 
		13.56 MHz, \newline 
		850 MHz\textendash \newline\phantom{I}960 MHz, \newline 
		2.45 MHz, \newline 
		5.8 GHz, \newline 
		3.1 GHz\textendash \newline\phantom{I}10.6 GHz, &
		4 kbit/s\textendash \newline\phantom{I}1.5 Mbit/s &
		10 m \newline (passive) \newline 
		\textcolor{white}{aaa} \newline
		\textcolor{white}{aaa} \newline
		100 m \newline (active) &
		\Circle \newline 
		\textcolor{white}{aaa} \newline
		\textcolor{white}{aaa} \newline 
		\textcolor{white}{aaa} \newline
		\Circle &
		\textcolor{white}{a}Medium \newline  
		\textcolor{white}{aaa} \newline
		\textcolor{white}{aaa} \newline 
		\textcolor{white}{aaa} \newline
		\textcolor{white}{a}High & 
    	\Circle \newline
    	\textcolor{white}{aaa} \newline
		\textcolor{white}{aaa} \newline 
		\textcolor{white}{aaa} \newline
		\Circle &
		\textcolor{white}{aaa}Eavesdropping \cite{Kasper:2011} \newline
		\textcolor{white}{aaa}Unauthorized access \newline
		\textcolor{white}{aa}\cite{Czeskis:2008, Zhang:2009} \newline
		\textcolor{white}{aaa}Relay \cite{Czeskis:2008, Francillon:2011} \newline
		\textcolor{white}{aaa}\gls{dos} \cite{Kasper:2011} & 
		\shortstack{\textcolor{white}{\textbf{aa}} \Circle} &
		Noisy tags \cite{Castelluccia:2006} \newline
		Adopted-Pet \cite{Amariucai:2011} \\ 
    	
    	\hline
    	
    	NFC \newline
		(\textsection \ref{subsubsec:nfc}, pp. \pageref{subsubsec:nfc}) & 
		Wireless Radio Comm. &
		13.56 MHz &
		424 kbit/s &
		10 cm &
		\Circle &
		\textcolor{white}{a}Medium & 
    	\Circle &
		\textcolor{white}{aaa}Eavesdropping\cite{Zhou:2014} \newline
		\textcolor{white}{aaa}Unauthorized access \cite{Zhang:2009} \newline 
		\textcolor{white}{aaa}Relay \cite{Francis:2010} & 
    	\shortstack{\textcolor{white}{\textbf{aa}} \CIRCLE} &
    	Near Field Communication \cite{WPS:NFC} \newline
    	Out-of-band \cite{SSP:OOB} \\
    	
		\hline    	
    	
    	VLC \newline
		(\textsection \ref{subsubsec:vlc}, pp. \pageref{subsubsec:vlc}) &  
		Wireless Visible Comm. &
		400 THz\textendash \newline\phantom{I}800 THz &
		11 kbit/s\textendash \newline\phantom{I}96 Mbit/s &
		10 m &
		\CIRCLE &
		\textcolor{white}{a}Low & 
    	\CIRCLE &
		\textcolor{white}{aaa}Eavesdropping\cite{Classen:2015} &
    	\shortstack{\textcolor{white}{\textbf{aa}} \Circle} &
    	KeyLED \cite{Roman:2008} \newline
    	Enlighten Me! \cite{Gauger:2009} \newline
    	Flashing displays \cite{Kovavcevic:2015} \\
    	
    	\hline
    	
    	Visual \newline
		(\textsection \ref{subsubsec:visual}, pp. \pageref{subsubsec:visual}) &  
		Wireless Visual Comm. &
		400 THz\textendash \newline\phantom{I}800 THz &
		12 Mbit/s, \newline
		324 kbit/s &
		10 m \newline
		20 cm &
		\CIRCLE &
		\textcolor{white}{a}Low & 
    	\CIRCLE &
		\textcolor{white}{aaa}Eavesdropping \cite{Zhang:2016} \newline 
		\textcolor{white}{aaa}Replay \cite{Rahman:2013} &
    	\shortstack{\textcolor{white}{\textbf{aa}} \CIRCLE} &
    	SBVLC \cite{Zhang:2016} \\ 
    	
		\hline 
    	
    	IrDA \newline
		(\textsection \ref{subsubsec:irda}, pp. \pageref{subsubsec:irda}) & 
		Wireless Infrared Comm. &
		334 THz\textendash \newline\phantom{I}353 THz &
		2.4 kbit/s\textendash \newline\phantom{I}1 Gbit/s &
		1 m &
		\Circle &
		\textcolor{white}{a}Low & 
    	\CIRCLE &
    	\textcolor{white}{aaa}Replay \cite{tv-be-gone:overview} \newline
		\textcolor{white}{aaa}Eavesdropping\cite{Classen:2015} &
    	\shortstack{\textcolor{white}{\textbf{aa}} \Circle} &
    	Talking to Strangers \cite{Balfanz:2002} \\
    	
		\hline    	
    	
    	Audio \newline
		(\textsection \ref{subsubsec:audio}, pp. \pageref{subsubsec:audio}) & 
		Wireless Acoustic Comm. &
		20 Hz\textendash \newline\phantom{I}20 kHz &
		20 bit/s, \newline
		4.7 kbit/s &
		\url{~}20 m, \newline
		\url{~}4 m &
		\CIRCLE &
		\textcolor{white}{a}Medium & 
    	\Circle &
		\textcolor{white}{aaa}Eavesdropping \cite{Halevi:2013} \newline
		\textcolor{white}{aaa}Relay \cite{Shrestha:2015} & 
    	\shortstack{\textcolor{white}{\textbf{aa}} \CIRCLE} &
    	Loud and Clear \cite{Goodrich:2006} \newline
    	HAPADEP \cite{Soriente:2008} \newline
    	Zero-Power pairing \cite{Halperin:2008} \\
    	
    	\hline
    	
    	Ultrasound \newline
		(\textsection \ref{subsubsec:ultrasound}, pp. \pageref{subsubsec:ultrasound}) &  
		Wireless Acoustic Comm. &
		20 kHz\textendash \newline\phantom{I}20 MHz &
		230 bit/s, \newline  
		2 kbit/s &
		11 m \newline
		2 m &
		\Circle &
		\textcolor{white}{a}Low & 
    	\Circle &
		\textcolor{white}{aaa}Eavesdropping \cite{Mayrhofer:2007} \newline
		\textcolor{white}{aaa}Relay \cite{Mayrhofer:2007, Clulow:2006} & %
    	\shortstack{\textcolor{white}{\textbf{aa}} \RIGHTcircle} &
    	Ultrasonic ranging \cite{Mayrhofer:2006} \\ 
    	
    	\hline
    	
    	Haptic \newline
		(\textsection \ref{subsubsec:haptic}, pp. \pageref{subsubsec:haptic}) &  
		Wireless Haptic Comm. &
		40 Hz\textendash \newline\phantom{I}800 Hz &
		200 bit/s &
		physical \newline
		contact &
		\CIRCLE &
		\textcolor{white}{a}Low & 
    	\Circle &
		\textcolor{white}{aaa}Eavesdropping \cite{Halevi:2013} &
    	\shortstack{\textcolor{white}{\textbf{aa}} \CIRCLE} &
    	Vibrate-to-unlock \cite{Saxena:2011vibrate} \newline
    	Shot \cite{Studer:2011} \newline
    	Vibreaker \cite{Anand:2016} \\
    	
    	\hline	
    	
    	Sensing \newline
		(\textsection \ref{subsubsec:sensing}, pp. \pageref{subsubsec:sensing}) &  
		Onboard Sensors &
		n/a &
		n/a &
		n/a &
		\RIGHTcircle &
		\textcolor{white}{a}n/a & 
    	\RIGHTcircle &
    	\textcolor{white}{aaa}Relay \cite{Shrestha:2015, Schultes:2016} \newline
		\textcolor{white}{aaa}Context-manipulation \newline
		\textcolor{white}{aa}\cite{Shrestha:2015} \newline
		\textcolor{white}{aaa}Reproducible readings \newline
		\textcolor{white}{aa}\cite{Studer:2011} \newline
		\textcolor{white}{aaa}GPS: Spoofing, jamming \newline
		\textcolor{white}{aa}\cite{Tippenhauer:2011} &
    	\shortstack{\textcolor{white}{\textbf{aa}} \RIGHTcircle} &
    	Amigo \cite{Varshavsky:2007} \newline
    	Good Neighbor \cite{Cai:2011} \newline
    	Wanda \cite{Pierson:2016} \newline
    	Ambient Audio pairing \cite{Schurmann:2013} \newline
    	Zero-interaction pairing \cite{Miettinen:2014} \newline
    	MagPairing \cite{Jin:2014} \newline
    	Touch-And-Guard \cite{Wang:2016} \\ 
    	
		\hline
		\hline
		\end{tabularx}
		}}
		\begin{tablenotes}[para,flushleft]
			\item \\
			\item 
			\item 
			\item 
			\item 
			\item 
			\item 
			\item 
			\item 
			\item \CIRCLE\textcolor{white}{-}= fulfills property; \RIGHTcircle\textcolor{white}{-}= partly fulfills property; \Circle\textcolor{white}{-}= does not fulfill property 
    	\end{tablenotes}
  \end{threeparttable}
\end{table*}

\subsection{Known Attacks}
To better understand the security implications of \gls{phy} channels in \gls{sdp}
we summarize the most prominent attacks that have been reported
on various physical media. This list of attacks is by no means exhaustive but provides
an overview of common possible attacks vectors. In the literature many attacks
on widespread wireless radio channels can be found. We summarize those and
additionally present security implications in other communication channels such
as visible light, audio, etc. Overall, our study provides a deep insight into
existing threats and outlines differences in vulnerabilities and security
properties among different \gls{phy} channels.
 

\subsection{Ease-of-adoption}
In order to evaluate how feasible it is for a specific channel to be adopted
in the context of \gls{sdp} we compare it with a set of available
interfaces on widespread hardware such as smartphones. We make this
particular choice because of the ubiquitous nature of smartphones which are
viably considered as a gateway for the personal \gls{iot} environment
\cite{Kim:2015} as well as the pairing mediator for the \gls{iot} devices
\cite{Suomalainen:2014}. In detail, we employ the hardware characteristics of
the fifth generation of Nexus devices \cite{Nexus:family} as a reference to identify the
common interfaces present on an average smartphone.  

\subsection{Survey of \gls{phy} Channels}
In the following, we survey \gls{phy} channels suitable for \gls{sdp}
by focusing on the previously described properties.

\subsubsection{Wi-Fi Channel}
\label{subsubsec:wifi}
Wi-Fi is a wireless communication technology based on a set of IEEE 802.11
standards and is used to connect devices within a wireless local area network. 
The most common Wi-Fi standards such as  802.11 a/b/g/n/ac operate in 2.4 and 5 GHz frequency bands. 
Other frequency bands, for example, around 60 GHz are also standardized (IEEE 802.11ad) but less frequently used. 
Due to different propagation characteristics at high frequencies, we discuss IEEE 802.11ad separately in the  
mm-wave section (\ref{subsubsec:mm-waves}).

Different modes of operation are available for Wi-Fi: infrastructure, direct, and ad-hoc. The infrastructure mode is
established with a centralized \gls{ap}, which handles all network traffic from
connected stations. The latter two modes are formed in a peer-to-peer fashion directly by
the devices. While Wi-Fi direct is intended to be applied in use-cases where ad-hoc Wi-Fi was previously envisaged for use, its implementation is very different. The primary difference is that Wi-Fi direct internally uses the infrastructure mode, while ad-hoc Wi-Fi remains a separate mode. Because the specification of ad-hoc Wi-Fi has not been substantially updated since the 802.11b standard, more recent Wi-Fi security and performance improvements have not necessarily been incorporated into ad-hoc Wi-Fi implementations. This would be of limited concern, were it not for the continued use of ad-hoc Wi-Fi in certain applications, particularly those where multi-hop mesh networking is required \cite{Adeyeye:2011, Gardner:2011}. While some technologies now widely adapt Wi-Fi direct \cite{Direct:Android, Shen:2016}, the pure ad-hoc mode is relatively rarely used despite the number of its advantages.

The data rates of Wi-Fi communication have increased significantly over the last
decade and can exceed 1 Gbit/s \cite{Gong:2015}. Wi-Fi coverage varies from 30
to 250 meters \cite{Banerji:2013, Atenas:2010} for indoor and outdoor
environments respectively. The 5 GHz band has a smaller communication range
due to a shorter wavelength and higher attenuation as compared to 2.4 GHz
\cite{Motorola:2009}. Wi-Fi communication is human-imperceptible and enables
omnidirectional transmission with signals propagating through most non-metal
objects such as walls, doors and windows.
 
Since Wi-Fi channels are inherently broadcast and have wide coverage, they are
susceptible to a number of threats \cite{Aime:2007}. Adversaries may, for example, 
obtain unauthorized access to intercept transmitted information,
inject and modify data in the air with surgical precision, 
reroute traffic for analysis with \gls{mitm} attacks \cite{Fund:2016}, or
efficiently jam the network to cause a \gls{dos} \cite{Lu:2010, Schulz:2017}.
Such attacks have been shown to be feasible with low effort \cite{Bayraktaroglu:2013}. 
Moreover, due to the widespread use of Wi-Fi, identity tracking might threaten user privacy \cite{Musa:2012}.

As Wi-Fi chips are ubiquitous
and integrated in a wide range of devices starting from powerful laptops to
resource-constrained sensors, the technology became a de-facto standard for
communication of mobile devices. 
In the following, we describe various pairing schemes which utilize the
Wi-Fi channel to accomplish pairing.

\textit{\gls{pbc}} was introduced as a part of standardized \gls{wps}
\cite{PBC:2006} which incorporates two other pairing schemes known as ``Pin
Entry'' and ``Near Field Communication''. The pairing is initiated when a user
presses a button on one device (enrollee) which starts searching for a
\gls{pbc}-enabled peer within its range to complete pairing. Once a button is
pressed on the second device (registrar) an unauthenticated \gls{dh} key
exchange is performed via the Wi-Fi channel. 

Despite the fact that the \gls{pbc} pairing
scheme is implemented on real devices and provides protection against passive
eavesdroppers, it is inherently vulnerable to active adversaries who can mount
\gls{mitm} attacks. 

Capkun et al. \cite{Capkun:2008} proposed \textit{integrity codes (I-codes)} a
security mechanism that enables authentication and integrity protection of
messages exchanged over a wireless radio channel. In order to achieve the
stated purpose, I-codes rely on three components: unidirectional message
coding, on-off keying communication, and the ability of the receiver to
determine if the transmitter is within its communication range. The authors
showed that authentication through presence can be achieved if communicating
devices are aware of each other's reception distance and are synchronized
with respect to the start of transmission. 

Security properties of I-codes were discussed in the presence of a powerful attacker who has full
control over a wireless channel except for her inability to disable the whole
communication, for example, remove the energy of a signal. Based on I-codes a
new version of the \gls{dh} protocol was proposed which was claimed to be optimal
in the sense of transmitted message length and the corresponding security level.

Gollakota et al. suggested \textit{\gls{tep}} \cite{Gollakota:2011}, a
scheme that utilizes on-off coding to prevent \gls{mitm} attacks on the
wireless channels. Specifically, the authors introduced a primitive called
\gls{tea} which completely prevents active attackers from either changing the
content of a transmitted message or hiding the fact that the message was sent.
To achieve the stated goal the \gls{tea} mechanism introduces silence periods.
Particularly, the payload of the \gls{tea} message is appended by a sequence
of short equal-sized packets called slots in which the transmitter chooses to
either send data (on-slot) or remain idle (off-slot). 

The \gls{tep} scheme uses a bit sequence produced by on-off slots to encode the hash of the
\gls{tea} payload. In this case, an adversary might tamper with the off-slots
by transmitting a signal, while she cannot remove energy from the on-slots.
Hence, attackers that have no physical access to the pairing devices are
prevented from tampering with the transmitted signal, they can neither
suppress the communication nor create a capture effect \cite{Ware:2000}.

\subsubsection{Bluetooth Channel}
\label{subsubsec:bluetooth}
Bluetooth is a wireless communication technology which operates in the 2.4 GHz
frequency band and is used to connect several devices in an ad-hoc manner,
thus forming a personal area network \cite{Bluetooth:101}. Typical data rates
for Bluetooth are 1{\textendash}3 Mbit/s but can reach 24 Mbit/s \cite{Scarfone:2008}.
Bluetooth coverage varies from 1 to 100 meters depending on the utilized
antennas \cite{Dunning:2010} and can be used in both indoor and outdoor
environments. Physical characteristics of Bluetooth communication are similar
to those of 2.4 GHz Wi-Fi, therefore it cannot be sensed by humans, achieves
relatively high penetration of solid objects and does not require \gls{los}
for data transmission. 

Bluetooth communication is vulnerable to similar security issues as Wi-Fi, 
despite the fact that a Bluetooth channel is more difficult to access due to its shorter range. 
In addition, attacks to extend over the nominal communication range, obtain unauthorized data access, 
or fuzz protocol implementations to reveal vulnerabilities have been shown to be feasible \cite{Dunning:2010} with low hardware requirements. 
Moreover \gls{mitm} attacks \cite{Haataja:2010}, as well as \gls{dos} \cite{Dunning:2010} are as practical as in Wi-Fi.

Currently, low-power Bluetooth chips are pervasive and can be found in
billions of devices. Further, we review a prominent pairing scheme
that relies on the Bluetooth channel. 

\textit{Secure Simple Pairing} \cite{Bluetooth:2007} proposed by the Bluetooth
SIG is a de facto standard for pairing multiple personal devices. With a
recent security enhancement \cite{Bluetooth:les} there are now four schemes
available for Bluetooth pairing: ``Just Works'', ``Numeric comparison'',
``Passkey Entry'' and ``Out-of-band''. However, only the first scheme
solely relies on the Bluetooth channel to achieve pairing, whereas others utilize
\gls{hci} or other \gls{phy} channels, for example, \gls{nfc}, to ensure
authenticity. 

The \textit{Just Works} scheme is used to perform pairing
with constrained devices, for example, a headset, which lack convenient
input/output capabilities such as a keyboard or a display. In essence, Just
Works is based on an unauthenticated \gls{dh} key exchange which provides
protection against passive eavesdroppers but is inherently vulnerable to
active \gls{mitm} attackers \cite{Bluetooth:2007}. 

\subsubsection{mm-Waves Channel}
\label{subsubsec:mm-waves}
mm-Wave wireless communications operate in a wide frequency band from 30 to
300 GHz. The lower part of the mm-Wave spectrum (30{\textendash}50 GHz) is considered to
be used in cellular and indoor environments with the coverage of up to 200
meters \cite{Rappaport:2013}, although high-speed outdoor point-to-point links
can work over longer distances \cite{Huang:2011}. At higher frequencies, an
unlicensed spectrum around 60 GHz is being standardized (IEEE 802.11ad
\cite{80211ad:2012}) and deemed to be actively used for a great variety of
indoor applications \cite{WiGig:2013}. 

With mm-Waves very high data rates are possible due the wide channel bandwidths available. 
For example, IEEE 802.11ad achieves transmission speed of up to 6.75 Gbit/s
within the coverage area of up to 10 meters \cite{Dash:2012}. 
Due to high attenuation and absorption rates mm-Waves at 60 GHz do not propagate through solid objects,
for example, walls, and the \gls{los} requirement is imposed on the mm-Wave communication \cite{Niu:2015}. 
Being a part of the microwave spectrum mm-Waves cannot be perceived by humans. 

The plausible properties of mm-Waves such as the short range, \gls{los}
transmission and no wall penetration were claimed to provide highly secure
operation \cite{Huang:2011}. However, as was recently shown
\cite{Steinmetzer:2015} eavesdropping is possible on a 60 GHz channel through
reflections caused by small-scale objects located within a transmission beam.
At the moment, 60 GHz chips can only be found in a few commercial products,
for example, \cite{60ghz:2015}, but the number of supported devices will
undoubtedly increase in the medium term. Next, we describe a pairing scheme
which uses the mm-Wave channel.   

Despite being a relatively new technology, 60 GHz communication has
already adopted pairing schemes from the standardized \gls{wps} such as
\textit{\gls{pbc}} \cite{PCB:60ghz}. Nevertheless, the \gls{pbc} pairing is
susceptible to \gls{mitm} attacks as stated above. However, due to the
short-range transmission with \gls{los}, an adversary would have to be
co-present, that is, in the same room, and interfere within a transmission
beam in order to mount such an attack. These actions are much harder to
perform stealthily without a benign user noticing them, which was not the case
for the legacy Wi-Fi \gls{pbc}. 

\subsubsection{\gls{rfid} Channel}
\label{subsubsec:rfid}
\gls{rfid} is a wireless communication technology which is used for automatic
identification in both indoor and outdoor environments. That is, an \gls{rfid}
system consists of tags (active or passive) which store the identification
information and readers that query the tags in order to extract and verify
that information \cite{Weis:2007}. More ubiquitous passive tags have to
harvest energy from nearby \gls{rfid} reader's interrogating radio waves,
whereas active tags have on-board power supply, for example, a battery. 

\gls{rfid} operates in several frequency bands \cite{Weis:2007}: Low Frequency
(120{\textendash}150 kHz), High Frequency(13.56 MHz), Ultra-High Frequency (860{\textendash}960 MHz),
Microwave (2.45 and 5.8 GHz), Ultra-Wide Band (3.1{\textendash}10.6 GHz). Typical data
rates vary from several to hundreds of kbit/s and depend on the utilized
spectra \cite{RFID:2010}. The coverage that was reported for the \gls{rfid}
technology ranges from 10 to 100 meters for passive and active tags
correspondingly \cite{Weis:2007, Sen:2009}. Regardless of the underlying
frequency, \gls{rfid} communication cannot be sensed by humans. However, the
capability of \gls{rfid} transmission to pass through solid objects depends on
the used spectrum as well as the employed antenna and is higher for active
\gls{rfid}. For sending and receiving data with \gls{rfid} \gls{los} is not
required. 

There are several security concerns regarding \gls{rfid} communication. First,
its wireless nature poses threats similar to Wi-Fi and Bluetooth which are
especially prominent for active tags operating over longer distances
\cite{Kasper:2011}. Second, passive tags are very constrained devices and can
promiscuously respond to any reader request \cite{Ma:2014} despite being
short-range. It was shown that \gls{rfid} channels are vulnerable to
eavesdropping, unauthorized access, relay and \gls{dos} attacks
\cite{Czeskis:2008, Kasper:2011, Francillon:2011, Zhang:2009}.
The relay attacks on contactless smart cards are especially severe, as they
can easily circumvent security of payment and access control systems \cite{Drimer:2007, Francillon:2011}, and 
defending against such relay attacks is an active research area \cite{Ma:2014, Ho:2016, Ranganathan:2017}.

\gls{rfid} tags are presently ubiquitous and can be found in many applications such as
logistics, tracking and access control \cite{RFID:2013}. Nevertheless, most
consumer devices, for example, smartphones, are not supplied with built-in
\gls{rfid} chips and use the \gls{nfc} technology instead. Several representative pairing
schemes which employ the \gls{rfid} channel are described below.  

Castelluccia and Avoine \cite{Castelluccia:2006} presented a paring scheme called
\textit{Nosy Tags} for secure key establishment over a wireless \gls{rfid}
channel between a passive tag and a reader. In essence, the pairing scheme relies
on noise injection into a public communication channel, which makes the actual
signal meaningless for an eavesdropping adversary, but allows the reader to
efficiently restore the original message. This idea is implemented by
introducing an extra \gls{rfid} tag (a nosy tag) which belongs to the reader
and shares a secret key with it. 

The proposed pairing scheme works as follows. When the reader
queries a passive tag within its proximity the nosy tag generates a sequence
of random bits, which prevents the eavesdropper from differentiating between
the original message sent by the queried tag and the one injected by the nosy
tag. On the reader's side the generated noise can be subtracted to recover the
actual signal. 

The authors provided three variants of their pairing scheme based on
the nosy tags and analyzed its security against passive attackers.
Nevertheless, the pairing scheme can still be circumvented by active adversaries. 

Amariucai et al. \cite{Amariucai:2011} suggested \textit{Adopted-Pet}, an
automatic time-based scheme for pairing a passive \gls{rfid} tag with a
reader without any human interaction or additional \gls{phy} channels, for
example, \gls{nfc}. The main idea is as follows. A tag can reassure that a
particular reader is trusted only if it spends a sufficient amount of
uninterrupted time within the proximity of this reader. Specifically, a tag
has to be interrogated only by a single reader (uninterrupted property) for a
time period during which the tag gradually transmits pieces of its secret
password which are accumulated by the reader in order to eventually restore
the secret. 

The authors implemented their pairing scheme using a linear-feedback shift
register and argued that it is robust against adversaries who can spend
numerous interrupted time intervals in the proximity of a victim tag.

\subsubsection{\gls{nfc} Channel}
\label{subsubsec:nfc}
\gls{nfc} is a wireless communication technology which is used to establish
point-to-point communication between two devices brought to close proximity.
\gls{nfc} is an offshoot of \gls{rfid} technology, thus \gls{nfc} devices can
similarly be active or passive \cite{NFC:101}. \gls{nfc} operates in 13.56 MHz
frequency band and supports data rates of up to 424 kbit/s
\cite{Francis:2010}. \gls{nfc} has very limited coverage of up to 10 cm
\cite{Francis:2010}. Similarly to \gls{rfid}, \gls{nfc} communication cannot
be perceived by humans, is able to penetrate solid object to a certain degree,
and does not require \gls{los} for data transmission. 

Initially, security assumptions about \gls{nfc} were based on its very short
range and, hence, severe difficulty for an attacker to access it. However,
recently it was shown that eavesdropping on \gls{nfc} channels is possible at
a distance of up to 240 cm \cite{Zhou:2014}. In addition, unauthorized
readings \cite{Zhang:2009} pose a real threat which can lead to practical
relay attacks on the \gls{nfc} communications \cite{Francis:2010}. 
With the advent of mobile \gls{nfc} payments, the relay attacks on such systems
have become a severe security threat \cite{Markantonakis:2012, Roland:2013, Maass:2015}, 
which has not yet been fully addressed \cite{Mehrnezhad:2015, Gurulian:2016}.  

\gls{nfc} chips are widely deployed, and can be found in numerous smartphones and other
devices. We present two pairing schemes which utilize the \gls{nfc}
channel.  

\textit{Near Field Communication} \cite{WPS:NFC} is a pairing scheme from the
standardized \gls{wps} mentioned previously. Specifically, the \gls{nfc}
channel can be used to transmit a hardware generated password from one device
that initiates pairing (enrollee) to another device (registrar) with which the
pairing should be performed. Another pairing setting available with
\gls{wps} \gls{nfc} is to exchange hashes of public keys between the enrollee
and registrar once they are brought to close proximity, that is, physical contact.

The security assumptions of \gls{wps} \gls{nfc} are based on the limited
communication range provided by the \gls{nfc} technology, which is much more
difficult for an adversary to eavesdrop. 

\textit{Out-of-band} \cite{SSP:OOB} is a pairing scheme provided by
standardized Bluetooth Secure Simple Pairing. It works as follows. Once two
devices have discovered each other via the Bluetooth channel, the \gls{nfc}
channel is used to exchange authentication information, for example, Hash C,
Randomizer R or TK-value, between the devices in order to accomplish pairing. 

The security arguments for the Out-of-band scheme rely on the restricted nature of the 
\gls{nfc} communication which cannot be easily accessed by an attacker.

\subsubsection{\gls{vlc} Channel}
\label{subsubsec:vlc}
\gls{vlc} is a wireless communication technology which carries information by
modulating light in the visible spectrum that is used for
illumination \cite{Arnon:2015}. \gls{vlc} operates in the 400{\textendash}800 THz frequency
band and is widely considered to be used for indoor short range communications
\cite{Arnon:2015}. The data rates that can be achieved by the existing
standard (IEEE 802.15.7 \cite{802157:2011}) vary from 11.67 kbit/s to 96
Mbit/s \cite{Rajagopal:2012}, although recent research demonstrated throughput
of up to 20 Gbit/s\cite{Hussein:2015}. Typically, \gls{vlc} has coverage of up
to 10 meters \cite{Pathak:2015} and it is perceived by humans via the sight
sense. \gls{vlc} transmission requires \gls{los} and cannot penetrate
non-transparent solid objects such as walls and doors. 

Therefore, \gls{vlc} communication is concealed, to some extent, from an
adversary who is not co-present. However, recently it was shown that \gls{vlc}
can be efficiently eavesdropped by the attacker located outside of the room
where communication happens \cite{Classen:2015}. Additionally, in
\cite{Perkovic:2012} it was discussed that the integrity of a \gls{vlc}
channel can be affected by an adversary using a directional light source, for
example, a laser. 

At present, there are no commercial devices, for example,
smartphones, that support a standardized \gls{vlc} technology (IEEE 802.15.7).
However, fully functional prototypes \cite{Rigg:2014} have been recently
demonstrated, which makes it feasible that the \gls{vlc}-enabled devices will
appear on the market soon. In the following, we review representative pairing schemes 
that rely on the \gls{vlc} channel. 

Roman and Lopez \cite{Roman:2008} studied the applicability of visible light
as \gls{oob} in the context of \gls{wsn} where two previously unknown devices
want to exchange sensitive information. The authors developed a scheme
called \textit{KeyLED} with which two constrained sensors can pair.
Particularly, two devices located in close proximity utilize a LED-photosensor
pair to set up a short distance communication channel (few cm) and transmit
their public keys using on-off keying. 

The security of the proposed pairing scheme was discussed with respect to 
eavesdropping, injection and \gls{dos} attacks. It was claimed that 
even though such threats are feasible, mounting them in
practice is difficult and a benign user who initiates communication between
two sensors can be easily alerted in case of the attack.   
    
The similar line of work by Gauger et al. \cite{Gauger:2009} investigated an
ad-hoc key assignment for devices in \gls{wsn}. The suggested \textit{Enlighten
Me!} scheme was considered for two application scenarios: \textit{a)} initial
key assignment as a part of the \gls{wsn} configuration \textit{b)} dynamic
key re-assignment of already deployed sensors. 

The proposed pairing scheme works as follows.
There is a master device (key sender), which provides a set of sensors (key
receiver) residing within its wireless range with secret keys using an
auxiliary light channel. During the key assignment procedure, the discovery of
a receiving device, secret transmission and verification is achieved with a
light source-sensor channel using Manchester coding. 

The authors implemented two
types of key senders: a sensor node lamp with a powerful LED and a smartphone
with a varying brightness level on a display. For both prototypes they argued
that eavesdropping the transmitted key is difficult to achieve in practice,
because the realized \gls{vlc} channel can be effectively concealed from an
outside observer. 

Kova{\v{c}}evi{\'c} et al. \cite{Kovavcevic:2015} proposed \textit{Flashing
displays}, two multichannel deployment schemes for secure initialization of
wireless sensors using only a multi-touch screen of a smartphone or a
tablet as a light source. Particularly, both schemes utilize two channels:
wireless radio and \gls{vlc}, where the former is considered as insecure and
the latter is used as \gls{oob}. 

The first scheme relies on a visible light channel that is established between a display of a smartphone and a light
sensor of a constrained device once it is put on top of the screen. In this
setting, several constrained devices can simultaneously receive secret keys which have
to be verified later on over a wireless radio channel. 

The second scheme was introduced in order to address a powerful adversary who can still eavesdrop on
the \gls{vlc} channel via electromagnetic emissions of a flashing display.
Specifically, the developed mechanism incorporated both the \gls{vlc} channel,
for synchronization purposes, together with customized integrity codes
\cite{Capkun:2008}. The authors showed that such a pairing scheme is secure against an
attacker who can read the content of the flashing screen at any moment in
time. 

\subsubsection{Visual Channel}
\label{subsubsec:visual}
A visual channel enables wireless communication in the visible light spectrum
(400{\textendash}800 THz) by utilizing currently abundant LCD-camera hardware. Such
real-time transmission was shown feasible at 12 Mbit/s within a distance of 10
meters using large displays and high-speed digital cameras \cite{Perli:2010}.
Another line of research \cite{Hao:2012, Wang:2014, Wang:2015rain}
investigated visual communication that can be established with the LCD-camera
found on commodity hardware such as smartphones. The results indicate that
data transmission at 324 kbit/s is possible in the vicinity of 20 cm. 

Such visual channels, whose properties include short range of communications
and interference-free operation, were claimed to provide secure transmission
\cite{Perli:2010, Niu:2015vince}. However, the fact that an LCD-camera channel
can be observed and easily interpreted by humans comes with a drawback.
Specifically, eavesdropping either in a from of shoulder surfing or ubiquitous
CCTV was shown to be a real threat \cite{Zhang:2016}, especially taking into
account the continuous increase of display sizes\cite{Barredo:2014,
Statista:dispaly} and recent advances in CCTV \cite{Dziech:2013}. Another
security issue that was raised is related to the ``liveness'' of the captured
video stream, which can lead to replay attacks \cite{Rahman:2013}. 

At present, camera-display peripherals are ubiquitous on numerous devices such as smartphones. 
Further, we describe a pairing scheme based on the visual channel. 

Zhang et al. \cite{Zhang:2016} investigated secure bar-code communication for
smartphones. They proposed \textit{SBVLC}, a novel approach for secure ad-hoc
interactions which can be established via a short-range LCD-camera channel on
mobile devices. 

The authors suggested a pairing scheme based on SBVLC
that works as follows. To pair, two parties utilize a full duplex LCD-camera
channel which is realized as a sequence of QR-codes displayed on the screen of
one device and captured by the camera of another device. Specifically, once
two smartphones are brought to physical proximity, that is, within a few
inches, they start to simultaneously exchange key material using the described
visual channel. Afterwards, one of the devices randomly picks a universal hash
function which is used to build a shared secret key from the material
accumulated by both parties. 

The security of the proposed approach was
formally analyzed against the eavesdropping adversary by employing 2D and 3D
geometric models. In addition, it was shown that proactive rotation of the
devices during pairing can enhance security, since the
attacker is forced to capture frames simultaneously from both displays to undermine the pairing scheme.
Moreover, the authors showed that the established key has enough entropy, that is,
cannot be recovered, if the adversary misses at least one frame during the key
exchange, which further improves security. 

\subsubsection{\gls{irda} Channel}
\label{subsubsec:irda}
\gls{irda} is a set of wireless communication technologies that uses the
infrared radio spectrum 334{\textendash}353 THz \cite{Won:2008} for point-to-point data
transmission. Since \gls{irda} is susceptible to interference from ambient
light sources \cite{Hallberg:2002} it is mostly considered for indoor
applications. With the \gls{irda} communication high data rates of up to 1
Gbit/s \cite{IrDA:2011} are possible. \gls{irda} has coverage of up to 1
meter, is human-imperceptible and requires direct \gls{los}. 

\gls{irda} was claimed \cite{IrDA:2011} to provide secure data transmission
due to its short range, directional operation (a 30{\degree} beamwidth) and
the fact that infrared communication cannot traverse through solid objects
such as walls and doors. However, such claims cause controversy because with
toolkits like \textit{TV-B-Gone} \cite{tv-be-gone:overview} a variant of
replay attacks can be mounted over relatively long distances
\cite{tv-be-gone:hackaday}. In addition, the eavesdropping attack through
reflections recently demonstrated on \gls{vlc} \cite{Classen:2015} can be
feasibly applied to the infrared channel since the two media have very similar
physical properties. 

Presently, many \gls{irda}-enabled devices can be found
in consumer electronics and household appliances, although such technology is
obsolete on modern smartphones. Nevertheless, there is a growing number of
personal devices supplied with IR-blasters \cite{IRB:2014} which indicates the
restored interest to the infrared communication. 
One such pairing scheme which makes use of the infrared channel is described next.

Balfanz et al. \cite{Balfanz:2002} suggested a pairing scheme known as
\textit{Talking To Strangers}. The core idea behind was to combine
demonstrative identification from the user perspective, for example, two
physical devices with which a user interacts, with location-limited channels
that aim to provide data authenticity. The latter concept denotes exactly the
\gls{oob} channel. 

The basic pairing scheme proposed by the authors works as follows.
First, two devices exchange commitments to their public keys, that is,
hashes, over an \gls{irda} communication channel which serves as the \gls{oob}
channel. Second, they transfer their corresponding public keys over a wireless
radio channel. The wireless communications is verified against the initial
commitment that was transmitted over the infrared channel. In addition,
several other schemes were developed upon the basic pairing approach which dealt
with constrained devices and group pairing. 

The authors discussed
security implications of the proposed pairing and pointed out that an adversary
would have to actively intercept the \gls{oob} channel in order to undermine
the pairing scheme.

\subsubsection{Audio Channel}
\label{subsubsec:audio}
Audio is a mechanical pressure wave caused by periodic vibrations within an audible
frequency range of 20 Hz{\textendash}20 kHz \cite{Rosen:2011}. An audio channel in this case
would be represented by a speaker-microphone pair, where the former generates
a sound and the latter records it. 

The line of research \cite{Madhavapeddy:2005, Hanspach:2014, Lee:2015,
Carrara:2014} investigated the throughput and coverage of the audio channel
utilizing different modulation schemes. The results indicate that with
inexpensive speakers and microphones found on commodity hardware, such as
laptops, data rates can go from 20 bit/s to 4.7 kbit/s over distances from
19.7 to 3.89 meters respectively. Naturally, the audio channel can be
perceived by humans via the hearing sense. The audio signal can, to a certain
extent, pass through solid objects, for example, walls, although the
penetration capability very much depends on the used frequency and environment
which determine pass loss factors \cite{Audio:101}. For transmission the audio
channel does not require \gls{los}, however, the signal reception is largely
affected by several aspects: \textit{a)} intensity of a sound source
\textit{b)} ambient noise \textit{c)} acoustic environment \textit{d)}
directionality and sensitivity of a microphone \cite{Shure:2014}. 

With respect to security, eavesdropping was shown to be easily achievable
using off-the-shelf equipment even for specifically designed short-range sound
waves \cite{Halevi:2013}. Moreover, relay attacks are possible since audio
streaming tools are highly available on mobile devices such as smartphones
\cite{Shrestha:2015}. 

Currently, microphones and speakers are pervasive
on many existing platforms ranging from simple sensors to powerful laptops. We
describe several pairing schemes which utilize the audio channel.   

Goodrich et al. \cite{Goodrich:2006} suggested an approach to human-assisted
authentication of previously unknown devices using the audio channel. 
The developed \textit{Loud and Clear} (LC) pairing scheme requires two devices to be equipped with 
speakers, or when one device does not have a speaker, it should be supplied with a display.

The LC pairing consists of two phases and works as follows. First, both devices exchange their
public keys over a wireless radio channel such as Wi-Fi or Bluetooth. 
Second, the audio channel is used to transmit the hashes of public keys
encoded as MadLib sentences which can be verified by the user.  
Specifically, in case of both devices having speakers the user has to
confirm the equality of the generated audio sequences. 
Whereas, for the speaker-display setting the user needs to ensure that a
sentence played by the first device is similar to the one displayed on the
screen of the second device. 

The authors analyzed the security of the LC pairing scheme and concluded that
\gls{mitm} attacks can be easily detected if the user is diligent when
comparing verification audio sequences. 

Soriente et al. \cite{Soriente:2008} proposed a pairing scheme called
\textit{HAPADEP} which relies on the audio channel to transmit both data and
verification information between previously unknown devices. 

The HAPADEP pairing scheme consists of two steps and works as follows. First, both devices
exchange their public keys over the audio channel using the fast codec which
allows higher transmission speed, but makes the signal meaningless for a user.
During the second phase two devices encode the hash of the exchanged
cryptographic material using the slow codec and play back the sound, for
example, a melody or a MadLib sentence, that can be recognized and verified by
the user. That is, if both audio sequences heard by the user match then the
pairing is considered to be successful. 

The authors provided the implementation of
the HAPADEP pairing and conducted a usability study, which revealed that the
scheme was generally accepted by the users. Moreover, they discussed the
resilience of the proposed pairing to \gls{mitm}, impersonation and \gls{dos} 
attacks. The HAPADEP scheme was cryptographically extended 
in the unified pairing framework \cite{Mayrhofer:2013} to provide \gls{pfs}, 
which further increases security against \gls{mitm} attackers.  

Halperin et al. \cite{Halperin:2008} investigated security and privacy
implications in \gls{imd}. Specifically, they revealed that communication
between the implant and the medical programmer, used for the collection of sensitive data and
\gls{imd} reprogramming, happened without encryption or authentication. This opened the
door for attacks such as eavesdropping, replay and \gls{dos}. To mitigate the
aforementioned threats, the authors proposed zero-power pairing, which can be
applied to batteryless constrained devices such as passive \gls{rfid} tags.

The suggested pairing scheme works as follows. The programmer initiating pairing sends
a RF signal to power the passive component of the \gls{imd}, which, in turn,
generates a session key and broadcasts it as a modulated sound wave that is
recorded by the programmer's microphone. 

The authors reasoned about the security of the
proposed pairing scheme based on two points. First, since the microphone is placed
within a few centimeters of a patient's chest it can easily receive the audio
signal. However, it is very difficult to obtain the same signal from farther
distance without dedicated hardware equipment. Second, by utilizing the audio
channel for the key exchange, the user is provided with audible and tactile
feedback, which brings her attention to the fact of pairing. 

\subsubsection{Ultrasound Channel}
\label{subsubsec:ultrasound}
Ultrasound refers to acoustic waves that lie within a frequency range above
audible sound (20 kHz{\textendash}20 MHz) \cite{Ultrasound:101}. In this spectrum,
frequencies higher than 250 kHz are strongly absorbed by the air and, thus,
mostly used for medical imaging rather than data transmission
\cite{Katho:2012}. As in case of audio, the ultrasound channel is formed by an
ultrasonic speaker-microphone pair, which are based on the piezoelectric
effect to produce high frequency waves \cite{Terzic:2013}. 

Similar to audio, the throughput and transmission range of the ultrasound
channel were evaluated in prior work \cite{Madhavapeddy:2005,
Hanspach:2014, Malley:2014, Lee:2015, Carrara:2014}. Specifically, data rates
can vary from 230 bit/s to 2 kbit/s at corresponding distances of 11 and 2
meters. Contrary to audio, the ultrasound communication cannot be sensed by
humans. When propagating through air, the ultrasound signal is subject to high
reflection and absorption rates caused by solid objects, for example, walls,
which makes the ultrasound communication limited to a single room
\cite{Mayrhofer:2007}. For data transmission with ultrasound \gls{los} is not required. 

As for security implications, it was shown that a co-present adversary can
eavesdrop and manipulate the ultrasound channel, although attacker
capabilities largely depend on her position relative to communicating parties
\cite{Mayrhofer:2007}. Moreover, relay attacks can be mounted on the
ultrasound channel when it is used for a distance estimation as a part of the
authentication procedure \cite{Mayrhofer:2007, Clulow:2006}. 

At the moment, few end-user devices are supplied with dedicated ultrasonic chips. However,
the lower part of the ultrasound spectrum can be generated and recorded by
non-specialized hardware present on existing smartphones \cite{Legendre:2015}
and laptops \cite{Malley:2014}. A pairing scheme that employs the ultrasound
channel is presented below.   

Mayrhofer et al. \cite{Mayrhofer:2006} studied how secure spontaneous
interactions can be established with spatial references. They developed
a pairing scheme that utilizes the ultrasound channel for initial device discovery and
then implicitly for authenticity verification. 

The proposed pairing scheme works as follows. First, two devices become aware of each other and learn
their corresponding distances and relative positions by employing the
ultrasound sensing. Second, both devices perform an unauthenticated \gls{dh}
key exchange over a wireless radio channel. Third, devices authenticate each
other by sending a nonce encrypted with a shared key over a wireless radio
channel and transmitting the plain nonce over the ultrasound channel using the
interlock protocol \cite{Rivest:1984}. Specifically, the nonce value is split
into pieces and each part is transmitted as a delayed ultrasound pulse to
another device, which results in a longer distance than the previously obtained
spacial reference. Thus, the receiving device can subtract the initially learnt
distance from the received measurement to acquire a part of the nonce and
gradually learn the full nonce. 

The authors discussed the security of the
pairing scheme with respect to eavesdropping, relay and \gls{mitm} attacks. It was
claimed that the suggested pairing can mitigate and detect those adversaries
even if they have access to both radio and ultrasound channels (assuming a
benign user is attentive).


\subsubsection{Haptic Channel}
\label{subsubsec:haptic}
A haptic channel is formed by low frequency waves within a range of 40{\textendash}800 Hz
that cause tactile sensations \cite{Bolanowski:1988}. For data transmission
such a channel can be represented by a vibrator-accelerometer pair, where the
former generates a set of pulses captured by the latter. Recently, it was
demonstrated that with advanced modulation and coding schemes, data rates of up
to 200 bit/s can be achieved over the haptic channel using off-the-shelf
hardware \cite{Roy:2015}. Obviously, haptic communication requires direct
physical contact between the sender and the receiver. By its nature haptic
transmission does not propagate well in the air and cannot pass through solid
objects, for example, walls. Furthermore, haptic communication is
human-perceptible and it does not require \gls{los}. 

Despite the restricted nature of the haptic channel it was shown that
eavesdropping is possible through acoustic side channels \cite{Halevi:2013}.

The haptic channel realized with a vibration motor and an accelerometer can
presently be found on numerous end-user devices such as smartphones. Several
pairing schemes that use the haptic channel are presented below. 

Saxena et al. \cite{Saxena:2011vibrate} proposed a pairing scheme called
\textit{Vibrate-to-Unlock} which is used to establish a shared secret between
an \gls{rfid} tag and a smartphone that belong to the same user. 

The suggested pairing scheme works as follows.
Initially, a smartphone selects a secret PIN (14-bits) and transmits it as an
on-off coded sequence of vibrations. An \gls{rfid} tag that is brought to
contact with the vibrating phone records the data with its accelerometer,
decodes the PIN and stores it. After the enrolling step, two devices share a
common secret. Later on, when an \gls{rfid} tag is powered once in the
vicinity of the reader it can only be unlocked if a user authenticates herself
by proving the possession of the pre-shared PIN with her phone, that is,
similarly as described above. 

The authors claimed that their scheme has a
corresponding security level of the 4-digit PIN prompted at the ATM with 3
attempts. Moreover, they argued that the suggested pairing can mitigate such
attacks as user tracking, impersonation and ghost-and-leech.    

Studer et al. \cite{Studer:2011} investigated security implications of the Bump
exchange protocol \cite{Bump:2014} and revealed that it is vulnerable to
\gls{mitm} attacks. Hence, the authors presented a new scheme known as
\textit{Shot} to pair two smartphones in a user-friendly manner.

The Shot scheme uses a server which is considered as an insecure channel
between two devices to be paired and works as follows. The first device
(endorser) hashes its public key and transmits the truncated version of the
hash (80-bits) to another device (verifier) as a sequence of vibrations. This
message serves as a pre-authenticator and is used by the verifier to bootstrap
communication with the server, that is, as a session identifier. By utilizing
the server two devices exchange their identities and public keys. With such
information at hand, the verifier can compute the hash of the endorser's public
key and compare it with the previously sent pre-authenticator. Once checked
the verifier informs the endorser about the success or failure of the pairing
via a binary vibration. 

The authors analyzed the security of Shot pairing and claimed that
in can withstand all types of active attacks on the insecure channel given an
adversary cannot inject messages into the haptic channel.     

Anand and Saxena \cite{Anand:2016} investigated how the previously proposed
Vibrate-to-Unlock scheme \cite{Saxena:2011vibrate} can be secured against
acoustic side channels \cite{Halevi:2013}. Their approach was to actively
inject noise in order to cloak the acoustic leakage emanating from the
vibrations. The enhanced pairing scheme called \textit{Vibreaker} utilized
a built-in speaker of a smartphone to generate a masking signal, which makes
the acoustic side channel indistinguishable for an eavesdropping adversary.
Specifically, the authors explored white noise and vibration noise, for example,
pre-recorded representation of audio leakage, as feasible candidates for
masking. In this case, the pairing procedure (Vibrate-to-Unlock) is
complemented by an extra step when a transmitter injects the masking signal during
the PIN transmission through vibrations. The results indicated
that both types of noise can efficiently conceal the acoustic side channel
even if the attacker applies filtering techniques. 

\subsubsection{Sensing Channel}
\label{subsubsec:sensing}
A sensing channel is used to obtain information about the ambient environment
as well as determine a device's location, position and orientation. Recently,
the use of built-in sensors was proposed for authentication purposes where
proximity detection was applied in order to mitigate relay attacks
\cite{Czeskis:2008, Halevi:2012, Shrestha:2014, Truong:2014}. There are
several types of sensor modalities that were discussed in prior research:

 \begin{itemize}
  \item Radio (Wi-Fi, Bluetooth).
  \item Audio.
  \item Motion and position (accelerometer, gyroscope, magnetometer).
  \item Location (GPS).
  \item Physical (temperature, pressure, luminosity, humidity, etc.).
\end{itemize}

\textit{Radio and audio} are used to obtain information about wireless radio
and acoustic channels described above. That is, with Wi-Fi and Bluetooth
antennas signals from \gls{ap}s and peer devices can be received, while
ambient audio can be captured with a microphone. In case of radio and audio
the sensing range cannot be delimited precisely, because it very much depends
on the receiving antenna or a microphone, transmitting power and the channel
quality.

\textit{Motion and position} is measured by a set of sensors that
allow a device to detect movement as well as determine its relative position
and orientation \cite{Sensors:101}. Readings from an accelerometer, a
gyroscope and a magnetometer are easily affected by user actions and the
ambient environment which makes the measurements obtained by similar sensors
within some distance highly uncorrelated \cite{Jin:2014}. 

\textit{Location} sensing is represented by the GPS system, which provides the worldwide outdoor
positioning within the accuracy of several meters \cite{Kaplan:2005}. The GPS
technology utilizes several frequency bands such as 1575.42 MHz and 1227.60
MHz for transmission. The data rates available with GPS can go up to 50 bit/s.
The GPS communication requires direct \gls{los} since the signals cannot
easily pass through non-transparent solid objects such as walls and doors.

\textit{Physical} sensing is used to capture information about the surrounding
environment such as temperature, pressure, luminosity, etc. Typically,
physical characteristics do not vary too much within close proximity, but
significantly differ for various locations, for example, indoor vs. outdoor
or neighboring rooms, etc. 

Previously, it was claimed \cite{Halevi:2012} that tampering with the ambient
environment is a hard task in which an adversary is unlikely to succeed.
However, a more recent study \cite{Shrestha:2015} revealed that it is feasible
to manipulate readings of different sensors such as as radio, audio and physical
in a controlled way using off-the-shelf hardware. With regard to motion and
position modalities, it was demonstrated that accelerometer readings can be
reproduced with sufficient precision \cite{Studer:2011}. This vulnerability
stems from the limited accuracy of built-in sensors and can be further
increased in case the attacker manages to observe specific user actions, for
example, shaking or bumping. As for outdoor location services, such as GPS, it
was shown to be susceptible to attacks such as spoofing and jamming
\cite{Tippenhauer:2011}. 

Currently, many devices are supplied with sensing
capabilities with smartphones having several different ones, although various
physical sensors are still not widely deployed. In the following, we review a
number of pairing schemes which utilize the sensing channel. 

Varshavsky et al. \cite{Varshavsky:2007} proposed a pairing scheme named
\textit{Amigo} to authenticate co-located devices without explicit user
involvement. Specifically, they suggested utilizing a common radio profile
which is location and time specific as the indicator of physical proximity.

The pairing scheme works as follows. First, two devices, brought to close proximity,
perform an unauthenticated \gls{dh} key exchange over a wireless radio channel
(Wi-Fi). Second, both devices start monitoring the ambient radio environment
for a short period of time and construct a signature containing identifiers
and signal strength of the packets received during the snapshot. Finally, two
devices exchange their signatures over a secure channel using a commitment
scheme in order to verify if the received and local measurements match.

The authors analyzed the security of their pairing scheme and reported that Amigo is
resilient to attacks such as eavesdropping, \gls{mitm} and impersonation.

Cai et al. \cite{Cai:2011} investigated how to establish secure communication between
previously unknown devices without any shared secrets and \gls{oob} channels.
They proposed a pairing scheme called \textit{Good Neighbor} which uses
\gls{rss} between multiple antennas of the same device (receiver) to
differentiate if another device (sender) is nearby or not. Specifically, if the
sender is in close proximity of the receiver, the \gls{rss} values measured by
two receiver's antennas would be substantially different, which is not the case
when the (malicious) sender is far away irrespective of its transmitting
power. 

The suggested pairing scheme relies on the correlation between the \gls{rss} and physical proximity and works as follows. First, the sending device initiates pairing
by requesting a public key of the receiving device once brought close to its
first antenna. Second, the sender generates a session key which is encrypted
with the receiver's public key and starts to repeatedly transmit the session
key to the receiver. Meanwhile, the sender needs to be moved to the second
antenna of the receiver. Finally, the receiver calculates the ratio of the
\gls{rss} values obtained from two antennas and checks if the number of
consecutive measurements are above a pre-defined threshold. 

The authors
evaluated their pairing scheme with respect to a powerful adversary who can
eavesdrop on the wireless channel, arbitrarily adjust the transmitting power
of her devices and gain knowledge about the location of receiver's antennas.
The results indicated that the proposed pairing can successfully mitigate such
an attacker.  

Pierson et al.\cite{Pierson:2016} proposed \textit{Wanda} a pairing scheme built
upon Good Neighbor pairing \cite{Cai:2011} to securely introduce mobile
devices. Conceptually, the ``Wand'' was realized as a portable hardware device
supplied with two antennas located half wavelength apart. Similarly to Good
Neighbor the scheme uses signal strength to determine if the Wand and a target
device are nearby (detect primitive). However, Wanda expands upon Good
Neighbor by utilizing wireless signal reciprocity to securely transmit data
between the Wand and the target device via the in-band channel (impart
primitive). 

The proposed pairing scheme works as follows. First, a user enables the target
device, for example, pressing a button, which starts broadcasting beacon
packets and points the Wand to it. Using the \gls{rss} ratio of the received
beacons from two antennas the Wand determines if the target device is in close
proximity. Second, to send a message the Wand encodes it as a binary string
and transfers one bit at a time. Particularly, a packet transmitted using the
closest antenna is considered as ``1'', and ``0'' if it is sent from the
farthest antenna. To decode the message the target device calculates the
average \gls{rss} from all received packets and checks if the \gls{rss} of a
specific packet is above or below the average, that is, either ``1'' or ``0''.
Finally, the Wand sends the hash of the transmitted message, which can be
verified by the target device. 

The authors evaluated the security of the Wanda
scheme against eavesdropping and malicious packet injection. Their findings
showed that the proposed pairing can withstand both types of attacks.

Sch{\"u}rmann and Sigg \cite{Schurmann:2013} studied how a secure
communication channel can be established between two devices in an ad-hoc
manner by utilizing ambient audio. Specifically, they proposed a mechanism
that uses audio fingerprints 
obtained by two devices from the shared ambient environment to derive a common secret
key without exchanging any information about the captured audio context. 

The suggested
pairing scheme works as follows. First, two devices synchronize their clocks
by running an NTP-based protocol. Second, two devices start simultaneously
recording the ambient audio with their local microphones. The obtained audio
fingerprints are very similar but not identical due to noise and sampling
effects. Finally, error correction codes (Reed-Solomon) are applied to obtain
identical codewords which are mapped to the unique secret key. 

The authors
analyzed the security of the proposed pairing scheme with respect to an adversary who is
not in the same context but can eavesdrop, as well as, mount \gls{mitm}, \gls{dos}
and audio amplification attacks. The experimental results confirmed
that such threats can be successfully mitigated.  

Miettinen et al. \cite{Miettinen:2014} proposed context-based zero-interaction
pairing for \gls{iot} devices which can happen without any user involvement.
Specifically, the notion of sustained co-presence was employed, meaning that
two devices would sense the same context over a substantial period of time if
they are in close proximity. 

The proposed pairing scheme works as follows. First, two devices
derive a shared secret key using an unauthenticated \gls{dh} key exchange.
Second, both devices continuously monitor ambient audio and luminosity in
order to obtain contextual fingerprints over time. Using these readings two
devices can iteratively evolve the initial secret key and obtain a new secret
key each time two fingerprints are sufficiently similar. Finally, after a
number of successful key evolution steps two devices can authenticate each
other and use the evolved secret key for secure communication. 

The authors
discussed the security of the suggested pairing scheme with regard to an adversary
being inside and outside the same context, as well as, examining context replay
attacks. Their findings implied that the proposed pairing can withstand both
types of adversaries and mitigate the replay attacks.  

Jin et al. \cite{Jin:2014} proposed a pairing scheme called
\textit{MagPairing} which requires minimum user interaction and, thus, yields
good usability. Particularly, they exploited magnetometer readings of two
smartphones brought to close proximity in order to establish pairing. 

The suggested pairing scheme works as follows. First, two devices are tapped which
triggers an authenticated \gls{dh} key exchange during which both devices
measure magnetic fields with their sensors. Second, two devices securely
exchange their magnetometer readings via the interlock protocol
\cite{Rivest:1984}. Finally, both devices can authenticate each other by
comparing if the received and local measurements match. 

For security analysis,
the authors considered attacks such as eavesdropping, \gls{mitm}, replay and
reflection. The results revealed that MagPairing is immune to the mentioned
threats even if a powerful active adversary is within a few centimeters from
the pairing devices.       

Wang et al. \cite{Wang:2016} suggested a pairing scheme known as
\textit{Touch-And-Guard (TAG)} for associating a wearable and another nearby
device by utilizing resonant properties of a human hand. Specifically, a
shared secret is obtained from a hand touch using vibration motors and
accelerometers. 

The proposed pairing scheme works as follows. First, a user
initiates pairing by touching a target device, for example, a payment
terminal, with the hand on which a wristband is worn. Second, the target device
generates vibrations which excite both the device itself and the hand. At this
point, both the wearable and the target device record vibrations with their
accelerometers. Finally, both devices process their accelometer data
separately without exchanging it, in order to extract reciprocal information to
eventually generate a shared secret. 

The security of the TAG scheme was
empirically evaluated against an eavesdropper acting via acoustic side
channels. It was shown that the proposed pairing can withstand such attackers
even if they are located in proximity. However, the authors pointed out that the
TAG scheme can still be susceptible to advanced visual eavesdroppers who
utilize high-speed cameras.  

\subsection{Discussion}
The results of our survey on \gls{phy} channels reveal interesting details.  
First, the literature makes it clear that there are no known confidential
channels, despite considerable efforts having been invested in pursuit of this
goal.  
Second, at the time of writing, the most promising communication channels,
in terms of security, were not present in the majority of devices.  
Third, the use of sensors to obtain a shared context has recently been
proposed as a new approach for \gls{sdp}, however it is not
without challenges.
We expand on these points in the following.

\subsubsection{There are no Confidential Channels}
Confidentiality cannot be guaranteed by any of the \gls{phy} channels surveyed,
even though this appears to be an explicit or implicit assumption in
a number of pairing schemes.
As shown in Table \ref{tab:phy_channels}, all \gls{phy} channels that we studied
are vulnerable to eavesdropping attacks, and those attacks have been
successfully mounted in the past.
Hence, none of these channels provides a secure transmission medium on its
own.

The problem here is two-fold.
First, the probability of ``off-the-shelf'' eavesdropping, that is, performed
without specialized equipment, has increased tremendously since numerous smart
devices nowadays are equipped with various peripherals, for example, cameras,
microphones, etc., and the sensing capabilities of commodity hardware
continue to grow \cite{Fitchard:2016}.
Second, as indicated by Halevi and Saxena \cite{Halevi:2013} side channels
pose a real threat because they can completely circumvent the security of the
pairing scheme.
Specifically, they showed that three pairing schemes (Zero-power pairing
\cite{Halperin:2008}, Vibrate-to-unlock \cite{Saxena:2011vibrate} and BEDA
\cite{Soriente:2007}), which assumed confidentiality of the \gls{oob} channel
were successfully broken by exploiting acoustic side channels.

The importance of side channels as a vital security issue has been recognized
by the research community and addressed in recent communication systems
\cite{Roy:2015} and pairing schemes \cite{Anand:2016, Wang:2016}.
Nevertheless, new sources of sensitive information leakage are being
continuously discovered \cite{Roy:2016}, which raises a fundamental question
whether it is feasible to identify and tackle all hidden channels in modern
systems.
Therefore, we argue that confidentiality of the \gls{phy} channel is very hard
to achieve and guarantee in practice.
Correspondingly, this property should be treated with a great deal of
attention when a \gls{phy} channel is considered as a candidate for the
\gls{oob} channel.

Regardless of this state of affairs where it is questionable that confidential
channels are possible in practice, pairing schemes continue to be proposed
that rely on secrecy of data transmission, for example, \cite{Wang:2015vbox}.

\subsubsection{Most Potentially Secure Channels Missing From Current Commodity Hardware}
The physical characteristics of various \gls{phy} channels provide different security
properties.
We have identified an important trade-off between security and ease of
adoption.
That is, a number of newer communication channels, such as mm-Waves and
\gls{vlc} can offer improved security, however they are not
yet ubiquitous.

In particular, mm-Waves and \gls{vlc} possess valuable security characteristics.
Their short-range of communications, together with \gls{los} requirements and low penetration rates,
make them ideal for deployment and use as \gls{oob} channels in the \gls{iot}
domain.
Research is, however, still ongoing to improve on both mm-Waves
\cite{Niu:2015} and \gls{vlc} \cite{Pathak:2015} communications.
A further advantage is that both technologies can be efficiently implemented
on constrained devices.
For example, the antennas required for mm-Wave transmission are very small,
and the \gls{vlc} building blocks such as diodes and photosensors are
inexpensive.
Hence, these technologies are worth considering for \gls{sdp}.

The challenge lies in the maturing of these newer channels, such that they
become widely available on commodity hardware. This requires the action on
both researchers and vendors.

\subsubsection{Using Environment Sensing}
A different approach to pairing is to utilize the sensing channel to obtain
the shared context, which can be used either as an indicator of physical proximity, or
as an entropy source to derive a shared secret key.  
The use of sensing information enables scalable pairing, which is crucial in a
distributed and diverse environment such as the \gls{iot}.
It also reduces or eliminates the user effort, which results in more
usable and less error-prone pairing.

For example, the use of physical environment sensing
\cite{Shrestha:2014} and GPS data \cite{Ma:2013}, as it widely explored in
schemes such as \gls{zia}, can provide a suitable base to increase security in
device pairing as well.
The key insights provided by \gls{zia} both those that are security-enhancing, such as fusing
multiple sensing modalities \cite{Truong:2014}, as well as those that are adversarial, such as
context-manipulation threats \cite{Shrestha:2015}, 
should also be taken into account in \gls{sdp}.    

\gls{csi} in radio communications provides a different use for sensing.
Reciprocal radio channels, meaning that the same antenna is used for
transmitting and receiving, lead to correlated channel observations at both
sides of a communication link.
This correlation of channel observations allows the transmitter and receiver
to obtain a common fingerprint of the radio environment, that can in turn be
used to mitigate various types of attacks, including \gls{mitm} and relay
attacks.

Pairing schemes such as Amigo \cite{Varshavsky:2007}, Good Neighbor
\cite{Cai:2011}, and Wanda \cite{Pierson:2016} and also \cite{Patwari:2007,
Kalamandeen:2010, Mathur:2011, Liu:2012, Shi:2013, Wang:2015wave, Wang:2015vbox, Xi:2016} rely on
such information to ensure that both pairing devices communicate over the
same channel.
However, the robustness of such channel fingerprinting schemes against
spoofing is still an open question under investigation. For example, Zafer et
al.
\cite{Zafer:2012fe} have demonstrated an active \gls{csi} spoofing attack.
Hence, pairing schemes leveraging the radio channel must account for
manipulated and forged channel states.

Environment and channel sensing can provide an additional layer of verification.
However, it still suffers from similar security limitations as \gls{phy} channels.
Therefore, the sensing channel cannot, by itself, guarantee that no one is
intermediating communications between the pairing devices, for example, through
an \gls{mitm} attack.

In this section, we have investigated and discussed \gls{phy} channels along with the \gls{sdp} schemes utilizing them.
In the next section, we review \gls{hci} channels and the corresponding pairing schemes. 

%% file: section/hci_channels.tex

\section{Human-computer Interaction Channels}
\label{sec:hci}
Modern information and communication technologies have become an indispensable 
part of the human society. The way people live, work and interact with each other and the 
environment has changed significantly with the advent of smart devices, social networking and cloud-based services. 
Various research and technologies have utilized \gls{hci} to provide
security in a wide range of applications such e-commerce, home automation,
and social networking \cite{Sasse:2001, Jaimes:2007, Das:2014}. With the upcoming \gls{iot} the importance of 
developing socially compatible security tools based on \gls{hci} is becoming more evident \cite{Das:2013, Das:2017}. 
However, relying on human interactions to achieve security often introduces vulnerabilities to the system. 
Bruce Schneier \cite{Schneier:2011} emphasized the relevance of the human factor in the system as follows: 
``... security is only as good as it's weakest link, and people are the weakest link in the chain.'' 
Hence, the security of the system where a user is involved depends not only on the technical aspects of the system, but also on how people understand and use it, in addition to the system's capability to mitigate threats and issues introduced by users themselves \cite{Besnard:2004, Sasse:2001, Liginlal:2009}.
From the pairing perspective, a user also plays an important role with regard to security. 
Traditionally, the security of pairing schemes has involved an aspect of human supervision, which can
take the form of perception, for example, image comparison \cite{Perrig:1999}, decision making, 
for instance, pressing a button \cite{Soriente:2007}, and other interactive techniques, for example, 
drawing a pattern \cite{Sethi:2014}.

We start our discussion by identifying three points which are the base for
rigorous \gls{hci} investigation. In particular, we specify several types of
\textit{\gls{hci} channels} which have been used in \gls{sdp} and
denote two sets of properties, namely \textit{security properties} and
\textit{usability properties} being studied. Afterwards, we review existing
pairing schemes that rely on various \gls{hci} channels to exhibit the
trade-offs between security and usability. Finally, we discuss the most
significant insights and implications, presented in Table \ref{tab:hci_channels}, that were identified in our survey on \gls{hci} channels.  

\subsection{\gls{hci} Channels in Device Pairing}
Recently, numerous devices with rich input/output capabilities and considerable
processing power have become widely available, which has significantly improved
the quality of \gls{hci} \cite{Jaimes:2007}. Correspondingly,
many pairing schemes proposed up-to-date rely on some form of user
involvement. Chong et al. \cite{Chong:2014} surveyed existing pairing
schemes by considering user actions required to establish a secure channel between two devices. We refine their
findings to obtain fine-grained categories of user interaction that have been
used in \gls{sdp}. 

Specifically, we define three \gls{hci}
channels that fully satisfy our definition given in Section \ref{subsec:sdp_def}: \textit{Data relay},
\textit{Data comparison} and \textit{Data generation}. In addition, we
consider \textit{Device handling}, which while not a conventional \gls{hci}
channel, represents a more passive form of user interaction that is often
(implicitly) present in device pairing:

\subsubsection{Data Relay}
A channel where a user is prompted to transfer data generated by one pairing
device onto another pairing device.

\subsubsection{Data Comparison}
A channel where a user is required to compare and analyze data produced by two pairing
devices, for example, to verify the correctness or consistency of the information.

\subsubsection{Data Generation}
A channel where a user provides common input to both pairing devices
simultaneously, for example, shaking, drawing, or first imposes (secret) input
on one device and then provides it again on a second device.

\subsubsection{Device Handling}
A form of user interaction where a human actor is required to bring pairing
devices in proximity, make physical contact, align them, or take similar
action. 

\subsection{Security Properties} 
\label{sub:security}
The security properties of the pairing schemes based on \gls{hci} channels are
quite different from the ones purely relying on \gls{phy} channels. First, users are the
unavoidable source of errors \cite{Liginlal:2009, Kraemer:2007} and their
behavior, as well as, attitude towards security sensitive tasks can vary
significantly \cite{Besnard:2004}. Second, user interaction is subject to
observation by both an internal participant, who is curious, and an external
adversary, who is malicious. To compile the list of representative security
properties we combined issues that have been raised in the pairing community
with respect to human factors \cite{Saxena:2009rushing, Gallego:2011} and
complemented them with the implications found in the authentication domain
\cite{Bonneau:2012}:

\subsubsection{Inattentive User}
Defines if a pairing scheme has certain tolerance to mistakes and errors
introduced by the user. In particular, a pairing mechanism that does not
verify user input for errors or provide corresponding feedback can be
circumvented by the attacker who can impersonate a legitimate device.

\subsubsection{Rushing Behavior}
Specifies if a pairing scheme accounts for rushing users who are willing to
skip certain steps of the pairing procedure or accept specific conditions
without verification, in order to speed up pairing. 

\subsubsection{Consent Tampering}
Determines if a pairing scheme is resilient to consent tampering by a
dishonest user. That is, if the user can accept pairing even if the data
exchanged between two devices mismatch, or conversely, reject pairing, even
though both devices successfully establish a connection. 

\subsubsection{User Observation}
Defines if a pairing scheme is resistant to an adversary who can observe
user actions during the pairing process. In other words, the attacker does not
benefit from learning user interactions, including
(secret) data exchanged on the \gls{hci} channel, 
and cannot compromise pairing with such information at hand. 

\subsubsection{Forward Secrecy}
Determines how resilient a pairing scheme from a cryptographic perspective to an eavesdropper 
who can leverage user observation and the compromise of the long-term keys. 
That is, if the underlying cryptographic protocol used in the pairing scheme
mitigates brute-force offline attacks aided by (secret) data observed on the \gls{hci} channel, 
and restricts an adversary to a one-off (online) guessing game. 
We evaluate this property under the assumption that \gls{dh} keys used by  
the underlying cryptographic protocols are ephemeral. 

\subsubsection{Honest-but-curious}
Specifies if a pairing scheme is susceptible to an honest-but-curious
adversary who legitimately participates in the pairing process but tries to
learn or infer more information about another pairing party. 

\begin{table*}[!htb]
	\caption{Summary of \gls{hci} channels}
	\label{tab:hci_channels}
	\centering
	\begin{threeparttable}
	\rotatebox{0}{\scriptsize{
		\begin{tabularx}{\textwidth}{p{3.9 cm}p{0.3cm}p{0.3cm}p{0.3cm}p{0.85cm}p{0.3cm}p{0.3cm}p{0.3cm}p{0.3cm}p{0.3cm}p{0.85cm}p{0.3cm}p{0.3cm}p{0.3cm}p{0.3cm}p{0.3cm}p{0.3cm}p{0.3cm}}

		\hline
		\hline
	
	    \multicolumn{1}{c}{\shortstack{\\ \textbf{Pairing Scheme} }} &
		\multicolumn{4}{c}{\shortstack{\\ \textbf{\gls{hci} Channel} }} &
		\multicolumn{6}{c}{\shortstack{\\ \textbf{Security Properties} }} &
		\multicolumn{7}{c}{\shortstack{\\ \textbf{Usability Properties} }} 
		 \\
		\hline
		
		&  
		\rotatebox{55}{\textbf{Data Relay       }}& 		
		\rotatebox{55}{\textbf{Data Comparison  }}& 		
		\rotatebox{55}{\textbf{Data Generation  }}&
		\rotatebox{55}{\textbf{Device Handling  }}&
		
		\rotatebox{55}{\textbf{Inattentive User}}& 		
		\rotatebox{55}{\textbf{Rushing Behavior}}&
		\rotatebox{55}{\textbf{Consent Tampering}}&	
		\rotatebox{55}{\textbf{User Observation}}&
		\rotatebox{55}{\textbf{Forward Secrecy}}&
		\rotatebox{55}{\textbf{Honest-but-curious}} &
		
		\rotatebox{55}{\textbf{Effortless Initialization}}&		
		\rotatebox{55}{\textbf{No Secret Relay                }}&		
		\rotatebox{55}{\textbf{Auto. Secret Generation}}&		
		\rotatebox{55}{\textbf{Auto. Consistency Check}}& 		
		\rotatebox{55}{\textbf{Environment Insensitivity}}& 	
		\rotatebox{55}{\textbf{Explicit User Feedback}}& 	
		\rotatebox{55}{\textbf{Familiarity}}	\\
		\hline

		\hline
		\gls{mana} I \cite{Gehrmann:2004}  & 
		\CIRCLE & 
		\raisebox{-0.75ex}{\APLminus} & 
		\raisebox{-0.75ex}{\APLminus} & 
		\raisebox{-0.75ex}{\APLminus} & 
		\Circle & 
		\Circle & 
		\Circle & 
		\Circle & 
		\Circle & 
		\Circle & 
	    \Circle & 
		\Circle & 
		\RIGHTcircle & 
		\RIGHTcircle & 
		\CIRCLE & 
		\RIGHTcircle & 
		\CIRCLE \\ 
		
		\gls{mana} II \cite{Gehrmann:2004}  & 
		\raisebox{-0.75ex}{\APLminus} & 
		\CIRCLE & 
		\raisebox{-0.75ex}{\APLminus} & 
		\raisebox{-0.75ex}{\APLminus} & 
		\Circle & 
		\Circle & 
		\Circle & 
		\Circle & 
		\Circle & 
		\Circle & 
	    \Circle & 
		\CIRCLE & 
		\CIRCLE & 
		\Circle & 
		\CIRCLE & 
		\Circle & 
		\CIRCLE \\ 
		
		\gls{mana} III \cite{Gehrmann:2004}  & 
		\raisebox{-0.75ex}{\APLminus} & 
		\RIGHTcircle & 
		\CIRCLE & 
		\raisebox{-0.75ex}{\APLminus} & 
		\Circle & 
		\Circle & 
		\Circle & 
		\Circle & 
		\CIRCLE & 
		\Circle &  
	    \Circle & 
		\CIRCLE & 
		\Circle & 
		\RIGHTcircle & 
		\CIRCLE & 
		\CIRCLE & 
		\CIRCLE \\ 

		Access point authentication \cite{Roth:2008} & 
		\raisebox{-0.75ex}{\APLminus} & 
		\CIRCLE & 
		\raisebox{-0.75ex}{\APLminus} & 
		\raisebox{-0.75ex}{\APLminus} & 
		\Circle & 
		\Circle & 
		\Circle & 
		\Circle & 
		\CIRCLE & 
		\Circle & 
		\CIRCLE & 
		\CIRCLE & 
		\CIRCLE & 
         \Circle & 
		\Circle & 
		\Circle & 
		\RIGHTcircle \\ 
		
		Shake Them Up! \cite{Castelluccia:2005}  & 
		\raisebox{-0.75ex}{\APLminus} & 
		\raisebox{-0.75ex}{\APLminus} & 
		\raisebox{-0.75ex}{\APLminus} & 
		\CIRCLE & 
		\CIRCLE & 
		\Circle & 
		\CIRCLE & 
		\Circle & 
		\Circle & 
		\Circle & 
	    \Circle & 
		\RIGHTcircle & 
		\CIRCLE & 
		\CIRCLE & 
		\CIRCLE & 
		\Circle & 
		\Circle \\ 
		
		Shake Well Before Use ShaVE \cite{Mayrhofer:2007sh} & 
		\raisebox{-0.75ex}{\APLminus} & 
		\raisebox{-0.75ex}{\APLminus} & 
		\CIRCLE & 
		\CIRCLE & 
		\CIRCLE & 
		\CIRCLE & 
		\CIRCLE & 
		\RIGHTcircle & 
		\CIRCLE & 
		\CIRCLE & 
		\CIRCLE & 
		\RIGHTcircle & 
		\RIGHTcircle & 
		\CIRCLE & 
		\CIRCLE & 
		\CIRCLE & 
		\RIGHTcircle \\ 
				
		Shake Well Before Use ShaCK \cite{Mayrhofer:2007sh} & 
		\raisebox{-0.75ex}{\APLminus} & 
		\raisebox{-0.75ex}{\APLminus} & 
		\CIRCLE & 
		\CIRCLE & 
		\CIRCLE & 
		\RIGHTcircle & 
		\CIRCLE & 
		\RIGHTcircle & 
		\Circle & 
		\CIRCLE & 
		\CIRCLE & 
		\RIGHTcircle & 
		\Circle & 
		\CIRCLE & 
		\CIRCLE & 
		\CIRCLE & 
		\RIGHTcircle \\ 

		SAPHE \cite{Groza:2012} & 
		\raisebox{-0.75ex}{\APLminus} & 
		\raisebox{-0.75ex}{\APLminus} & 
		\CIRCLE & 
		\CIRCLE & 
		\CIRCLE & 
		\CIRCLE & 
		\CIRCLE & 
		\RIGHTcircle & 
		\Circle & 
		\CIRCLE & 
		\CIRCLE & 
		\RIGHTcircle & 
		\Circle & 
		\CIRCLE & 
		\CIRCLE & 
		\CIRCLE & 
		\RIGHTcircle \\ 

		Authentication using ultrasound \cite{Kindberg:2003bh} & 
		\raisebox{-0.75ex}{\APLminus} & 
		\CIRCLE & 
		\raisebox{-0.75ex}{\APLminus} & 
		\CIRCLE & 
		\Circle & 
		\Circle & 
		\Circle & 
		\RIGHTcircle & 
		\Circle & 
		\Circle & 
		\Circle & 
		\CIRCLE & 
		\CIRCLE & 
		\Circle & 
		\Circle & 
		\Circle & 
		\RIGHTcircle \\ 
		
		Beep - Blink \cite{Prasad:2008} & 
		\raisebox{-0.75ex}{\APLminus} & 
		\CIRCLE & 
		\raisebox{-0.75ex}{\APLminus} & 
		\raisebox{-0.75ex}{\APLminus} & 
		\Circle & 
		\Circle & 
		\Circle & 
		\RIGHTcircle & 
		\CIRCLE & 
		\Circle & 
		\CIRCLE & 
		\CIRCLE & 
		\CIRCLE & 
		\Circle & 
		\Circle & 
		\RIGHTcircle & 
		\RIGHTcircle \\ 

	    Blink - Blink \cite{Prasad:2008} & 
		\raisebox{-0.75ex}{\APLminus} & 
		\CIRCLE & 
		\raisebox{-0.75ex}{\APLminus} & 
		\raisebox{-0.75ex}{\APLminus} & 
		\Circle & 
		\Circle & 
		\Circle & 
		\RIGHTcircle & 
		\CIRCLE & 
		\Circle & 
		\RIGHTcircle & 
		\CIRCLE & 
		\CIRCLE & 
		\Circle & 
		\RIGHTcircle & 
		\Circle & 
		\RIGHTcircle \\ 
		
		RhythmLink \cite{Lin:2011rlink} & 
		\raisebox{-0.75ex}{\APLminus} & 
		\raisebox{-0.75ex}{\APLminus} & 
		\CIRCLE & 
		\raisebox{-0.75ex}{\APLminus} & 
		\RIGHTcircle & 
		\RIGHTcircle & 
		\CIRCLE & 
		\Circle & 
		\Circle & 
		\Circle & 
		\Circle & 
		\Circle & 
		\Circle & 
		\CIRCLE & 
		\RIGHTcircle & 
		\CIRCLE & 
		\Circle \\ 

		Seeing-Is-Believing \cite{Mccune:2005se}  & 
		\CIRCLE & 
		\raisebox{-0.75ex}{\APLminus} & 
		\raisebox{-0.75ex}{\APLminus} & 
		\CIRCLE & 
		\RIGHTcircle & 
		\RIGHTcircle & 
		\RIGHTcircle & 
		\RIGHTcircle & 
		\RIGHTcircle & 
		\RIGHTcircle & 
 	 	\RIGHTcircle & 
		\Circle & 
		\CIRCLE & 
		\CIRCLE & 
		\RIGHTcircle & 
		\CIRCLE & 
		\CIRCLE \\ 
		
		Visible Laser Light \cite{Mayrhofer:2007kd}  & 
		\raisebox{-0.75ex}{\APLminus} & 
		\raisebox{-0.75ex}{\APLminus} & 
		\raisebox{-0.75ex}{\APLminus} & 
		\CIRCLE & 
		\CIRCLE & 
		\Circle & 
		\Circle & 
		\Circle & 
		\CIRCLE & 
        \Circle & 
	    \Circle & 
		\CIRCLE & 
		\CIRCLE & 
		\RIGHTcircle & 
		\Circle & 
		\CIRCLE & 
		\Circle \\ 
		
		VIC (mutual authentication) \cite{Saxena:2011visual} & 
		\raisebox{-0.75ex}{\APLminus} & 
		\CIRCLE & 
		\raisebox{-0.75ex}{\APLminus} & 
		\CIRCLE & 
		\Circle & 
		\Circle & 
		\Circle & 
		\RIGHTcircle & 
		\RIGHTcircle & 
		\RIGHTcircle & 
 		\Circle & 
		\CIRCLE & 
		\CIRCLE & 
		\Circle & 
		\RIGHTcircle & 
		\CIRCLE & 
		\CIRCLE \\ 
	
		BEDA (B2B) \cite{Soriente:2007} & 
		\raisebox{-0.75ex}{\APLminus} & 
		\raisebox{-0.75ex}{\APLminus} & 
	     \CIRCLE & 
		\raisebox{-0.75ex}{\APLminus} & 
		\CIRCLE & 
		\RIGHTcircle & 
		\CIRCLE & 
		\Circle & 
		\CIRCLE & 
		\Circle &  
		\CIRCLE & 
		\CIRCLE & 
		\Circle & 
		\CIRCLE & 
		\CIRCLE & 
		\Circle & 
		\Circle \\ 
		
		BEDA (D2B, SV2B, LV2B) \cite{Soriente:2007} & 
		\CIRCLE & 
		\raisebox{-0.75ex}{\APLminus} & 
		\raisebox{-0.75ex}{\APLminus} & 
		\raisebox{-0.75ex}{\APLminus} & 
		\CIRCLE & 
		\CIRCLE & 
		\CIRCLE & 
		\Circle & 
		\CIRCLE & 
		\Circle & 
		\CIRCLE & 
		\Circle & 
		\CIRCLE & 
         \CIRCLE & 
		\RIGHTcircle & 
		\Circle & 
		\Circle \\ 

		Playful Security \cite{Gallego:2011} & 
		\CIRCLE & 
		\raisebox{-0.75ex}{\APLminus} & 
		\raisebox{-0.75ex}{\APLminus} & 
		\raisebox{-0.75ex}{\APLminus} & 
		\CIRCLE & 
		\CIRCLE & 
		\CIRCLE & 
		\CIRCLE & 
		\raisebox{-0.75ex}{\APLminus} & 
		\Circle & 
		\CIRCLE & 
		\Circle & 
		\CIRCLE & 
		\CIRCLE & 
		\CIRCLE & 
		\CIRCLE & 
		\RIGHTcircle \\ 
				
		Safeslinger \cite{Farb:2013} & 
		\raisebox{-0.75ex}{\APLminus} & 
		\CIRCLE & 
		\RIGHTcircle & 
		\raisebox{-0.75ex}{\APLminus} & 
		\CIRCLE & 
		\CIRCLE & 
		\CIRCLE & 
		\CIRCLE & 
		\Circle & 
		\RIGHTcircle & 
	    \Circle & 
		\CIRCLE & 
		\CIRCLE & 
		\RIGHTcircle & 
		\CIRCLE & 
		\RIGHTcircle & 
		\RIGHTcircle \\ 
				
		Synchronized Drawing \cite{Sethi:2014}  & 
		\raisebox{-0.75ex}{\APLminus} & 
		\raisebox{-0.75ex}{\APLminus} & 
		\CIRCLE & 
		\CIRCLE & 
		\RIGHTcircle & 
		\RIGHTcircle & 
		\RIGHTcircle & 
		\Circle & 
		\CIRCLE & 
		\CIRCLE & 
		\CIRCLE & 
		\CIRCLE & 
		\Circle & 
		\CIRCLE & 
		\CIRCLE & 
		\Circle & 
		\RIGHTcircle \\ 
		
		Proximity Authentication \cite{Li:2015dr}  & 
		\raisebox{-0.75ex}{\APLminus} & 
		\raisebox{-0.75ex}{\APLminus} & 
		\CIRCLE & 
		\CIRCLE & 
		\RIGHTcircle & 
		\CIRCLE & 
		\CIRCLE & 
		\RIGHTcircle & 
		\CIRCLE & 
		\RIGHTcircle & 
	    \CIRCLE & 
		\CIRCLE & 
		\Circle & 
		\CIRCLE & 
		\CIRCLE & 
		\Circle & 
		\RIGHTcircle \\ 
		
		Checksum Gestures \cite{Ahmed:2015}  & 
		\CIRCLE & 
		\raisebox{-0.75ex}{\APLminus} & 
		\raisebox{-0.75ex}{\APLminus} & 
		\RIGHTcircle & 
		\CIRCLE & 
		\CIRCLE & 
		\CIRCLE & 
		\RIGHTcircle & 
		\CIRCLE & 
		\RIGHTcircle & 
	    \CIRCLE & 
		\Circle & 
		\CIRCLE & 
		\CIRCLE & 
		\CIRCLE & 
		\RIGHTcircle & 
		\RIGHTcircle \\ 
		
		\hline
		\hline
		
		\end{tabularx}
		}}
		\begin{tablenotes}[para,flushleft]
			\item \\
			\item 
			\item 
			\item 
			\item 
			\item 
			\item 
			\item 
			\item \CIRCLE\textcolor{white}{-}= fulfills property; \RIGHTcircle\textcolor{white}{-}= partly fulfills property; \Circle\textcolor{white}{-}= does not fulfill property; \raisebox{-0.75ex}{\APLminus}\textcolor{white}{-}= n/a
    	\end{tablenotes}
  \end{threeparttable}		
\end{table*}

\subsection{Usability Properties}
\label{sub:usability}
As mentioned previously, usability of pairing schemes has been a
subject in several studies and a number of works investigated how usability
can be enhanced in case of device pairing \cite{Hsiao:2009, Kray:2010,
Kainda:2010}. However, many works applied mostly quantitative metrics to
evaluate usability such as completion time and error rate
\cite{Kumar:2009} which are implementation dependent. In addition, subjective
characteristics such as personal preferences vary with context, as has been
previously demonstrated \cite{Ion:2010}. Thus, there is a lack of a common
baseline approach which would allow usability evaluation of pairing schemes
more qualitatively and coherently. We aim to remedy that situation by
presenting a set of usability properties, which we derived by studying the
usability implications in general human-device interaction \cite{Grubert:2016},
as well as, authentication techniques \cite{Bonneau:2012} and projecting the
findings onto the pairing domain:

\subsubsection{Effortless Initialization} 
Defines minimal user effort during the discovery phase of the pairing process.
For example, a user is not required to provide any additional information, such as
a number of participants, or pre-configure devices prior to pairing. 

\subsubsection{No Secret Relay} 
Does not prompt users to transfer any (secret) information from one pairing
device to another or if it is required the length of the relayed data should be
minimal.

\subsubsection{Automatic Secret Generation}
Specifies that the data used for authentication, for example, cryptographic keys,
is generated by pairing devices without requiring any user input or
assistance, such as shaking, drawing, etc.

\subsubsection{Automatic Consistency Check}
Determines user effort necessary for verifying that information exchanged
between pairing devices is similar.

\subsubsection{Environmental Insensitivity}
Defines applicability of the pairing schemes with respect to the ambient
environment. For example, a pairing scheme may lead to high error rates or
even fail if the environment is too noisy, crowded or has poor illumination.

\subsubsection{Explicit User Feedback} 
Specifies if a pairing scheme provides meaningful feedback to the user
during and upon the completion of the pairing process. For example, two
pairing devices can indicate success by making an appropriate sound,
and provide explanatory, actionable, feedback if pairing fails.

\subsubsection{Familiarity}
Determines if the user actions imposed by the pairing scheme correspond to the
daily user experience \cite{Sasse:2001, Ion:2010}. That is, if a pairing
scheme relies on well-established interaction patterns, for example,
smartphone usage, and requires no extra training for an average user in order
to be adopted. 

\subsection{Survey of \gls{hci} Channels} 
\label{sub:evaluation_hci}
In this section we review representative pairing schemes which rely on
\gls{hci} by focusing on the properties given above. 

\subsubsection{\gls{mana}} 
Gehrmann et. al. \cite{Gehrmann:2004} presented several \textit{MAN}ual
\textit{A}uthentication (\gls{mana}) schemes for authenticating 
\gls{dh} public keys. They assumed that devices have at least one input and/or output interface, for example, a display and/or a keypad. From the user perspective, a human operator plays a crucial role in pairing. Three variants of the \gls{mana} scheme were proposed which work as follows:

\begin{itemize}
\item {\gls{mana} I: One device has a display and a simple input interface, for example, a button, while another device has a keypad and a simple output interface, for instance, an LCD panel. The first device computes a random key and a checksum value and displays this data. The user reads the checksum value and the random key from the screen of the first device and inputs this information into the second device. Then, the second device computes the checksum value using the provided random key and compares the two checksums. The outcome of the comparison is indicated as an accept or reject message to the user. Finally, the user enters the result back into the first device.}

\item {\gls{mana} II: Both devices have a display but neither of them a keypad, although they are supplied with a simple input interface, for example, a button. Similar to \gls{mana} I, the first device computes the random key and the checksum and displays two values. In addition, the first device sends the random key to the second device over an insecure channel, for example, wireless radio. Afterwards, the second device computes the checksum value and outputs it together with the key. By comparing values displayed by both devices a user has to either accept the connection if they are equal or reject it otherwise.} 

\item {\gls{mana} III: Both devices are assumed to have a keypad. The user enters a short random bit-string \textbf{R} into both devices. Then, each device generates a random \gls{mac} key and calculates a \gls{mac} value over \textbf{R} concatenated with a device identifier and the \gls{dh}-public keys. Afterwards, both devices exchange their corresponding \gls{mac} values via a wireless radio channel. Only upon receiving the \gls{mac} value from the pairing peer each device reveals its \gls{mac} key. Finally, both devices verify the received \gls{mac} values and indicate the result to the user who is required to compare and confirm it. A simpler variant exists in case one of the devices has only a display, that is, no means of input.}
\end{itemize}

The authors argued that \gls{mana}-schemes are robust against \gls{mitm}
attacks, given user diligence in verifying calculated hash values.

\subsubsection{Access point authentication}
Roth et. al. \cite{Roth:2008} suggested a pairing scheme to protect the
connection between an \gls{ap} and a client device against evil twin attacks.

The proposed pairing scheme uses \gls{sas} for key establishment and consists of two phases.
In the setup phase, both devices exchange their public keys and a nonce value
over an insecure wireless channel. During the authentication phase, a user is
required to compare a certain number of color sequences (minimum two) in order
to verify that pairing was performed with the intended \gls{ap}. In
detail, each sequence is comprised of two colors and represents a \gls{sas}.
Both devices display the sequence of colors, that is, one color at a time, and
the user has to verify their equality by pressing a button and proceeding to
the next sequence. The number of sequences shown depends on the desired level
of security and eventually the user is prompted to either accept or reject
pairing. 

The authors discussed the security of the proposed pairing scheme and
concluded that it can withstand evil twin attacks.

\subsubsection{Shake Them Up!}
Castelluccia et al. \cite{Castelluccia:2005} proposed a pairing scheme for
CPU-constrained devices, for example, sensors, that do not have enough
computational power to perform public key cryptography. 

The proposed pairing scheme utilizes
the anonymous broadcast channel and works as follows. In order to derive a
shared secret key, two devices are held together and shaken, either by a single
user, or by two users in close proximity. Meanwhile, both devices broadcast
empty packets over an insecure wireless channel. The anonymous broadcast implies
that each device sends a packet by setting its own identifier or the
identifier of the pairing peer as the source of the message. In this case, an
adversary can read the transmitted packets but cannot distinguish the source.
In contrast, each pairing device knows if it has sent a particular message or not, which
is interpreted by the device as a secret bit 1 or 0, and the shared key can be obtained by observing a
pre-defined number of packets. The shaking is done to thwart signal strength
analysis by an attacker to identify the actual sender. 

The authors analyzed the 
security of their pairing scheme against an adversary who can read all packets but
cannot distinguish the source of the packet and reported that it is resilient against
\gls{mitm} and \gls{dos} attacks. However, Rasmussen et. al.
\cite{Rasmussen:2007} showed the vulnerability of this scheme by using
radio fingerprinting to identify the sender.      

\subsubsection{Shake Well Before Use} 
Mayrhofer et al. \cite{Mayrhofer:2007sh} suggested a pairing approach which
utilizes accelerometer data generated from distinct movement patterns.
Specifically, they proposed two schemes to securely pair devices where
a user is required to hold them together and them shake simultaneously. 

The first
scheme (ShaVE) uses the \gls{dh} key exchange to derive a shared key over an
insecure wireless channel followed by the exchange of accelerometer readings
via the interlock protocol \cite{Rivest:1984} to verify authenticity of
pairing devices. 

The second scheme (ShaCK) relies on the data captured by the
accelerometer to derive a shared secret key. In detail, two devices hash their
synchronized feature vectors obtained from the sensor readings and accumulate
them until the entropy is sufficient to produce the shared secret key. 

The
authors discussed the security of the proposed pairing with regard to an active
adversary and concluded that both schemes can withstand \gls{mitm} attacks.
However, they conceded that the ShaCK variant does not provide forward secrecy
and is vulnerable to offline guessing attacks. 

\subsubsection{SAPHE}
Groza and Mayrhofer \cite{Groza:2012} proposed a pairing scheme based on shaking, which improved upon the previous works, for example, ShaCK \cite{Mayrhofer:2007sh}, by devising a more lightweight approach to securely exchange low entropy vectors obtained from accelerometer data.

The suggested pairing scheme employs a hashed heuristic tree and works as follows. 
First, the commitments between two devices are exchanged in the form of hashes of randomly generated values. 
Second, accelerometer data produced by shaking two devices together is recorded and used to obtain a unique secret key on each device. The unique secret keys are extracted by comparing the accelerometer readings to the threshold values obtained from the initial commitments by means of the Euclidian distance. 
The key extraction algorithm relies on a hashed heuristic tree, which is essentially a search tree, where the accelerometer readings are first sorted in a descending order with respect to the distance from the threshold values, and then bit-by-bit hashing is applied to retrieve the unique secret key. 
Third, both devices exchange challenges which are nonces encrypted with the individual secret keys, and each device proofs the possession of the peer's key by verifying the challenge. 

The authors analyzed the security of the proposed pairing scheme and claimed that their approach provides better resilience to \gls{mitm} attackers, who try to guess the low entropy vectors obtained from accelerometer data. 
However, the authors conceded that further research is required to evaluate resilience of the SAPHE scheme against the adversaries who can observe user interaction.

\subsubsection{Authentication using Ultrasound} 
Kindberg et. al. \cite{Kindberg:2003bh} presented a pairing scheme which
utilized ultrasound to physically validate two devices and establish a secure channel between them.

The proposed pairing scheme consists of two phases and works as follows. In the
locate phase, a user selects a target device to communicate with, and makes
sure that her personal device (client) is in LoS with the target. Then the
client sends a message to locate the target, which replies with its designated
identifier, for example, network address, over RF and ultrasound channels. The
client receives those messages, matches the identifier and is able to
calculate the approximate distance to the target device, which is displayed to
a user for verification. During the associate phase, the user points the client
device to the target and initiates pairing. The target device replies with
the RF message containing its public key together with a random number and
simultaneously emits the ultrasound message with the same random number. Upon
receipt, the client checks if random numbers from RF and ultrasound
messages match and asks the user to confirm the relative position of the
target device. Finally, the client encrypts a session key with the target's public
key and sends it along with a random number back to the target. 

The authors
argued that the proposed pairing scheme is robust against various spoofing and replay
attacks given the adversary is unable to counterfeit ultrasound messages. 

\subsubsection{Synchronized Audio-Visual Patterns} 
Prasad and Saxena \cite{Prasad:2008} presented two pairing schemes suitable
for devices with only basic interfaces such as a pair of LEDs and/or speakers.
Specifically, both schemes rely on \gls{sas}s transmitted by two devices in
the form of synchronized audiovisual patterns, for example, blinking LEDs,
which have to be compared by a user for equality. 

In the first scheme
(blink-blink) two devices encode their \gls{sas}s as sequences of blinking
LEDs and the user is required to compare these sequences and determine if they
are synchronous on both devices, for example, green or red LEDs. 

In the second
scheme (beep-blink) one device transmits its \gls{sas} as a sequence of
blinking LEDs, while another device encodes the \gls{sas} as a series of
beeping sounds and silence periods. The user has to verify if these two
patterns match, such as the LED light corresponds to the sound. 

The authors
analyzed the security of the proposed pairing with regard to a \gls{mitm} adversary and
concluded that both schemes can withstand such attacks, yet security depends
on user diligence when comparing two audiovisual sequences. 

\subsubsection{RhythmLink} 
Lin et. al. \cite{Lin:2011rlink} proposed a pairing scheme based on rhythm
tapping.
 
Initially, a user inputs a song rhythm several times on her personal device,
for example, a smartphone, to provide some training data and eventually obtain
a tapped password, referred to as a tapword. 
Afterwards, this generated tapword is stored on the user device and used further for pairing. 

To pair with a target device, the user inputs the same tapped rhythm into it. 
Therefore, the target device can compute a tapword and compare it with the
pattern stored on the user device by means of the Euclidean distance. 
The protocol uses elliptic curve cryptography to calculate the Euclidean
distance between the tapwords, without either device revealing its tapword. 
To generate a session key, password authenticated key exchange is used in
order to avoid \gls{mitm} attacks. 
A device encrypts its model information with this session key and sends the
encrypted data to the other device, which decrypts this information and
computes the Euclidean distance. 
Afterwards both distances are compared. If the distances match, the devices
accept pairing.

\subsubsection{Seeing-Is-Believing (SiB)} 
McCune et. al. \cite{Mccune:2005se} proposed a pairing scheme, based on
taking a snapshot of a two-dimensional barcode displayed on
the screen of one device by the camera of another device. 
The two-dimensional barcodes are generated by the devices automatically
without any human effort. 
A user is required to configure the camera and take the snapshot of the 2-D
barcode. 

To perform pairing, one device sends its public key to another device over an
insecure channel, for example, WiFi, and displays a two-dimensional barcode.
This barcode represents a visual encoding of the public key sent over the
insecure channel. 
The second device, supplied with the camera, takes a snapshot of the barcode and
runs a barcode recognition algorithm in order to process the image and
extract the public key. Afterwards, this device compares the
data obtained from the barcode with the data received over the insecure channel. If they match, the second device can trust the first device. 
The barcode-scanning procedure has to be executed by both devices for
bidirectional authentication.

The security assumption made by this pairing scheme is that mounting active
attacks is difficult without being detected.
The authors further analyzed the security of their pairing scheme against passive
attacks and proposed to additionally use the \gls{dh} session key exchange
protocol to protect against brute-force attacks. 

\subsubsection{Visible Laser Light} 
Mayrhofer et. al. \cite{Mayrhofer:2007kd} described a pairing scheme based on 
visible laser light for personal mobile devices equipped with a laser diode. 
These personal devices interact with another remote device, which is able to detect the laser light. 

The proposed pairing scheme works as follows. First, a user presses a button and turns on the
laser on the personal device. This causes the device to begin continuously
transmitting messages. When the remote device detects these messages, it
generates a response and broadcasts it over a wireless radio channel. 
Second, both devices start a key agreement protocol, and the target turns on a LED to
identify itself. 
Third, if the LED is activated on the target device expected by the user,
she presses a second button triggering an autonomous phase.  
During the autonomous phase the derived secret key is verified 
by sending a series of cryptographic challenges via the wireless radio channel, and
requiring that the responses to the challenges to be transmitted via the laser.

The authors evaluated their pairing scheme in the face of an active adversary
attempting to mount a \gls{mitm} attack.
They reported that the attack would only succeed
if the adversary can compromise the integrity and confidentiality
of the laser and wireless radio channels at the same time. 

\subsubsection{Visual authentication based on Integrity Checking (VIC)} 
Saxena et. al. \cite{Saxena:2011visual} improved the SiB pairing scheme by
providing mutual authentication between devices to be paired using only a
unidirectional visual channel, that is, requiring that only one of the two
devices has a camera, instead of both. 

The proposed pairing scheme employs short authenticated integrity checksums for key agreement 
and works as follows. First, each pairing device 
exchanges its public data, a public key and a random bit string, over
an insecure channel. Second, each device calculates a checksum, in
practice a cryptographic hash-function, over this public data, that is, both
public keys and random bit strings. 
Third, one of the devices sends its results to
the other device using the visual channel for comparison, that is, the second
device uses its camera to read the 2-D barcode displayed by the first device.
Fourth, the second device compares the hash transmitted over a display-camera channel by
the first device with the locally computed value. If the two values match, the
second device accepts the connection, and displays a confirmation message to
the user. Finally, the first device prompts the user to indicate if the second device
accepted the connection or not. 

The authors discussed the security of their
pairing scheme and indicated that it is resilient to \gls{mitm} attacks, only if the
hash function used in the scheme is collision-resistant.

\subsubsection{BEDA} 
Soriente et. al. \cite{Soriente:2007} investigated how to pair devices with
very limited interface capabilities such as a single button. 
They proposed a pairing scheme which first performs a \gls{dh} key agreement and then executes the
pairing procedure to authenticate the \gls{dh} public keys. 

The suggested pairing scheme consists of two phases and works as follows. In the first phase, a short 21-bit secret
is distributed between the devices with user assistance. Depending on the
available hardware interfaces this initial secret can either be obtained via
the user input provided to both devices (Button-to-Button) or by relaying the
data generated by one device to another device (Display-, Short Vibration-, Long
Vibration-to-Button). In the second phase, the authenticity of the exchanged
public key is incrementally verified in a 21-round procedure by using the
initial secret. 

The security of the proposed pairing depends on the confidentiality
of the channel. The authors discussed that their pairing scheme is secure against
\gls{mitm} attacks only if the data exchanged between the devices cannot be
eavesdropped. The BEDA scheme was cryptographically extended 
in the unified pairing framework \cite{Mayrhofer:2013} to provide \gls{pfs}, 
which further increases security against \gls{mitm} attackers.  

\subsubsection{Playful Security (Alice says)} 
Gallego et. al. \cite{Gallego:2011} proposed a pairing scheme based on the
memory game Simon. 
The suggested scheme uses \gls{sas}s computed by each device individually,
and a user is required to transmit these strings from one device to the
another device. 

The proposed pairing scheme works as follows. One device displays several audiovisual
patterns and the user relays these patterns to another device supplied with
the input interface. 
The first pattern consists of a single color and tone that encodes the first
two bits of the \gls{sas}. 
For the next round two bits will be concatenated to the first pattern. 
This data forms a new pattern that needs to be similarly transmitted by the
user. 
This iterative process continues until a sufficient number of bits have been
successfully exchanged between two devices. 
If an error occurs in a round, a new pattern will be concatenated with the
previous patterns that were exchanged successfully.  
To avoid synchronization issues the first device is equipped with two buttons. 
If an error occurs, the user selects previous button to repeat the exchange of the
\gls{sas}s between the devices. 

The authors argued that the proposed pairing scheme is robust to human errors and,
therefore, can mitigate \gls{mitm} attacks caused by such errors. 
 
\subsubsection{Safeslinger} 
Farb et. al. \cite{Farb:2013} presented a pairing scheme for data exchange
with smartphones.
That is, users upon a physical encounter can initiate the exchange of their
public keys, as well as, selected contact information and communicate securely
afterwards. 
The SafeSlinger scheme is built upon two cryptographic mechanisms, namely
multi-value commitments and group \gls{dh} key agreement. 
The pairing scheme requires active user interaction, which includes entering
the number of participating devices, selecting the data to be exchanged, and
finally, comparing a 3-word phrase which has to be commonly chosen by all
users. 

The authors analyzed the security of their pairing scheme and argued that SafeSlinger
mitigates attacks such as \gls{mitm}, group-in-the-middle, impersonation and
sybil attacks, by involving the user in the security chain and accounting for
user misbehavior.   

\subsubsection{Synchronized Drawing} 
Sethi et. al. \cite{Sethi:2014} presented a pairing scheme based on physical
proximity and commitment-based cryptographic primitives.

The proposed pairing scheme consists of four phases and works as follows.  
In the first phase, two devices attempt to establish a shared secret using \gls{dh} or a
similar protocol over an insecure channel.
In the second phase, fuzzy secrets are extracted from user input produced by simultaneously
drawing the same pattern with two fingers of the same hand, for example, a
thumb and index finger, on two touchscreens or surfaces of two devices to be
paired.
In the third phase, each device sends an unencrypted commitment message to another
device, which contains a hash of: \textit{(a)} the device's identifier,
\textit{(b)} the fuzzy secret derived from the drawing, \textit{(c)} a random
number, and \textit{(d)} the \gls{dh}-shared key.
In the fourth phase, each device encrypts its random number and fuzzy secret obtained in the 
third phase using the shared secret calculated in the first phase.

By carefully ensuring that both devices complete the third phase before
entering the fourth phase, authors argued that \gls{mitm} attacks can be
prevented. 

\subsubsection{Proximity Authentication} 
Li et. al. \cite{Li:2015dr} presented a pairing scheme which uses proximity to
perform mutual authentication between two devices without using NFC chips. 

The suggested pairing scheme works as follows. 
First, a user draws a zigzag pattern simultaneously on both devices to be
paired, using two fingers of the same hand. 
Second, each device individually derives a set of common features obtained from the
drawing.     
Third, the private set intersection approach \cite{Jarecki:2010} is applied to the
feature vectors of both devices in order to generate a shared secret key. 

The authors discussed security implications of their pairing scheme and claimed that
it is secure against dictionary and \gls{mitm} attacks. 

\subsubsection{Checksum Gestures}
Ahmed et. al. \cite{Ahmed:2015} proposed a pairing scheme based on \gls{sas}s,
where a continuous gesture is required for encoding authentication
information. 

The suggested pairing scheme works as follows. 
First, the user and target devices execute a key exchange protocol based on \gls{sas}s to 
obtain a checksum string (at least 20 bits) stored on both devices.
Second, the user device transforms this checksum string into a motion pattern, which is displayed
to the user, who is required to reproduce this motion pattern as a continuous gesture on
the target device.
Third, the input gesture is captured and processed by the target device, which then
compares the obtained data with the motion pattern derived locally from the shared
checksum string. 
If both match, the unidirectional communication channel is authenticated between the user and target devices. 
The security of the proposed pairing scheme is based on the feasibility of gesture
recognition technologies, particularly in maintaining sufficiently low
false-positive and false-negative error rates.

The authors analyzed the security of their pairing scheme based on the probability
of interpreting a false input
of an attacker as a correct gesture and reported that the probability of success
of a relay attack is under 5.5\%.  


\subsection{Discussion} 
\label{sub:discussion_hci}
The results of our \gls{hci} study are summarized in Table
\ref{tab:hci_channels}, from which we identify and discuss four
key points which have important security and usability implications for \gls{sdp}. 
First, we identify an important trade-off that exists between passive and
active \gls{hci} channels.
Second, the significance of usability properties, including the provision of
explicit user feedback and insensitivity of \gls{hci} input to environmental
conditions is considered.
Third, security issues resulting from various forms of intentional and
unintentional, as well as, benevolent and malicious user misbehavior are
explored.
Finally, the vital problem of observation threats for \gls{hci} channels is
presented, that is, the situation where an attacker can observe and exploit
human interaction.

\subsubsection{Trade-offs Between Passive and Active \gls{hci} Channels}
The \textit{handling} channel yields the best results in terms of usability
because it requires the minimum amount of user effort.
However, such pairing schemes do not give a user fine-grained control over the
pairing process and provide less assurance that pairing was established with
the intended device.
In contrast, data \textit{relay}, \textit{comparison} and \textit{generation}
require more user involvement but provide better control and assurance of
pairing.
Yet, these types of interaction are susceptible to user misbehavior and errors,
which makes it necessary for users to adequately understand the impact of
their actions.
For example, if the generation channel is involved it is not sufficient to
only incorporate common user experience. It is additionally required that the
user is alerted if the generated secret lacks sufficient entropy for its
intended use, so that the user can take appropriate action.

Hence, we identify an important trade-off between different \gls{hci}
channels.
While passive user interaction can be viably used for pairing in situations
where no sensitive information, such as financial or personal
data, is involved.
Active user participation should be used for more critical applications, for
example, bank transactions, where user awareness can be leveraged to increase
security in device pairing.

\subsubsection{Usability Properties}
Two usability properties which are crucial to augment both usability and
security in pairing are providing explicit user feedback, and ensuring
insensitivity to environmental conditions.

First, the importance of \textit{explicit user feedback} was outlined
previously \cite{Ion:2010, Gauger:2009}, yet only a few pairing schemes
provide it in a meaningful way.
However, the user feedback can not only mitigate input errors, and present the
evidence of pairing devices, for example, that pairing with the intended
device was successful, but also assist a human operator with security advice.
For instance, if the user generates data to produce a secret, the pairing
mechanism can notify the user if the provided input has sufficient entropy for
the intended application, or not.

Second, \textit{environment insensitivity} is also vital for maximizing user
experience.
That is, a pairing scheme should work for the intended use-case, irrespective
of the ambient conditions that might be reasonably expected to occur.
Section \ref{sec:app_classes} examines a range of specific use-cases exploring
this topic further.
The key point is that these two factors interact, for example, a pairing
scheme that requires audio comparison and confirmation from the user should
not be expected to be used in public scenarios.

\subsubsection{Security Issues}
The prior work emphasized the security issues in pairing
stemming from unintentional or deliberate user misbehavior \cite{Kumar:2009,
Uzun:2007, Saxena:2009rushing}.
Interestingly, only two pairing schemes (Playful Security \cite{Gallego:2011}
and Safeslinger \cite{Farb:2013}) accounted for such properties as
\textit{inattentive user}, \textit{rushing behavior} and \textit{consent
tampering} by design.
Table \ref{tab:hci_channels} clearly indicates that human interaction by
itself does not bring any security benefit if it does not consider threats
posed by the user behavior, for example, \gls{mana} \cite{Gehrmann:2004}.
Additionally, in our analysis we introduced an \textit{honest-but-curious}
participant who tries to obtain more information about the pairing party.
The motivation for this stems from a number of application classes that we
considered.
Since social and public pairing are in scope (Section \ref{sec:app_classes}),
it cannot be assumed that all pairing parties are benign and collaborative.
For example, social engineering can be used to infer extra information about
another user or if the sensing channel is involved another device or
participant can leverage this sensor data to violate privacy.
Moreover, \textit{human observation} has not been well addressed in the
pairing literature.
However, as we show under observation threats, the situation is dire and
this point must be taken into account if the pairing scheme relies on human
interaction.
 
\subsubsection{Observation Threats}
Regarding observation threats, we focus on security implications in
authentication techniques as the adversary similarly tries to circumvent
security by examining user interaction.
The examples of malicious observation include, but are not limited to,
shoulder surfing, audio or video analysis of the keyboard utilization and
voice recognition.
Specifically, Halevi and Saxena \cite{Halevi:2015} showed that keyboard
acoustic emanations can be used to successfully retrieve (even random)
passwords prompted with different typing styles.
Similarly, Davis et al. \cite{Davis:2014} proposed a method to extract audio
data from the high-speed video analysis in order to perform acoustic
eavesdropping without having a microphone.
More sophisticated attacks \cite{Backes:2009} exploited reflection from the
objects to reconstruct any confidential data displayed on the screen of a
device. Yue et al. \cite{Yue:2014} applied computer vision techniques to show
that it is possible, with 95\% probability, to reconstruct user input on the
touchscreen of a mobile device using a low resolution video of user
interaction.
Recent attacks against voice verification \cite{Mukhopadhyay:2015}
demonstrated that voice impersonation is achievable with the success rate of
90\% using only a limited number of victim's voice samples.
In short, observational threats are increasingly easy to achieve, and
therefore this risk should be taken into account when designing \gls{sdp} schemes.

In this section, we have investigated and discussed \gls{hci} channels along with the \gls{sdp} schemes utilizing them.
In the next section, we review the application classes and classify the surveyed \gls{sdp} schemes accordingly. 

%% file: section/application_classes.tex

\section{Application Classes}
\label{sec:app_classes}
In this section, we explore and analyze the four application classes 
introduced in Section \ref{sec:background}. 
First, we describe each application class
with respect to typical interactions, as well as its security and usability
insights. 
Second, we categorize the pairing
schemes covered in Sections \ref{sec:phy} and \ref{sec:hci} with regard to
their application classes and discuss the most interesting results of this classification.
Finally, we highlight important open issues in \gls{sdp} that have been identified in our study of application classes.

\begin{figure*}[htb!]
	\begin{minipage}{.5\textwidth}
	\centering
		\subfloat[Private]{
		\begin{tikzpicture}
		
			\node[inner sep=0pt] (device1) at (-1.6,0){\includegraphics[scale=2]{gfx/device}};
			\draw (-1.58, 0) node {\textit{D1}};
			
			\node[inner sep=0pt] (device2) at (1.6,0){\includegraphics[scale=2]{gfx/device}};
			\draw (1.58, 0) node {\textit{D2}};
			
   			\draw (0, 0.85) node [inner sep=0pt] {\textsc{\footnotesize{Physical}}};
   			
   			\draw [black, thick, <->,>=latex] (-1,0.65) -- (0.95,0.65);
   			\draw [black, thick, <->,>=latex] (-1,0.25) -- (0.95,0.25);
   			\node[text width=1cm, black] at (0.35,-0.15) { \textbf{...}};
   			\draw [black, thick, <->,>=latex] (-1,-0.55) -- (0.95,-0.55);
			
			\node[inner sep=0pt] (user) at (0,-2.3) {\includegraphics[scale=0.11]{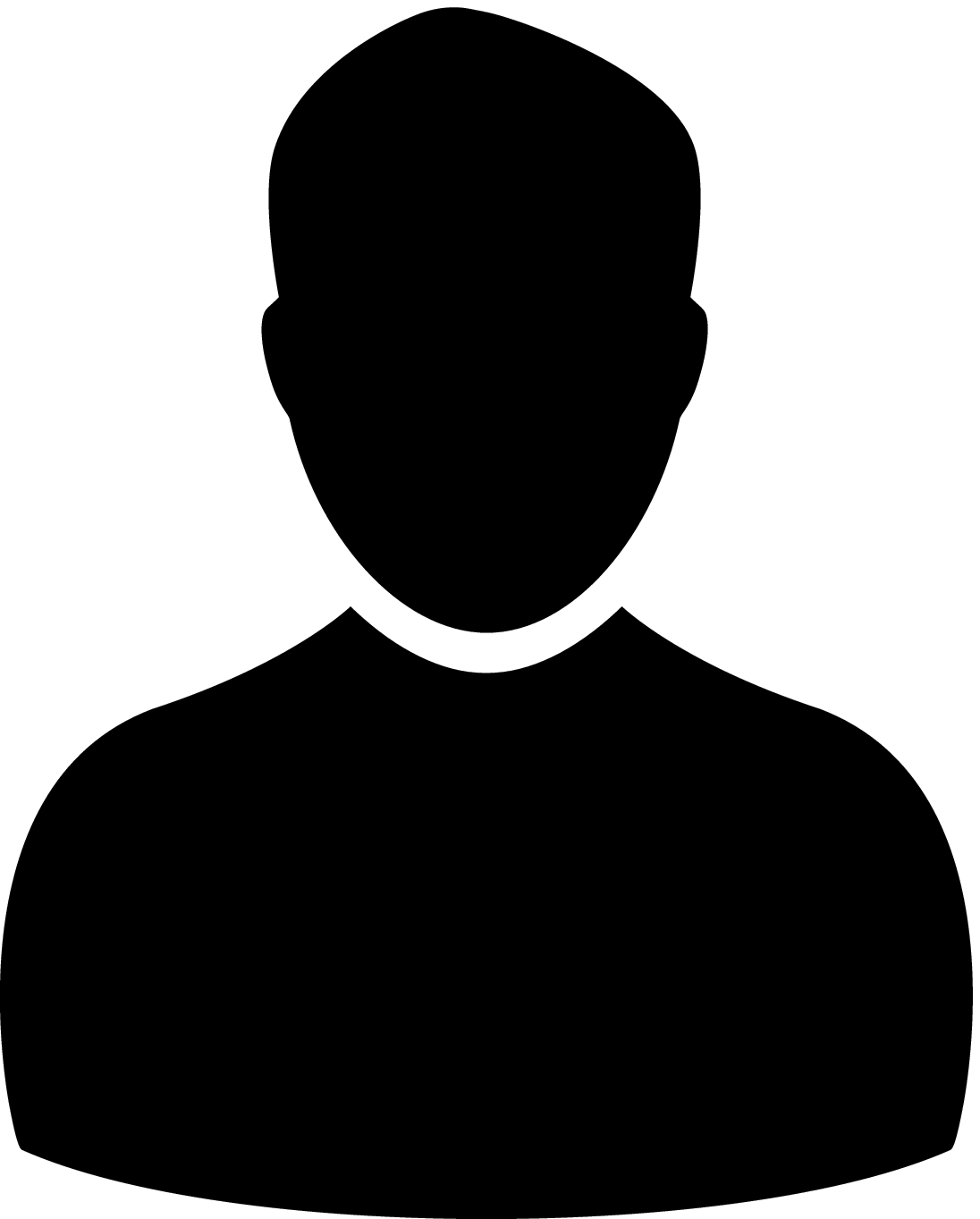}};		
			
			\draw (-1.4, -1.75) node [inner sep=0pt] {\textsc{\footnotesize{HCI}}};  
			\draw (1.4, -1.75) node [inner sep=0pt] {\textsc{\footnotesize{HCI}}}; 	
			
			\draw [black, double, thick, <->,>=stealth] (-1.6,-1.05) -- (-0.37,-2.05); 
   			\draw [black, double, thick, <->,>=stealth] (1.6,-1.05) -- (0.37,-2.05);  
   			
   			\node[
    			xshift = -3.5cm,
   				inner sep=0pt,
        		cloud, 
        		cloud puffs = 10,
        		thick,
    			minimum width=2cm,
   				minimum height=1.5cm,
        		draw,
        		scale=0.85,
    		] (env1) at (0,-0.05) {}; 
    	
    		\node[inner sep=0pt] at (env1.center){\includegraphics[scale=0.35]{gfx/wifi}};
    		\node[inner sep=0pt] at ([xshift=1.2em, yshift=1em]env1.south){\includegraphics[scale=0.5]{gfx/bluetooth}};
    		\node[inner sep=0pt] at ([xshift=-1.1em, yshift=-1.05em]env1.north){\includegraphics[scale=0.03]{gfx/gps}};
    		\node[inner sep=0pt] at ([xshift=-1.12em, yshift=0.87em]env1.east){\includegraphics[scale=0.3]{gfx/audio}};
    		\node[inner sep=0pt] at ([xshift=1.1em, yshift=-0.75em]env1.west){\includegraphics[scale=0.027]{gfx/temperature}};

    		\node[
    			xshift = 3.5cm,
   				inner sep=0pt,
       			cloud, 
        		cloud puffs = 10,
        		thick,
    			minimum width=2cm,
   				minimum height=1.5cm,
        		draw,
        		scale=0.85,
    		] (env2) at (0,-0.05) {}; 
    		
    		\node[inner sep=0pt] at (env2.center){\includegraphics[scale=0.35]{gfx/wifi}};
    		\node[inner sep=0pt] at ([xshift=1.2em, yshift=1em]env2.south){\includegraphics[scale=0.5]{gfx/bluetooth}};
    		\node[inner sep=0pt] at ([xshift=-1.1em, yshift=-1.05em]env2.north){\includegraphics[scale=0.03]{gfx/gps}};
    		\node[inner sep=0pt] at ([xshift=-1.12em, yshift=0.87em]env2.east){\includegraphics[scale=0.3]{gfx/audio}};
    		\node[inner sep=0pt] at ([xshift=1.1em, yshift=-0.75em]env2.west){\includegraphics[scale=0.027]{gfx/temperature}};
    		
			\draw (-3.05, 0.85) node [inner sep=0pt] {\textsc{\footnotesize{Physical}}};
   			\draw (3, 0.85) node [inner sep=0pt] {\textsc{\footnotesize{Physical}}};   	
   	 
   			\draw [black, densely dotted, thick, ->, >=triangle 60] (-2.7,-0.05) -- (-2.25,-0.05);
   			\draw [black, densely dotted, thick, ->, >=triangle 60] (2.7,-0.05) -- (2.2,-0.05);
   			
   			// Dummy nodes for security domains
   			\node[inner sep=0pt, opacity=0] (nd1) [inner sep=0pt] at (-1.6,0.985) {\textcolor{red}{\textsc{{.}}}};
   			\node[inner sep=0pt, opacity=0] (wd1) [inner sep=0pt] at (-2.25,-0.05) {\textcolor{red}{\textsc{{.}}}};
   			\node[inner sep=0pt, opacity=0] (sd1) [inner sep=0pt] at (-1.6, -2.1) {\textcolor{red}{\textsc{{.}}}};
   			\node[inner sep=0pt, opacity=0] (ed1) [inner sep=0pt] at (-0.99, -0.05) {\textcolor{red}{\textsc{{.}}}};
   			
   			\node[inner sep=0pt, opacity=0] (tlw) [inner sep=0pt] at (0, -3.2) {\textcolor{red}{\textsc{{.}}}}; 
   			\node[inner sep=0pt, opacity=0] (thw) [inner sep=0pt] at (0, 1.2) {\textcolor{red}{\textsc{{.}}}}; 
   			\node[inner sep=0pt, opacity=0] (mid) [inner sep=0pt] at (0, 0.985) {\textcolor{red}{\textsc{{.}}}}; 
   			
   			\node[inner sep=0pt, opacity=0] (nd2) [inner sep=0pt] at (1.6,0.985) {\textcolor{red}{\textsc{{.}}}};
   			\node[inner sep=0pt, opacity=0] (wd2) [inner sep=0pt] at (2.2,-0.05) {\textcolor{red}{\textsc{{.}}}};
   			\node[inner sep=0pt, opacity=0] (sd2) [inner sep=0pt] at (1.6, -2.1) {\textcolor{red}{\textsc{{.}}}};
   			\node[inner sep=0pt, opacity=0] (ed2) [inner sep=0pt] at (0.94, -0.05) {\textcolor{red}{\textsc{{.}}}};
   	
   			// Highlight: one security domain
   		 	\begin{pgfonlayer}{background}
   		 	
   		 	   \draw[black, dashed, semithick](nd1.west) to[closed,curve through={ 
  				(wd1.west) .. (sd1.south) .. (tlw.north) .. (sd2.south)
  				(wd2.east) .. (nd2.west) .. (mid.west) .. }] (nd1.west);
   		
  			\end{pgfonlayer}
   	
		\end{tikzpicture}
		\label{subfig:private}
		}
	\end{minipage}
	\begin{minipage}{.5\textwidth}
	\centering
		\subfloat[Public]{
		\begin{tikzpicture}
		
			\node[inner sep=0pt] (device1) at (-1.6,0){\includegraphics[scale=2]{gfx/device}};
			\draw (-1.58, 0) node {\textit{D1}};
			
			\node[inner sep=0pt] (device2) at (1.6,0){\includegraphics[scale=2]{gfx/device}};
			\draw (1.58, 0) node {\textit{D2}};
			
   			\draw (0, 0.85) node [inner sep=0pt] {\textsc{\footnotesize{Physical}}};
   			
   			\draw [black, thick, <->,>=latex] (-1,0.65) -- (0.95,0.65);
   			\draw [black, thick, <->,>=latex] (-1,0.25) -- (0.95,0.25);
   			\node[text width=1cm, black] at (0.35,-0.15) { \textbf{...}};
   			\draw [black, thick, <->,>=latex] (-1,-0.55) -- (0.95,-0.55);
			
			\node[inner sep=0pt] (user) at (-1.6,-2.3) {\includegraphics[scale=0.11]{gfx/user}};
			
			\draw (-1.97, -1.3) node [inner sep=0pt] {\textsc{\footnotesize{HCI}}};  
			\draw (0.3, -1.9) node [inner sep=0pt] {\textsc{\footnotesize{HCI}}}; 	
			
			\draw [black, double, thick, <->,>=stealth] (-1.6,-1.02) -- (-1.6,-1.52); 
   			\draw [black, double, thick, <->,>=stealth] (1.6,-1.07) -- (-1.25,-2); 
			
   			\node[
    			xshift = -3.5cm,
   				inner sep=0pt,
        		cloud, 
        		cloud puffs = 10,
        		thick,
    			minimum width=2cm,
   				minimum height=1.5cm,
        		draw,
        		scale=0.85,
    		] (env1) at (0,-0.05) {}; 
    	
    		\node[inner sep=0pt] at (env1.center){\includegraphics[scale=0.35]{gfx/wifi}};
    		\node[inner sep=0pt] at ([xshift=1.2em, yshift=1em]env1.south){\includegraphics[scale=0.5]{gfx/bluetooth}};
    		\node[inner sep=0pt] at ([xshift=-1.1em, yshift=-1.05em]env1.north){\includegraphics[scale=0.03]{gfx/gps}};
    		\node[inner sep=0pt] at ([xshift=-1.12em, yshift=0.87em]env1.east){\includegraphics[scale=0.3]{gfx/audio}};
    		\node[inner sep=0pt] at ([xshift=1.1em, yshift=-0.75em]env1.west){\includegraphics[scale=0.027]{gfx/temperature}};

    		\node[
    			xshift = 3.5cm,
   				inner sep=0pt,
       			cloud, 
        		cloud puffs = 10,
        		thick,
    			minimum width=2cm,
   				minimum height=1.5cm,
        		draw,
        		scale=0.85,
    		] (env2) at (0,-0.05) {}; 
    		
    		\node[inner sep=0pt] at (env2.center){\includegraphics[scale=0.35]{gfx/wifi}};
    		\node[inner sep=0pt] at ([xshift=1.2em, yshift=1em]env2.south){\includegraphics[scale=0.5]{gfx/bluetooth}};
    		\node[inner sep=0pt] at ([xshift=-1.1em, yshift=-1.05em]env2.north){\includegraphics[scale=0.03]{gfx/gps}};
    		\node[inner sep=0pt] at ([xshift=-1.12em, yshift=0.87em]env2.east){\includegraphics[scale=0.3]{gfx/audio}};
    		\node[inner sep=0pt] at ([xshift=1.1em, yshift=-0.75em]env2.west){\includegraphics[scale=0.027]{gfx/temperature}};
    		
			\draw (-3.05, 0.85) node [inner sep=0pt] {\textsc{\footnotesize{Physical}}};
   			\draw (3, 0.85) node [inner sep=0pt] {\textsc{\footnotesize{Physical}}};   	
   	 
   			\draw [black, densely dotted, thick, ->, >=triangle 60] (-2.7,-0.05) -- (-2.25,-0.05);
   			\draw [black, densely dotted, thick, ->, >=triangle 60] (2.7,-0.05) -- (2.2,-0.05);
   			
   			// Dummy nodes for security domains
   			\node[inner sep=0pt, opacity=0] (nd1) [inner sep=0pt] at (-1.6,1.1) {\textcolor{red}{\textsc{{.}}}};
   			\node[inner sep=0pt, opacity=0] (wd1) [inner sep=0pt] at (-2.25,-0.05) {\textcolor{red}{\textsc{{.}}}};
   			\node[inner sep=0pt, opacity=0] (wld1) [inner sep=0pt] at (-2.25,-1.7) {\textcolor{red}{\textsc{{.}}}}; 
   			\node[inner sep=0pt, opacity=0] (ed1) [inner sep=0pt] at (-0.99, -0.05) {\textcolor{red}{\textsc{{.}}}};
   			\node[inner sep=0pt, opacity=0] (eld1) [inner sep=0pt] at (-0.99, -1.7) {\textcolor{red}{\textsc{{.}}}}; 
   			
   			\node[inner sep=0pt, opacity=0] (thw) [inner sep=0pt] at (0, 1.2) {\textcolor{red}{\textsc{{.}}}}; 
   			\node[inner sep=0pt, opacity=0] (tlw) [inner sep=0pt] at (-1.6, -3.2) {\textcolor{red}{\textsc{{.}}}}; 
   			
   			\node[inner sep=0pt, opacity=0] (nd2) [inner sep=0pt] at (1.6,1.1) {\textcolor{red}{\textsc{{.}}}};
   			\node[inner sep=0pt, opacity=0] (wd2) [inner sep=0pt] at (2.2,-0.05) {\textcolor{red}{\textsc{{.}}}};
   			\node[inner sep=0pt, opacity=0] (sd2) [inner sep=0pt] at (1.6, -1.1) {\textcolor{red}{\textsc{{.}}}};
   			\node[inner sep=0pt, opacity=0] (ed2) [inner sep=0pt] at (0.94, -0.05) {\textcolor{red}{\textsc{{.}}}};
   			
   			// Highlight: two security domain
   		 	\begin{pgfonlayer}{background}
   		 	
   		 		\draw[black, dashed, semithick](nd1.west) to[closed,curve through={ 
  				(wd1.west) .. (wld1.west) .. (tlw.north) .. (eld1.east)
  				(ed1.east)}] (nd1.west);
				
				\draw[black, dashed, semithick](nd2.west) to[closed, curve through={
				(wd2.east) .. (sd2.south) .. (ed2.west) }] (nd2.west);
				
  			\end{pgfonlayer}
  			
		\end{tikzpicture}
		\label{subfig:public}
		}
	\end{minipage}
	
	\begin{minipage}{.5\textwidth}
	\centering
		\subfloat[Social]{
		\begin{tikzpicture}
			
			\node[inner sep=0pt] (device1) at (-1.6,0){\includegraphics[scale=2]{gfx/device}};
			\draw (-1.58, 0) node {\textit{D1}};
			
			\node[inner sep=0pt] (device2) at (1.6,0){\includegraphics[scale=2]{gfx/device}};
			\draw (1.58, 0) node {\textit{D2}};
			
   			\draw (0, 1.5) node [inner sep=0pt] {\textsc{\textcolor{white}{Physical}}};
   			\draw (0, 0.85) node [inner sep=0pt] {\textsc{\footnotesize{Physical}}};
   			
   			\draw [black, thick, <->,>=latex] (-1,0.65) -- (0.95,0.65);
   			\draw [black, thick, <->,>=latex] (-1,0.25) -- (0.95,0.25);
   			\node[text width=1cm, black] at (0.35,-0.15) { \textbf{...}};
   			\draw [black, thick, <->,>=latex] (-1,-0.55) -- (0.95,-0.55);;
			
			\node[inner sep=0pt] (user1) at (-1.6,-2.3) {\includegraphics[scale=0.11]{gfx/user}};	
		
			\node[inner sep=0pt] (user2) at (1.6,-2.3) {\includegraphics[scale=0.11]{gfx/user}};	
			
			\draw (-1.97, -1.3) node [inner sep=0pt] {\textsc{\footnotesize{HCI}}};  
			\draw (1.95, -1.3) node [inner sep=0pt] {\textsc{\footnotesize{HCI}}}; 	
			
			\draw [black, double, thick, <->,>=stealth] (-1.6,-1.02) -- (-1.6,-1.52); 
			\draw [black, double, thick, <->,>=stealth] (1.6,-1.02) -- (1.6,-1.52); 
   			
   			\draw (0, -1.95) node {\textsc{\footnotesize{H2H}}};
   		    \draw [black, dashed, thick, <->, >=angle 90] (-1.3,-2.15) -- (1.25,-2.15);  
   		    
   			\node[
    			xshift = -3.5cm,
   				inner sep=0pt,
        		cloud, 
        		cloud puffs = 10,
        		thick,
    			minimum width=2cm,
   				minimum height=1.5cm,
        		draw,
        		scale=0.85,
    		] (env1) at (0,-0.05) {}; 
    	
    		\node[inner sep=0pt] at (env1.center){\includegraphics[scale=0.35]{gfx/wifi}};
    		\node[inner sep=0pt] at ([xshift=1.2em, yshift=1em]env1.south){\includegraphics[scale=0.5]{gfx/bluetooth}};
    		\node[inner sep=0pt] at ([xshift=-1.1em, yshift=-1.05em]env1.north){\includegraphics[scale=0.03]{gfx/gps}};
    		\node[inner sep=0pt] at ([xshift=-1.12em, yshift=0.87em]env1.east){\includegraphics[scale=0.3]{gfx/audio}};
    		\node[inner sep=0pt] at ([xshift=1.1em, yshift=-0.75em]env1.west){\includegraphics[scale=0.027]{gfx/temperature}};

    		\node[
    			xshift = 3.5cm,
   				inner sep=0pt,
       			cloud, 
        		cloud puffs = 10,
        		thick,
    			minimum width=2cm,
   				minimum height=1.5cm,
        		draw,
        		scale=0.85,
    		] (env2) at (0,-0.05) {}; 
    		
    		\node[inner sep=0pt] at (env2.center){\includegraphics[scale=0.35]{gfx/wifi}};
    		\node[inner sep=0pt] at ([xshift=1.2em, yshift=1em]env2.south){\includegraphics[scale=0.5]{gfx/bluetooth}};
    		\node[inner sep=0pt] at ([xshift=-1.1em, yshift=-1.05em]env2.north){\includegraphics[scale=0.03]{gfx/gps}};
    		\node[inner sep=0pt] at ([xshift=-1.12em, yshift=0.87em]env2.east){\includegraphics[scale=0.3]{gfx/audio}};
    		\node[inner sep=0pt] at ([xshift=1.1em, yshift=-0.75em]env2.west){\includegraphics[scale=0.027]{gfx/temperature}};
    		
			\draw (-3.05, 0.85) node [inner sep=0pt] {\textsc{\footnotesize{Physical}}};
   			\draw (3, 0.85) node [inner sep=0pt] {\textsc{\footnotesize{Physical}}};   	
   	 
   			\draw [black, densely dotted, thick, ->, >=triangle 60] (-2.7,-0.05) -- (-2.25,-0.05);
   			\draw [black, densely dotted, thick, ->, >=triangle 60] (2.7,-0.05) -- (2.2,-0.05);
   			 			
   			// Dummy nodes for security domains
   			\node[inner sep=0pt, opacity=0] (nd1) [inner sep=0pt] at (-1.6,1.1) {\textcolor{red}{\textsc{{.}}}};
   			\node[inner sep=0pt, opacity=0] (wd1) [inner sep=0pt] at (-2.25,-0.05) {\textcolor{red}{\textsc{{.}}}};
   			\node[inner sep=0pt, opacity=0] (wld1) [inner sep=0pt] at (-2.25,-1.7) {\textcolor{red}{\textsc{{.}}}}; 
   			\node[inner sep=0pt, opacity=0] (ed1) [inner sep=0pt] at (-0.99, -0.05) {\textcolor{red}{\textsc{{.}}}};
   			\node[inner sep=0pt, opacity=0] (eld1) [inner sep=0pt] at (-0.99, -1.7) {\textcolor{red}{\textsc{{.}}}}; 
   			\node[inner sep=0pt, opacity=0] (sd1) [inner sep=0pt] at (-1.6, -3.2) {\textcolor{red}{\textsc{{.}}}};
   			
   			\node[inner sep=0pt, opacity=0] (thw) [inner sep=0pt] at (0, 1.2) {\textcolor{red}{\textsc{{.}}}}; 
   			
   			\node[inner sep=0pt, opacity=0] (nd2) [inner sep=0pt] at (1.6,1.1) {\textcolor{red}{\textsc{{.}}}};
   			\node[inner sep=0pt, opacity=0] (wd2) [inner sep=0pt] at (2.2,-0.05) {\textcolor{red}{\textsc{{.}}}};
   			\node[inner sep=0pt, opacity=0] (wld2) [inner sep=0pt] at (2.2,-1.7) {\textcolor{red}{\textsc{{.}}}}; 
   			\node[inner sep=0pt, opacity=0] (ed2) [inner sep=0pt] at (0.94, -0.05) {\textcolor{red}{\textsc{{.}}}};
   			\node[inner sep=0pt, opacity=0] (eld2) [inner sep=0pt] at (0.94, -1.7) {\textcolor{red}{\textsc{{.}}}}; 
   			\node[inner sep=0pt, opacity=0] (sd2) [inner sep=0pt] at (1.6, -3.2) {\textcolor{red}{\textsc{{.}}}};
   			
   			// Highlight: two security domain
   		 	\begin{pgfonlayer}{background}
   		 	
   		 		\draw[black, dashed, semithick](nd1.west) to[closed,curve through={ 
  				(wd1.west) .. (wld1.west) .. (sd1.north) .. (eld1.east)
  				(ed1.east)}] (nd1.west);
				
				\draw[black, dashed, semithick](nd2.east) to[closed,curve through={ 
  				(wd2.east) .. (wld2.east) .. (sd2.north) .. (eld2.west)
  				(ed2.west)}] (nd2.east);
				
  			\end{pgfonlayer}
   			
		\end{tikzpicture}
		\label{subfig:social}
		}
	\end{minipage}
	\begin{minipage}{.5\textwidth}
	\centering
		\subfloat[Unattended]{
		\begin{tikzpicture}
		
			\node[inner sep=0pt] (device1) at (-1.6,0){\includegraphics[scale=2]{gfx/device}};
			\draw (-1.58, 0) node {\textit{D1}};
			
			\node[inner sep=0pt] (device2) at (1.6,0){\includegraphics[scale=2]{gfx/device}};
			\draw (1.58, 0) node {\textit{D2}};
			
   			\draw (0, 1.5) node [inner sep=0pt] {\textsc{\textcolor{white}{Physical}}};
   			\draw (0, 0.85) node [inner sep=0pt] {\textsc{\footnotesize{Physical}}};
   			
   			\draw [black, thick, <->,>=latex] (-1,0.65) -- (0.95,0.65);
   			\draw [black, thick, <->,>=latex] (-1,0.25) -- (0.95,0.25);
   			\node[text width=1cm, black] at (0.35,-0.15) { \textbf{...}};
   			\draw [black, thick, <->,>=latex] (-1,-0.55) -- (0.95,-0.55);
			
			\node[inner sep=0pt, opacity=0] (user) at (1.6,-2.3) {\includegraphics[scale=0.11]{gfx/user}};
			
   			\node[
    			xshift = -3.5cm,
   				inner sep=0pt,
        		cloud, 
        		cloud puffs = 10,
        		thick,
    			minimum width=2cm,
   				minimum height=1.5cm,
        		draw,
        		scale=0.85,
    		] (env1) at (0,-0.05) {}; 
    	
    		\node[inner sep=0pt] at (env1.center){\includegraphics[scale=0.35]{gfx/wifi}};
    		\node[inner sep=0pt] at ([xshift=1.2em, yshift=1em]env1.south){\includegraphics[scale=0.5]{gfx/bluetooth}};
    		\node[inner sep=0pt] at ([xshift=-1.1em, yshift=-1.05em]env1.north){\includegraphics[scale=0.03]{gfx/gps}};
    		\node[inner sep=0pt] at ([xshift=-1.12em, yshift=0.87em]env1.east){\includegraphics[scale=0.3]{gfx/audio}};
    		\node[inner sep=0pt] at ([xshift=1.1em, yshift=-0.75em]env1.west){\includegraphics[scale=0.027]{gfx/temperature}};

    		\node[
    			xshift = 3.5cm,
   				inner sep=0pt,
       			cloud, 
        		cloud puffs = 10,
        		thick,
    			minimum width=2cm,
   				minimum height=1.5cm,
        		draw,
        		scale=0.85,
    		] (env2) at (0,-0.05) {}; 
    		
    		\node[inner sep=0pt] at (env2.center){\includegraphics[scale=0.35]{gfx/wifi}};
    		\node[inner sep=0pt] at ([xshift=1.2em, yshift=1em]env2.south){\includegraphics[scale=0.5]{gfx/bluetooth}};
    		\node[inner sep=0pt] at ([xshift=-1.1em, yshift=-1.05em]env2.north){\includegraphics[scale=0.03]{gfx/gps}};
    		\node[inner sep=0pt] at ([xshift=-1.12em, yshift=0.87em]env2.east){\includegraphics[scale=0.3]{gfx/audio}};
    		\node[inner sep=0pt] at ([xshift=1.1em, yshift=-0.75em]env2.west){\includegraphics[scale=0.027]{gfx/temperature}};
    		
			\draw (-3.05, 0.85) node [inner sep=0pt] {\textsc{\footnotesize{Physical}}};
   			\draw (3, 0.85) node [inner sep=0pt] {\textsc{\footnotesize{Physical}}};   	
   	 
   			\draw [black, densely dotted, thick, ->, >=triangle 60] (-2.7,-0.05) -- (-2.25,-0.05);
   			\draw [black, densely dotted, thick, ->, >=triangle 60] (2.7,-0.05) -- (2.2,-0.05);

			// Dummy nodes for security domains
   			\node[inner sep=0pt, opacity=0] (nd1) [inner sep=0pt] at (-1.6,1.1) {\textcolor{red}{\textsc{{.}}}};
   			\node[inner sep=0pt, opacity=0] (wd1) [inner sep=0pt] at (-2.25,-0.05) {\textcolor{red}{\textsc{{.}}}};
   			\node[inner sep=0pt, opacity=0] (sd1) [inner sep=0pt] at (-1.6, -1.1) {\textcolor{red}{\textsc{{.}}}};
   			\node[inner sep=0pt, opacity=0] (ed1) [inner sep=0pt] at (-0.99, -0.05) {\textcolor{red}{\textsc{{.}}}};
   	
   			\node[inner sep=0pt, opacity=0] (thw) [inner sep=0pt] at (0, 1.2) {\textcolor{red}{\textsc{{.}}}}; 
   			\node[inner sep=0pt, opacity=0] (tlw) [inner sep=0pt] at (-1.6, -3.2) {\textcolor{red}{\textsc{{.}}}}; 
   			
   			\node[inner sep=0pt, opacity=0] (nd2) [inner sep=0pt] at (1.6,1.1) {\textcolor{red}{\textsc{{.}}}};
   			\node[inner sep=0pt, opacity=0] (wd2) [inner sep=0pt] at (2.2,-0.05) {\textcolor{red}{\textsc{{.}}}};
   			\node[inner sep=0pt, opacity=0] (sd2) [inner sep=0pt] at (1.6, -1.1) {\textcolor{red}{\textsc{{.}}}};
   			\node[inner sep=0pt, opacity=0] (ed2) [inner sep=0pt] at (0.94, -0.05) {\textcolor{red}{\textsc{{.}}}};
   			
   			// Highlight: two security domain
   		 	\begin{pgfonlayer}{background}
   		 
		   		\draw[black, dashed, semithick](nd1.east) to[closed, curve through={
				(wd1.west) .. (sd1.south) .. (ed1.east) }] (nd1.east);
				
				\draw[black, dashed, semithick](nd2.west) to[closed, curve through={
				(wd2.east) .. (sd2.south) .. (ed2.west) }] (nd2.west);
				
  			\end{pgfonlayer}
   			
		\end{tikzpicture}
		\label{subfig:unattended}
		}
	\end{minipage}
\caption{The four application classes. Each application class consists of two
devices to be paired, each from a distinct security domain, except for the
private application class (a).  The boundries of security
domains are indicated by a dashed line.}
\label{fig:application_classes}
\end{figure*}

\subsection{Overview of Application Classes}
An application class covers a set of similar \gls{sdp} use-cases,
each of which involves a similar degree of involvement and level of user control over the pairing process.
 We recall the
four application classes introduced earlier:
\textit{a)} private, \textit{(b)} public, \textit{(c)} social, and \textit{(d)} unattended. 
The private class corresponds to a ``classic pairing'' case, where a single user
either owns or directly controls two devices that ought to be
paired. The public class is related to a single user possessing
one device, where the user performs the pairing with some third party
infrastructure, for example, a payment terminal, over which
she has no control. The social class incorporates two users
who would like to securely pair their corresponding devices.
The unattended class deals with the case where two devices
belonging to the same ownership domain, for example, owned by the same person
or organization, pair with no user involvement.
Figure \ref{fig:application_classes} depicts the four application classes, 
instantiated from our system model, to provide a better understanding of the 
typical interactions for each application class.  

The ownership of the devices being paired plays a critical role in \gls{sdp},
necessitating its explicit consideration when describing application classes.
We recall that a \textit{security domain}
is the set of devices, data, policies and intentions that a single party controls. 
That is, a security domain refers to the limit
of enforcement of security policy by a particular owner or
controller of one or more devices. These security domains
are especially significant when more than one exists,
as it allows for security requirements of pairing devices to be
differentially achieved or undermined, either by the pairing
process, or subsequent actions of one of the pairing parties.

For example, consider Figures \ref{subfig:private} and \ref{subfig:social}. In Figure \ref{subfig:private}, a
single user controls all devices, and so a single security domain
exists. Therefore, following a successful pairing procedure,
there are only two possibilities: either, the policy requirements
of the single security domain are met, or not. In contrast, for
Figure \ref{subfig:social}, there are two users each controlling a separate
device, D1 and D2 respectively. In this case, if the security
policy requirements of each user differ, it may be possible that
the security policy of one user is satisfied, but not for the other.
Similarly, one of the users may later reveal information that,
without violating their own security policy, may violate that of
the other. That is, the presence of the second security policy
allows for a more complex set of outcomes, as compared to if there were only a
single security domain.

In the following, we expand on the four application classes under consideration. 

\subsubsection{Private}
Figure \ref{subfig:private} depicts the well-known private class,
which applies when a single user either owns or controls both devices. A good
example of this scenario is pairing smart devices that belong to the same
person. In such a setting, a rich set of \gls{hci} interactions are possible
since a user can freely communicate with and handle her portable devices in
many ways. The physical interactions between the devices, as well as, with the
ambient environment are user-enabled, and only limited by the availability of
hardware interfaces on the devices. 

From a security perspective, private pairing is often performed
in a rather restricted environment, for example, home premises
or a workplace, where such threats as external observation and
communication interception are reduced. The private class
consists of a single security domain, that is, all devices are
subject to the same security policy requirements, because they
are controlled by a single party. In this context, the focus of
the user tends towards usability, due to the combination of
reduced perceived threats and the relative frequency of pairing
that may occur, especially given the increasing numbers of
devices that people own. Hence, usability must be preserved
and emphasized, even in the face of numerous devices to
be paired with one another. Despite the lower perception of
risk, it remains important to maintain security. 

\subsubsection{Public}
The public class, shown in Figure \ref{subfig:public}, corresponds
to the case where a single user possesses one device
but has no control over another device to be paired with.
For example, the user wants to pair her personal device, for instance, a
smartphone, with a third party infrastructure such as a public
\gls{ap}, a printer or a payment terminal.

In terms of \gls{hci} interactions, a human operator has fewer
options as compared to the private class, because the public
infrastructure typically has only a few common user interfaces
and cannot be moved, shaken or handled in a convenient
way. Similarly, physical interfaces used for communication
between the devices, as well as, with the ambient environment
are restricted and typically cannot be invoked by the user.

From a security perspective, the public class implies a
more hostile environment, that is, public places, as compared to
the private class. Thus, user actions during the pairing procedure
are subject to external observation, which can come in
the form of shoulder surfing or ubiquitous CCTV. Additionally,
an attacker can stealthily install rouge devices in the public
premises to interfere or hijack the pairing process.

The public class incorporates two distinct security domains,
namely the user with her device and the infrastructure,
which opens a door to a number of threats outlined in the
following discussion. In comparison with the private class,
users are likely to have an increased perception of security
risks in such public scenarios. Therefore, users may reasonably
accept some shift in the balance away from usability in order
to improve security. However, care must be taken not to reduce
usability to the point where users' tolerance is exhausted.

\subsubsection{Social}
The social class, illustrated in Figure \ref{subfig:social}, represents a case 
where two different users would like perform pairing between their personal devices \cite{Uzun:2011}. 
Pairing two smartphones that belong to different people is a good example of such a scenario. 
It is obvious from the given example that the social class implies two distinct security
domains, that is, two users with their devices.
The presence of multiple security domains can result in complicated security
outcomes, as previously described.

The reality of users' concerns regarding these complications
can be observed, for example, through users' reluctance to
hand their personal devices over to others. 
Explicit human-to-human (H2H) interaction can be used to resolve this concern,
that is, to allow users to pair their devices, without losing
physical possession of them at any point during the pairing process. 
Since users are interfacing with their devices individually,
numerous \gls{hci} and physical interactions can be enabled similarly to the private class.

With regard to security, social pairing is vulnerable to external observation. 
On the one hand, the typical environment for the social class may present lower inherent risk as
compared to the public class, for example, by occurring in a private house instead of in public places. 
On the other hand, the social pairing may still occur in a public place. 
Also, social pairing procedures typically involve user interaction, which is particularly at risk
of observation attacks. 
Thus, while the social class can suffer from a similar level of risk to the public class, users'
perception of the risk may be lower, potentially reducing their tolerance for security measures that harm usability. Therefore, considerable attention should be given to optimizing user experience for social pairing procedures, 
while still ensuring adequate security.

\subsubsection{Unattended}
Figure \ref{subfig:unattended} depicts the unattended class, which applies
when two devices perform pairing without any user involvement. 
For example, two \gls{iot} devices, for instance, sensors, located nearby can
pair and similarly wearables, as well as, \gls{imd}s can be paired. 
In the case of wearables and \gls{imd}s, a user is present, but
acts only as a carrier of the devices, and does not consciously participate
in the pairing process. 

Since no user is involved, unattended pairing relies solely on various
physical interactions, especially those used for data acquisition, that is,
sensing. 
The key approach employed in unattended pairing is to utilize various sensor
capabilities to measure the ambient environment over time. 
Thus, if two devices continuously sense sufficiently similar contexts, they
interpret this as evidence of their physical proximity. When two devices believe
that they are in physical proximity, they may then attempt to pair \cite{Schurmann:2013,
Miettinen:2014}. 
The ambient environment does not only refer to physical characteristics such
as wireless radio, audio, luminosity, humidity, etc. 
It can also correspond to measuring the human body, for example, a heartbeat
rate \cite{Rostami:2013} or muscle contraction \cite{Yang:2016}, as well as, 
capturing user specific actions, for example, a gait \cite{Shrestha:2016, Schurmann:2017}, an approach trajectory \cite{Juuti:2016} 
or a head movement pattern \cite{Li:2016}.

In terms of security, the unattended class significantly differs from other
application classes since the pairing devices communicate in a standalone
fashion without explicit user control. 
This poses major security challenges, such as physical access to the devices
by an adversary, in addition to her ability to efficiently monitor
\cite{Mehrnezhad:2015}, disturb or even manipulate \cite{Shrestha:2015} the
pairing environment without being noticed. 
Moreover, it is not straightforward to unambiguously define a number of
security domains in the unattended class. For example, the proposed pairing
schemes \cite{Schurmann:2013, Miettinen:2014} assumed that devices
originated from the same ownership, for example, either a user or the
infrastructure, and, thus, security domain. 
An open question is the pairing of \gls{iot} devices which belong to different
security domains. 

The unattended pairing is by definition an autonomous process, removing all user
interaction. 
It can be viewed as pushing the usability-security trade-off completely in the
direction of usability. 
It is, therefore, not surprising that the security properties of unattended
pairing schemes are often weaker as compared to the other application classes. 
Thus, more research is required to devise more secure unattended pairing schemes.

\subsection{Classification of Pairing Schemes}
To categorize different pairing schemes with respect to their application
classes, we used the following approach. First, we considered pairing
schemes that were surveyed in the physical and \gls{hci} sections. 
Second, for each pairing scheme we sought a particular use-case discussed by
the authors, or looked at the specific setting in which the implemented
pairing scheme was tested and evaluated. 
Using this information, we explicitly assigned each pairing scheme to one or
more of the application classes. 
Finally, we considered for each pairing scheme, whether it could be
extended to other application classes, either by an implicit reference in the
paper, or by considering the physical and \gls{hci} interactions necessary for
a specific pairing scheme, and comparing them with interactions possible in each
application class.
Then pairing schemes that rely on biometry and have been used in the field of
\gls{imd}s, for example, \cite{Rostami:2013}, are outside of the scope of this
article, and thus, are not included in these results, and are mentioned here only
for completeness. 
The results of our classification are presented in Table \ref{tab:appclass},
and are discussed below.

In line with the prior research we see that most of the proposed pairing
schemes are aimed at the private application class. 
The public class is the second most targeted application scenario, followed
by the social and unattended classes respectively. 
It was also observed that many pairing schemes could be extended to other
application classes, especially schemes that implement security mechanisms on the
physical layer \cite{Capkun:2008, Gollakota:2011} or utilize contextual
sensing \cite{Schurmann:2013, Miettinen:2014}. 
An interesting trade-off exists between those two groups of pairing schemes. 
While the former can offer provable security guarantees, it requires low-level
changes of the communication stack which hinders the wide-spread adoption. 
In contrast, the latter group can be more easily deployed, but lacks clear
security guarantees \cite{Shrestha:2015}.

Another observation is related to pairing schemes deployed in commercial
products, for example, \cite{PBC:2006, Bluetooth:2007, WPS:NFC, SSP:OOB}. 
Often these schemes are claimed to be applicable to multiple of the
application classes, irrespective of whether they are suitable on the basis of
their security properties. 
For example, the \gls{pbc} scheme \cite{PBC:2006} is available in both the
infrastructure mode, as well as for Wi-Fi direct \cite{Wifi:direct}. 
However, \gls{pbc} is known to be vulnerable to \gls{mitm} attacks, and the
exposure is much greater in public and social contexts as compared to the
private application class. 
Similar arguments apply to Just Works \cite{Bluetooth:2007} which is the
Bluetooth pairing scheme. 
Two other pairing schemes provided by the standardized bodies, namely Near Field Communication
\cite{WPS:NFC} and Out-of-band \cite{SSP:OOB} rely on the \gls{nfc} technology
to transmit sensitive data, for example, a device generated password, in plain
text. 
Despite being difficult, eavesdropping the \gls{nfc} channel is not impossible
and the chance of successful attack is much higher in public and social
scenarios.

\begin{table}[!htb]
	\caption{Application classes - classification of pairing schemes}
	\label{tab:appclass}
	\centering
	\begin{threeparttable}
	\rotatebox{0}{
		\begin{tabularx}{\columnwidth}{p{4.8 cm}p{0.40cm}p{0.40cm}p{0.40cm}p{0.40cm}}

		\hline
		\hline
	
	    \multicolumn{1}{c}{\shortstack{\\ \textbf{Pairing Scheme} }} &
		\multicolumn{4}{c}{\shortstack{\\ \textbf{Application classes} }} \\
		\hline
		
		&  
		\rotatebox{55}{\textbf{Private  }}& 		
		\rotatebox{55}{\textbf{Public   }}& 		
		\rotatebox{55}{\textbf{Social   }}& 		
		\rotatebox{55}{\textbf{Unattended   }}\\
		\hline


		\hline
		Push Button Configuration \cite{PBC:2006} & 
		\CIRCLE & 
		\CIRCLE & 
		\CIRCLE & 
		\Circle  \\ 
		
    	Integrity codes \cite{Capkun:2008}& 
		\RIGHTcircle & 
		\CIRCLE & 
		\RIGHTcircle & 
		\RIGHTcircle  \\ 
		
    	Tamper-evident pairing \cite{Gollakota:2011}& 
		\CIRCLE & 
		\RIGHTcircle & 
		\RIGHTcircle & 
		\RIGHTcircle  \\ 
		
		Just Works \cite{Bluetooth:2007} & 
		\CIRCLE & 
		\CIRCLE & 
		\CIRCLE & 
		\Circle  \\ 
		
		Noisy tags \cite{Castelluccia:2006} & 
		\RIGHTcircle & 
		\CIRCLE & 
		\Circle & 
		\Circle  \\ 
		
		Adopted-Pet \cite{Amariucai:2011} & 
		\CIRCLE & 
		\RIGHTcircle & 
		\Circle & 
		\CIRCLE  \\ 

		Near Field Communication \cite{WPS:NFC} & 
		\CIRCLE & 
		\CIRCLE & 
		\CIRCLE & 
		\Circle \\ 
		
	    Out-of-band \cite{SSP:OOB} & 
		\CIRCLE & 
		\CIRCLE & 
		\CIRCLE & 
		\Circle  \\ 
		
		KeyLED \cite{Roman:2008} & 
		\CIRCLE & 
		\RIGHTcircle & 
		\RIGHTcircle & 
		\Circle  \\ 
		
     	Enlighten Me! \cite{Gauger:2009} & 
		\RIGHTcircle & 
		\CIRCLE & 
		\Circle & 
		\Circle  \\ 
		
     	Flashing displays \cite{Kovavcevic:2015} & 
		\CIRCLE & 
		\RIGHTcircle & 
		\RIGHTcircle & 
		\Circle  \\ 
		
		Talking to Strangers \cite{Balfanz:2002} & 
		\RIGHTcircle & 
		\CIRCLE & 
		\RIGHTcircle & 
		\Circle  \\ 
		
		Loud and Clear \cite{Goodrich:2006} & 
		\CIRCLE & 
		\RIGHTcircle & 
		\RIGHTcircle & 
		\Circle  \\ 

	    HAPADEP \cite{Soriente:2008} & 
		\CIRCLE & 
		\RIGHTcircle & 
		\RIGHTcircle & 
		\Circle  \\ 
    		
        Zero-Power pairing \cite{Halperin:2008} & 
		\Circle & 
		\Circle & 
		\Circle & 
		\CIRCLE  \\ 
		
		Ultrasonic ranging \cite{Mayrhofer:2006} & 
		\RIGHTcircle & 
		\RIGHTcircle & 
		\CIRCLE & 
		\Circle  \\ 
		
		SBVLC \cite{Zhang:2016} & 
		\CIRCLE & 
		\CIRCLE & 
		\CIRCLE & 
		\Circle  \\ 
		
		Vibrate-to-unlock \cite{Saxena:2011vibrate} & 
		\CIRCLE & 
		\Circle & 
		\Circle & 
		\Circle \\ 
		
		Shot \cite{Studer:2011} & 
		\RIGHTcircle & 
		\Circle & 
		\CIRCLE & 
		\Circle  \\ 
    	    
    	Vibreaker \cite{Anand:2016} & 
		\CIRCLE & 
		\Circle & 
		\Circle & 
		\Circle  \\ 
		
		Amigo \cite{Varshavsky:2007}  & 
		\RIGHTcircle & 
		\CIRCLE & 
		\RIGHTcircle & 
		\RIGHTcircle \\ 
		
	    Good Neighbor \cite{Cai:2011}  & 
		\CIRCLE & 
		\RIGHTcircle & 
		\RIGHTcircle & 
		\Circle \\ 
		
    	Wanda \cite{Pierson:2016}  & 
		\CIRCLE & 
		\CIRCLE & 
		\RIGHTcircle & 
		\Circle \\ 
		
    	Ambient Audio pairing \cite{Schurmann:2013}  & 
		\RIGHTcircle & 
		\RIGHTcircle & 
		\RIGHTcircle & 
		\CIRCLE  \\ 
		
    	Zero-interaction pairing \cite{Miettinen:2014}  & 
		\RIGHTcircle & 
		\RIGHTcircle & 
		\RIGHTcircle & 
		\CIRCLE  \\ 
		
        MagPairing \cite{Jin:2014}  & 
		\RIGHTcircle & 
		\RIGHTcircle & 
		\CIRCLE & 
		\Circle  \\ 
		
		Touch-and-Guard \cite{Wang:2016}  & 
		\CIRCLE & 
		\CIRCLE & 
		\Circle & 
		\Circle  \\ 
					
		MANA  \cite{Gehrmann:2004}  & 
		\CIRCLE & 
		\RIGHTcircle & 
		\CIRCLE & 
		\Circle  \\ 
		
		Access point authentication \cite{Roth:2008}  & 
		\RIGHTcircle & 
		\CIRCLE & 
		\Circle & 
		\Circle  \\ 
		
		Shake Them Up! \cite{Castelluccia:2005}  & 
		\CIRCLE & 
		\Circle & 
		\RIGHTcircle & 
		\Circle  \\ 
		
		Shake Well Before Use \cite{Mayrhofer:2007sh} & 
		\CIRCLE & 
		\Circle & 
		\Circle & 
		\Circle  \\ 
		
		SAPHE  \cite{Groza:2012} & 
		\CIRCLE & 
		\Circle & 
		\Circle & 
		\Circle  \\ 

		Authentication using ultrasound \cite{Kindberg:2003bh} & 
		\RIGHTcircle & 
		\CIRCLE & 
		\CIRCLE & 
		\Circle \\ 
		
		Synchronized Audio-Visual Patterns \cite{Prasad:2008} & 
		\CIRCLE & 
		\CIRCLE & 
		\CIRCLE & 
		\Circle  \\ 
		
		RhythmLink \cite{Lin:2011rlink} & 
		\CIRCLE & 
		\CIRCLE & 
		\Circle & 
		\Circle  \\ 

		Seeing-Is-Believing \cite{Mccune:2005se}  & 
		\RIGHTcircle & 
		\RIGHTcircle & 
		\CIRCLE & 
		\Circle  \\ 
				
		Visible Laser Light \cite{Mayrhofer:2007kd}  & 
		\RIGHTcircle & 
		\CIRCLE & 
		\RIGHTcircle & 
		\Circle  \\ 
			
		VIC (mutual authentication) \cite{Saxena:2011visual} & 
		\RIGHTcircle & 
		\RIGHTcircle & 
		\CIRCLE & 
		\Circle  \\ 
	
		BEDA \cite{Soriente:2008} & 
		\CIRCLE & 
		\RIGHTcircle & 
		\RIGHTcircle & 
		\Circle  \\ 

		Playful Security \cite{Gallego:2011} & 
		\RIGHTcircle & 
		\Circle & 
		\CIRCLE & 
		\Circle  \\ 
				
		Safeslinger \cite{Farb:2013} & 
		\RIGHTcircle & 
		\Circle & 
		\CIRCLE & 
		\Circle  \\ 
		
		Synchronized Drawing \cite{Sethi:2014}  & 
		\CIRCLE & 
		\Circle & 
		\Circle & 
		\Circle  \\ 

		Proximity Authentication \cite{Li:2015dr}  & 
		\CIRCLE & 
		\CIRCLE & 
		\Circle & 
		\Circle  \\ 
				
		Checksum Gestures \cite{Ahmed:2015}  & 
		\RIGHTcircle & 
		\CIRCLE & 
		\Circle & 
		\Circle  \\ 
		\hline
		\hline
	        \\	
		\end{tabularx}
		}
		\center{\CIRCLE\textcolor{white}{-}= explicitly applies; \RIGHTcircle\textcolor{white}{-}= can be applied; \Circle\textcolor{white}{-}= does not apply}
  \end{threeparttable}		
\end{table}

\subsection{Discussion}
Based on the investigation of the application classes, we discuss three
open issues that have not been resolved by the prior research in \gls{sdp}.
First, how the presence of multiple security domains introduces
complications. Second, what privacy issues arise in the respective
application classes. Finally, whether pairing of devices should be valid
indefinitely, or only for a finite time.

\subsubsection{Multiple Security Domains}
Issues arise when pairing devices belong to different security domains.
The goals of two pairing parties, and the assets they protect can vary. This
leads to security, privacy and usability implications that can affect the
adoption of a given pairing scheme. 
For example, in the public application class the infrastructure side can provide
acceptable user experience, and a certain level of security, but ignore users'
privacy. 
Since privacy awareness is growing \cite{Privacy:2016}, many users may be
reluctant to adopt a pairing scheme with such a drawback. 
The opposite situation is also feasible, when the infrastructure side aims to
enhance security and privacy, but this occurs at the expense of usability. 
In this case, users may become confused, as they seek to understand how
pairing works. Such confusion could result in high error rates, that can
negatively affect both security and privacy, as well as jeopardize the
acceptance of the pairing scheme. 
Similarly, in the social application class two users may have completely different
attitudes towards security and privacy. 
Therefore, it should not be assumed that both participants are always
attentive, collaborative and security-motivated. 
A pairing scheme that is designed to operate in the presence of several
security domains should take into account the possible inconsistencies
existing between them, and the impacts that this can have on user behavior and
resulting security.

\subsubsection{Privacy Issues}
Each application class differs from the others in terms of the privacy risks
and their potential impact.  The key privacy issues regarding each application
class are summarized below.
 
The private class is the least problematic, since only a single user is
involved, who directly controls both devices. Therefore, all private
information remains within the sphere of control of the user involved.  
Nonetheless, there exists the potential risk of observation attacks
exfiltrating private information.

The public class introduces the risk of user tracking. 
Consider, for example, a distributed service that allows paying for the petrol
in some area. 
Initially, a user pairs with the terminal on a petrol station. Behind the
scenes, the user is being enrolled in the service, so that she can easily pay
at other stations without the need to pair again. 
This example is both simple and realistic, and would allow the service to
track the users, significantly impacting their privacy.

The social class is exposed to the risk of honest-but-curious participants. 
Such a threat can come in different forms, for example, peeking at another
person's screen or observing her actions, or making a deliberate
mistake to get physical access to the peer pairing device or retrieve extra data. 
None of the surveyed pairing schemes considered this type of attack. This
is, therefore, a topic that justifies attention.

The unattended class is also prone to privacy leakage. 
The surveyed unattended pairing schemes rely on contextual sensing, which was
shown to be plagued with privacy issues \cite{Christin:2011}. 
Since \gls{iot} devices at home or wearables can disclose a great deal of
private information about the user and/or their environment, unattended
pairing schemes must account for privacy protection during pairing. 
This presents, perhaps, the most critical privacy issue uncovered during this
survey. That is, devices which can pair autonomously and may have access
to the considerable amount of private data currently rely on the pairing mechanisms that do not
take privacy into account, and the current state of the art does not yet offer any solution.
 
\subsubsection{Pairing Validity} 
Historically the norm for device pairing has been to establish a ``once and
forever'' pairing.
However, there are good reasons why this is not always the most sensible
approach, when instead the alternative may be more appropriate, that is, a
temporary or transient pairing.
In the private class, once-and-forever makes sense, where, for example, a
user wishes to pair her smartphone with her car's entertainment system.  
In such cases, there exists an expectation of a long-term relationship between
the devices, and that the devices will continue to belong to a single, common
security domain.
In contrast, many pairing scenarios in the public class are more sensibly
handled by creating transient relationships between devices, for example, when
paying for a parking ticket, printing or some other short-lived, transient
activity.
In such situations the devices belong to separate security domains, and the
owner of one device has no control over the behavior of the other, or its
handling of any potentially private data. 
It, therefore, makes no sense for the pairing relationship to endure
indefinitely.
Indeed, there may be additional advantages to transient pairing, for example,
by preventing the user tracking.
An open question is how one should implement short-term pairing in the public
application class.

One approach would be to un-pair the devices after the necessary operation has
been completed. 
However, it should be seamless and require no human effort, otherwise the
usability will be jeopardized. 
Recently, a similar problem was addressed with respect to de-authentication
\cite{Huhta:2015}, exposing the non-triviality of designing such schemes in a
secure way.

Regarding the social class, both transient and long-term pairing may be
applicable, depending on the social context and the amount of trust two people
put into each other. 
For encounters of naturally limited scope or duration, for example, the exchange of
contact details at a conference, pairing two devices permanently may be
excessive.
Furthermore, the level of trust between people can degrade which is another
argument against pairing once-and-forever. 
Short-term pairing can also provide users with better security and privacy
assurances, as the pairing is established only on an as-needed basis. This is
in stark contrast to long-term pairing, which can be abused by another person
or her device, for example, if the other person's device were to be
compromised. 
However, if two users communicate regularly, for example, colleagues, having
to repeatedly pair the same devices may be inconvenient.

Finally, considering the unattended class, the once-and-forever paradigm does
not take into account the highly dynamic nature of \gls{iot} environments. 
In such environments it is already common to pair devices only if they are physically co-located. 
It may, therefore, make sense to un-pair devices whenever they conclude that
they are no longer in close proximity. 
Yet, it remains unclear how to handle such un-pairing events, including how to
determine when confidence of physical proximity reduces such that un-pairing
is justified.

In this section, we have discussed the application classes and provided the classification of existing 
\gls{sdp} schemes. 
In the next section, we outline open research challenges and future perspectives in the field of \gls{sdp}. 

%% file: section/challenges.tex
\section{Future challenges and perspective}
\label{sec: challenges}
In order to design and build viable pairing schemes a wide range of
challenges and open issues need to be resolved.
We discuss several prominent challenges and provide a broad outlook for future research.
We begin by explaining the need for creating adaptable \gls{sdp}
schemes, that are independent of specific \gls{phy} and \gls{hci} channels.  
The importance of including human interaction in the security chain is then
discussed in terms of its potential to improve both security and usability. 
Following this, we explain why it is critical that the design process of a
pairing scheme begins with the target use-case or application class, so
that, again, security and usability can be maximized for each application.
Fourth, we emphasize that \gls{sdp} schemes currently lack ease of
comparability, which hampers evidence-driven improvement of the state of the
art for such pairing schemes.
Finally, we highlight the problem that user privacy is rarely considered by
the current cohort of \gls{sdp} schemes.

\subsection{Adaptable Secure Device Pairing}
As has been shown through the course of this work it is impossible to find a
universal pairing solution.
The selection of both \gls{phy} and \gls{hci} channels highly depends on a
number of factors, including: application classes, the environment and
(social) context, potential attacks, the data to be exchanged and availability
of the channel.
Thus, we argue that future research should be conducted towards a more general
framework for pairing, which would take the aforementioned factors into
account, and develop dynamic and customized pairing schemes built upon
various \gls{phy} and \gls{hci} channels.
In this case, the best security-usability trade-off can be obtained for a
given situation.
Such a framework should offer a higher level of abstraction, which would
account for adding new factors, for example, in a form of ``rules'', that
influence pairing, as well as, \gls{phy} and \gls{hci} channels seamlessly.
Finally, we stress that the current design flow in pairing which starts with
the hardware capabilities should be fundamentally rethought.

\subsection{Including Human Interaction in the Security Chain}
So far, the role of human interaction in \gls{sdp} has not been
fully acknowledged as fundamentally important.
Yet, human interaction is unavoidable in device pairing, for example, when a user
wants to have more control and assurance of the pairing process. 
In our study, we have shown that human interaction can be used to
improve security if properly utilized.
However, users' incentives for pairing, and the common \gls{hci}
practices in pairing have not been well-studied.
Surprisingly, few pairing schemes we reviewed accounted for mitigating user
misbehavior, or actually leveraging human involvement to achieve better
security.
Thus, we advocate for making the \gls{hci} component an indispensable part of
the pairing design and outline several points that are subject to future
investigation.
First, having a continuous and transparent feedback loop between a user and a
pairing mechanism is crucial.
As we stated before, feedback to the user can mitigate many aspects of user
misbehavior.
Also, the prior research relied heavily on human-perceptible \gls{phy}
channels, but the full potential of this property has not been yet realized.
For example, with the feedback loop, both security and usability benefits can
be obtained, such as leveraging user perception to locate the source of the
attack to improve security, or making a human-device link more interactive to
improve usability.
Second, more research on basic user experience and its applicability to
pairing should be carried out to facilitate the creation of more usable and
error-resilient pairing schemes.
Finally, we highlighted several issues with regard to \gls{hci} observation
attacks, however more sophisticated analysis is required to evaluate security
of \gls{hci} channels.    

\subsection{Application Class Driven Design}
Many of the pairing schemes surveyed were designed without a particular
application class or use-case in mind.
However, our findings have shown that each application class has
unique and often highly-divergent security and usability requirements.
Similarly, the sensitivity of the data being exchanged varies considerably
among use-cases \cite{Ion:2010}, ranging from negligible, for example,
exchanging contact information at a conference, to critical, such as performing 
internet banking transactions.
Therefore, it makes sense to begin the design process of a \gls{sdp}
scheme with the target data, use-case, application class in mind.
Only in this way can the resulting design be optimized to the particular needs
and opportunities afforded by the target use-case.
This optimization of the security-usability trade-off is critical to ensure
the best possible outcome.

\subsection{Improving Comparability of \gls{sdp} Schemes}
A sound comparative analysis of different \gls{sdp} schemes was previously
impractical, given the current design approach that starts from hardware
capabilities, instead of the target application class or use-case.
While the contributions of this paper have facilitated comparison of \gls{sdp}
schemes, complications remain, for example, 
 due to the lack of distinction between \gls{phy} and
\gls{hci} channels in most of the \gls{sdp} schemes surveyed.
By shifting the focus to the target use-cases and application classes, it
becomes possible to identify a set of implementation-independent
security and usability metrics. Those metrics could then be used to provide
qualitative or quantitative comparison between different pairing schemes within
an application class.
Building a more generalized attacker model within an application class would
assist in defining such security metrics.
Derivation of specific threat-models for each of the application classes would
be a particularly valuable contribution, as it would allow more objective
assessment and comparison of the security properties of proposed pairing
schemes.

\subsection{Considering User Privacy}
Prior research has not adequately addressed privacy issues in \gls{sdp}.
Increasing numbers of user devices store sensitive information and have
sophisticated sensing capabilities with which many aspects of users' daily
life can be directly measured or inferred \cite{Rachuri:2015}.
Privacy concerns relating to this exist, and attacks that can obtain private
data are feasible in the public, social and unattended application classes. 
Several channels by which users' privacy can be readily violated were revealed
in the process of this survey. 
While not necessarily new information, it is a clear reminder of the attention
required to devise systems that are privacy-preserving.
That is why \gls{sdp} schemes should be designed with user
privacy and the specific target use-cases as the starting
point, rather than physical hardware capabilities or other factors taking the leading role.
Further research is also required to uncover hitherto undetected channels by
which privacy may be violated, so that they can be taken into account in
future \gls{sdp} schemes.

In this section, we have discussed open research challenges and future perspectives in the field of \gls{sdp}. 
In the next section, we provide the concluding remarks of our work.

%% file: section/conclusion.tex
\section{Conclusion}
\label{sec:conclusion}
In this survey, we proposed a system model and consistent terminology to
facilitate meaningful comparison and analysis of \gls{sdp} schemes.
Our system model is based on the three key components drawn from the design
space of \gls{sdp}: physical channels, \gls{hci} channels and application classes.

With regard to \gls{phy} channels, the survey revealed that data confidentially
of the physical medium is very hard to guarantee in practice. Emerging
communication technologies such \gls{vlc} and mm-Waves offer improved security
properties.
Other opportunities arise from the use of sensing of the shared environment by
nearby devices.
 
With regard to \gls{hci} channels, the importance was highlighted of building
pairing schemes resilient to: \textit{(a)} user misbehavior, \textit{(b)}
observation of user actions during the pairing process and \textit{(c)}
honest-but-curious adversaries. It is only when these potential threats are properly
considered, that \gls{hci} channels can play a trusted role in \gls{sdp}
schemes.

We also introduced application classes as a means of classification of \gls{sdp} use-cases.
Through the identification of the target application
class, considerable insight can be gained that can be used to guide the design of 
\gls{sdp} schemes to optimize the security-usability trade-off for a particular use-case.
This stands in contrast to the current practice of beginning with physical
hardware capabilities, instead of with the target use-cases.
This shift to use-case oriented design was also identified as a necessary advance
of the art. 
It is only by making this change, that \gls{sdp} schemes within an application can be better
compared in the future, whether qualitatively or quantitatively, 
allowing for evidence-based design and comparison of \gls{sdp} schemes. 
Until this occurs, \gls{sdp} schemes will likely continue to
fail to address the security, privacy and usability requirements of the
various use-cases.

%% file: section/acknowledgment.tex

\section*{Acknowledgment}
The authors would like to thank Jacqueline Brendel for her insightful discussion on the security of cryptographic protocols used in \gls{sdp}.
This work has been co-funded by the DFG within the CROSSING project and the RTG 2050 ``Privacy and Trust for Mobile User'', the BMBF within the smarter project, the LOEWE initiative (Hessen, Germany) within the NICER project,
and by the BMBF/HMWK CRISP.
Dr. Gardner-Stephen acknowledges the support of the DFG Mercator Fellowship program,
the Shuttleworth Foundation, the NLnet Foundation and the Australian
Department of Foreign Affairs and Trade.